\documentclass[prd,superscriptaddress,twocolumn,aps,floatfix,showpacs,nofootinbib,preprintnumbers]{revtex4-1}
\usepackage{graphicx}
\usepackage{latexsym}
\usepackage{revsymb}
\usepackage{amsfonts}
\usepackage{amsmath}
\usepackage{amssymb}
\usepackage{bm}

\newcommand{\vev}[1]{\langle #1\rangle}
\newcommand{\ket}[1]{\left| #1 \right\rangle}
\newcommand{\bra}[1]{\left\langle #1 \right|}
\newcommand{\braket}[2]{\langle #1 | #2 \rangle}
\newcommand{\Ai}{\mathrm{Ai}}
\newcommand{\Bi}{\mathrm{Bi}}

\begin{document}

\title{Position and momentum uncertainties of a particle in a V-shaped potential\\
under the minimal length uncertainty relation
}

\author{Zachary Lewis}\email{zlewis@vt.edu}
\affiliation{Department of Physics, Virginia Tech, Blacksburg, VA 24061, USA}

\author{Ahmed Roman}\email{mido@vt.edu}
\affiliation{Department of Physics, Virginia Tech, Blacksburg, VA 24061, USA}
\affiliation{Department of Organismic and Evolutionary Biology, Harvard University, Cambridge 02138, MA, USA}

\author{Tatsu Takeuchi}\email{takeuchi@vt.edu}
\affiliation{Department of Physics, Virginia Tech, Blacksburg, VA 24061, USA}
\affiliation{Kavli Institute for the Physics and Mathematics of the Universe (WPI), The University of Tokyo, Kashiwa-shi, Chiba-ken 277-8583, Japan}

\date{\today}

\begin{abstract}
We calculate the uncertainties in the position and momentum of a particle
in the 1D potential $V(x)=F|x|$, $F>0$, when the
position and momentum operators obey the deformed commutation relation 
$[\hat{x},\hat{p}]=i\hbar(1+\beta\hat{p}^2)$, $\beta>0$.
As in the harmonic oscillator case, which was investigated in a previous publication, the Hamiltonian $\hat{H}_1 = \hat{p}^2/2m + F|\hat{x}|$
admits discrete positive energy eigenstates for both positive and negative mass.
The uncertainties for the positive mass states behave as
$\Delta x \sim 1/\Delta p$ as in the $\beta=0$ limit.
For the negative mass states, however, in contrast to the harmonic
oscillator case where we had $\Delta x\sim \Delta p$,
both $\Delta x$ and $\Delta p$ diverge.
We argue that the existence of the negative mass states
and the divergence of their uncertainties can be understood 
by taking the classical limit of the theory.
Comparison of our results is made with previous work by Benczik.
\end{abstract}


\preprint{IPMU13-0139}

\maketitle
\section{Introduction}

One of the expected consequences of quantum gravity is the
deformation of the canonical uncertainty relation between
position and momentum to the form \cite{Amati:1988tn,Witten:2001ib}
\begin{equation}
\Delta x \;\ge\; \frac{\hbar}{2}
             \left( \frac{1}{\Delta p} + \beta\,\Delta p
             \right)\;.
\label{MLUR}
\end{equation}
This relation, called the minimal length uncertainty relation (MLUR) or
the generalized uncertainty relation (GUP) in the literature, 
implies the existence of a minimal length scale
\begin{equation}
\Delta x_{\min} \;=\; \hbar\sqrt{\beta}\;,
\label{MinLength}
\end{equation}
below which the uncertainty in position, $\Delta x$, cannot be reduced.
In the context of quantum gravity $\Delta x_{\min}$ is
identified with the Planck length $\ell_P=\sqrt{\hbar G_N/c^3}$.

The above expectation is based on generic
Heisenberg-microscope-like arguments \cite{Mead:1964zz,Maggiore:1993rv,Maggiore:1993kv,Maggiore:1993zu,Garay:1994en,Adler:1999bu}, which 
demonstrate the impossibility of reducing $\Delta x$ 
below the right-hand-side of Eq.~(\ref{MLUR}).  
Simply put, the gravitational attraction of the probing particle
perturbs the position of the measured particle leading to the
extra uncertainty proportional to $\Delta p$.
While the uncertainties involved in Heisenberg-microscope-like arguments
are distinct from quantum mechanical uncertainties \cite{Ozawa:2003A,Ozawa:2003B,Ozawa:2004},
the latter being independent of any influence of the measurement process,
they do suggest that defining the corresponding physical observables,
spacetime distances in the case of $\Delta x$, to better
accuracy may be conceptually meaningless.  
Thus, the expectation expressed in Eq.~(\ref{MLUR}) may be fairly robust.

Indeed, in string theory, the most prominent candidate theory for
quantum gravity, Eq.~(\ref{MLUR}) has been found to hold in
perturbative string-string scattering amplitudes \cite{Gross:1987kza,Gross:1987ar,Amati:1987wq,Amati:1987uf,Konishi:1989wk}
where $\Delta x_{\min}$ is identified with the string length scale
$\ell_s =\sqrt{\alpha'}$.
It should be noted, however, that within string theory, D-brane scattering
can probe distances shorter than the string scale \cite{Polchinski:1995mt,Douglas:1996yp},
and non-perturbative effects could also modify Eq.~(\ref{MLUR}) \cite{Yoneya:2000bt}.
See Ref.~\cite{Hossenfelder:2012jw} for a review of 
the various arguments and studies which either support or suggest modifications 
to Eq.~(\ref{MLUR}).
Such modifications are to be expected of a full theory of quantum gravity,
given that Eq.~(\ref{MLUR}) is clearly non-relativistic.

Assuming that quantum gravity would lead to Eq.~(\ref{MLUR})
in the non-relativistic regime, 
the relation demands that that the canonical commutation relation between
the position and momentum operators in quantum mechanics also be modified to
reflect the existence of the minimal length, e.g.
\begin{equation}
[\, \hat{x},\, \hat{p} \,] \;=\; i\hbar (1 + \beta \hat{p}^2) \;,\qquad\beta\,>\,0\;.
\label{CommutationRelation}
\end{equation}
The consequences of both Eqs.~(\ref{MLUR}) and (\ref{CommutationRelation}),
and also their various modifications,
have been studied by many authors in many different contexts
and a vast literature on the subject exists \cite{Kempf:1994su,Hinrichsen:1995mf,Kempf:1996ss,Kempf:1996mv,Kempf:1996nk,Kempf:1996fz,Kempf:1999xt,Brau:1999uv,Brau:2006ca,Chang:2001kn,Chang:2001bm,Benczik:2002tt,Benczik:2002px,Chang:2010ir,Chang:2011jj,Benczik:2005bh,Benczik:2007we,Lewis:2011fg,Brout:1998ei,Scardigli:1999jh,Scardigli:2003kr,Adler:2001vs,Custodio:2003jp,Hossenfelder:2003jz,Harbach:2003qz,Harbach:2005yu,Hossenfelder:2004up,Hossenfelder:2005ed,Hossenfelder:2006cw,Hossenfelder:2007re,AmelinoCamelia:2005ik,Nouicer:2005dp,Nouicer:2006xua,Nouicer:2007jg,Setare:2005sj,Moayedi:2010vp,Moayedi:2013nxa,Moayedi:2013nba,Fityo:2005xaa,Fityo:2005lwa,Fityo:2008xpa,Fityo:2008zz,Stetsko:2006jna,Stetsko:2007ze,Stetsko:2007yaa,Quesne:2006fs,Quesne:2006is,Quesne:2009vc,Maslowski:2012aj,Zhao:2006ri,Vakili:2007yz,Vakili:2008zg,Vakili:2008tt,Vakili:2012qt,Battisti:2007jd,Battisti:2007zg,Kim:2007hf,Das:2008kaa,Das:2010sj,Das:2009hs,Ali:2009zq,Das:2010zf,Basilakos:2010vs,Ali:2011fa,Bagchi:2009wb,Fring:2010pw,Fring:2010pn,Dey:2012dm,Dey:2012tv,Dey:2012tx,Dey:2012pq,Dey:2013hda,Myung:2009ur,Banerjee:2010sd,Ghosh:2012cv,
Pedram:2010hx,Pedram:2010zz,Pedram:2012ub,Pedram:2012zz,Pedram:2011gw,Pedram:2012my,Pedram:2011aa,Pedram:2012dm,Nozari:2011gj,Nozari:2012gd,
Nozari:2010qy,Pedram:2011xj,Pedram:2012np,Kober:2010sj,Bina:2010ir,Rashidi:2012cj,Menculini:2013ida,Frassino:2011aa,Hassanabadi:2013mq,Hassanabadi:2013jga}.
The hope is that such endeavors would shed light on how quantum gravity may manifest itself in the infrared, and provide us with observable handles on the existence of 
the fundamental length scale.

\begin{figure}[t]
\includegraphics[width=8cm]{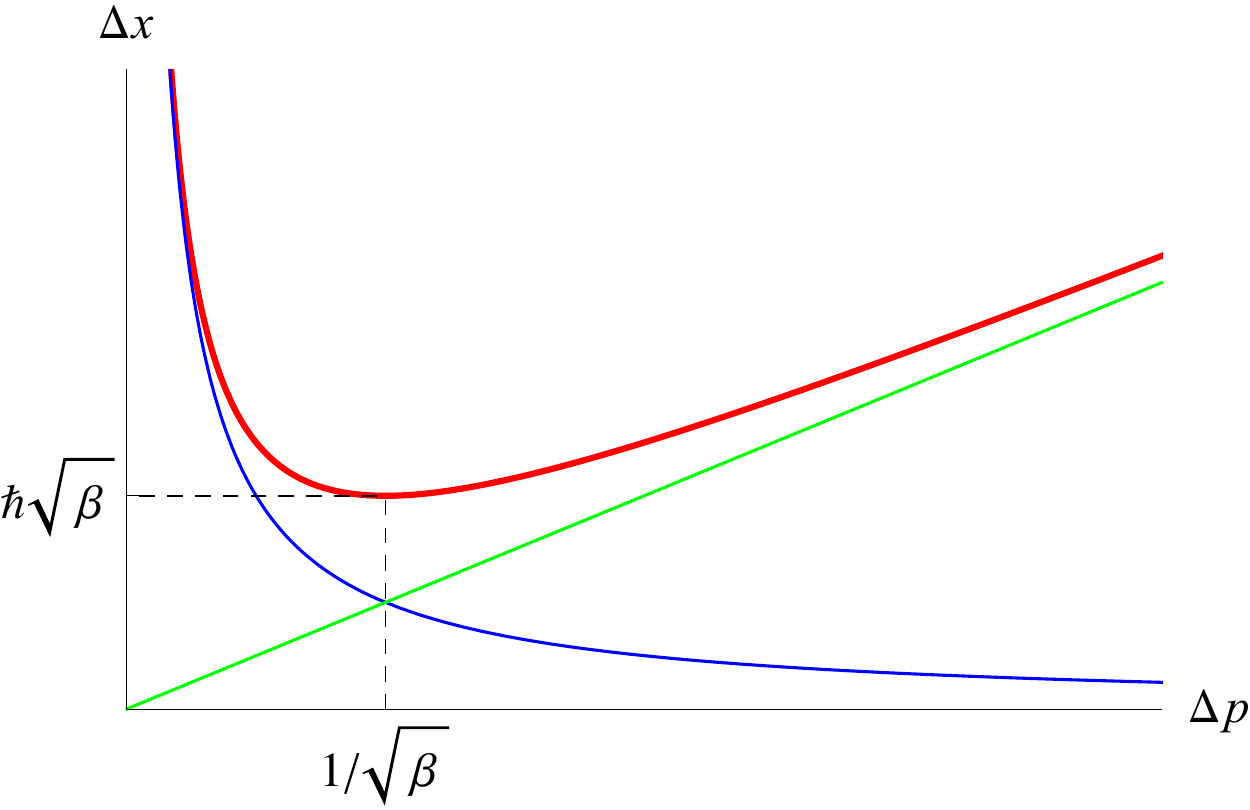}
\caption{The Minimal Length Uncertainty Relation.}
\label{Fig01}
\end{figure}

Despite the volume of works on this subject, to our knowledge, few actually consider how
$\Delta x$ and $\Delta p$ would behave in a deformed
quantum mechanics obeying Eq.~(\ref{CommutationRelation}).
Note that the uncertainty in position $\Delta x$ which saturates the equality in Eq.~(\ref{MLUR}) for a given $\Delta p$
behaves as $\Delta x \sim 1/\Delta p$ for $\Delta p < 1/\sqrt{\beta}$, 
while $\Delta x \sim \Delta p$ for $\Delta p > 1/\sqrt{\beta}$,
as shown in FIG.~\ref{Fig01}.
$\Delta x \sim 1/\Delta p$ is the standard behavior seen in
canonical quantum mechanics, while the $\Delta x \sim\Delta p$
behavior would be quite novel. 
Such behavior could be indicative of quantum gravitational effects and
it behooves us to understand when and how it would set in.

In a previous paper \cite{Lewis:2011fg}, we looked at a particle in a harmonic oscillator
potential, \textit{i.e.}
\begin{equation}
\hat{H}_2 \;=\; \dfrac{\hat{p}^2}{2m} + \dfrac{1}{2}k\hat{x}^2\;,\qquad k\;>\;0\;,
\label{H2}
\end{equation}
where $\hat{x}$ and $\hat{p}$ obeyed Eq.~(\ref{CommutationRelation}),
to see how the $\Delta x\sim\Delta p$ behavior may come about.
There, it was discovered that:
\begin{enumerate}
\item When Eq.~(\ref{CommutationRelation}) is assumed, the harmonic oscillator
Hamiltonian, Eq.~(\ref{H2}), admits an infinite ladder of eigenstates with 
discrete positive eigenvalues not only when the mass $m$ is
positive, but also when the mass $m$ is negative provided that
\begin{equation}
\hbar\sqrt{\beta} \;>\; \sqrt{2}\left[-\dfrac{\hbar^2}{km}\right]^{1/4}\;.
\label{HOcondition}
\end{equation}

\item The uncertainties in position and momentum of the energy eigenstates
behave as $\Delta x\sim 1/\Delta p$ in the positive mass case, while
the $\Delta x \sim \Delta p$ behavior is observed in the negative mass case.

\item The negative mass case effectively inverts the harmonic oscillator potential
and the particle is allowed to zoom off to infinity when the system is treated classically.
However, the time it takes for the particle to travel back and forth between the turning points and infinity is finite (\textit{i.e.} non-infinite). Consequently, 
the amount of time that the particle spends near the turning points is also finite (\textit{i.e.} non-zero),
making bound states with discrete energy eigenvalues possible.

\end{enumerate}
Thus, seeing the $\Delta x\sim \Delta p$ behavior in the harmonic oscillator
required the mass $m$ to be negative, and the particle to be able to reach
super-luminal speeds in the classical limit.
The latter, or course, is forbidden in relativistic contexts,
but given that Eq.~(\ref{CommutationRelation}) is non-relativistic, it may not be
hardly surprising.
It does, however, bring into question whether Eq.~(\ref{CommutationRelation}) correctly accounts for quantum gravitational effects in the infrared, and further investigation 
of the $\Delta x\sim \Delta p$ behavior is called for.
In particular,
a natural question which arises from the above results is whether
similar properties can be observed universally for particles in other potentials as well.

In this paper, we will look at a particle in a 1D 
left-right symmetric V-shaped potential
\begin{equation}
\hat{H}_{1} \;=\; \dfrac{\hat{p}^2}{2m} + F|\hat{x}|\;,\qquad F\,>\,0\;,
\label{H1}
\end{equation}
where $\hat{x}$ and $\hat{p}$ obey Eq.~(\ref{CommutationRelation}).
As in Ref.~\cite{Lewis:2011fg},
we will allow the particle mass $m$ to be either positive or negative.
The operator $|\hat{x}|$ is defined via its action on the eigenstates of $\hat{x}^2$:
for an eigenstate of $\hat{x}^2$ with eigenvalue $\sigma^2$ ($\sigma>0$),
\textit{i.e.}
$\hat{x}^2\ket{\sigma^2}\;=\;\sigma^2\ket{\sigma^2}$,
the action of $|\hat{x}|$ on $\ket{\sigma^2}$ is given by
\begin{equation}
|\hat{x}|\ket{\sigma^2}\;=\;\sigma\ket{\sigma^2}\;.
\label{absxdef}
\end{equation}

Note that 
in the infinite mass limit, $|m|\rightarrow\infty$,
the Hamiltonian $\hat{H}_1$ reduces to
$F|\hat{x}|$, so the eigenstates of $\hat{H}_1$
would reduce to the eigenstates of $|\hat{x}|$,
which are simultaneously eigenstates of $\hat{x}^2$.
So for infinite mass, the eigenstates of $\hat{H}_1$ will
be simply those of $\hat{x}^2$.
For finite mass, we can expand the
eigenstates of $\hat{H}_1$ in terms of the eigenstates of $\hat{x}^2$, 
with the eigenvalues determined from the requirement that
the states be normalizable.
This is the main method used in this paper to determine the eigenvalues and eigenstates of
$\hat{H}_1$, and calculate the uncertainties $\Delta x$ and $\Delta p$ for those states.

Recall, however, that this is not the standard method used in the canonical $\beta=0$ case.
There, the eigenvalues and eigenstates of $\hat{H}_{1}$ are obtained by solving the
Schr\"odinger equation for
\begin{equation}
\hat{H}_{1}'\;=\;\dfrac{\hat{p}^2}{2m}+F\hat{x}
\label{H1prime}
\end{equation}
in the region $x\ge 0$,
and then imposing the boundary condition
$\psi'(0)=0$ or $\psi(0)=0$ at $x=0$
to obtain the parity even and parity odd states, respectively.
We find that a similar technique works in the $\beta\neq 0$ case
for the parity odd states, but not for the parity even states
partly due to the difficulty in identifying what is meant by `the derivative of the wave-function
at $x=0$' when $\Delta x_{\min}=\hbar\sqrt{\beta}$ is non-zero. 
Note also that the location of the boundary at $x=0$ itself is blurred out 
in the presence of a minimal length.
Since the odd-parity wave-functions vanish at $x=0$ whereas the even-parity ones do not,
the odd-parity states are less sensitive to this blurring out than the even-parity ones.

The parity odd eigenstates of $\hat{H}_{1}$
are essentially the same as those
considered previously by several authors \cite{Brau:2006ca,Benczik:2007we,Nozari:2010qy}
in the context of applying Eq.~(\ref{CommutationRelation}) to a particle in the potential
\begin{equation}
V(x) \;=\;
\begin{cases}
Fx\;\; & \mbox{for $x>0$} \\ \infty & \mbox{for $x\le 0$}
\end{cases} 
\;,\qquad F\,>\,0\;.
\label{HalfPotential}
\end{equation}
This system would correspond to a particle bouncing in a uniform gravitational
field in which $F=mg$ with a rigid floor at $x=0$, and can, in principle, be compared to experimental
results \cite{Nesvizhevsky:2002ef,Nesvizhevsky:2003ww,Nesvizhevsky:2005ss} to constrain the deformation parameter $\beta$.
We will be utilizing some of these previous results, in particular that of
Benczik \cite{Benczik:2007we}.

This paper is organized as follows.
In section~\ref{sec2}, we set up the Schr\"odinger equation for $\hat{H}_1$,
and solve it by expanding the eigenstates of $\hat{H}_1$
in terms of the eigenstates of $\hat{x}^2$.
It is discovered that, just as in the harmonic oscillator case,
energy eigenstates with discrete positive energy eigenvalues exist for both
the positive and negative mass cases.
The uncertainties in position and momentum, $\Delta x$ and $\Delta p$,
are calculated for these states and we find that $\Delta x \sim 1/\Delta p$
in the positive mass case, but
both $\Delta x$ and $\Delta p$ are divergent in the negative mass case.
In section~\ref{alternative}, we approach the problem from a different angle
by solving the Schr\"odinger equation for $\hat{H}_1'$ directly in terms of the Bateman function \cite{Bateman:1931,Wolfram}.
It is found that for the odd-parity states
the energy eigenvalues found in Section~\ref{sec2} 
agree with those obtained by demanding that the wave-function vanish at $x=0$.
On the other hand, for the even-parity states the
energy eigenvalues from Section~\ref{sec2}
do not agree with those obtain by demanding 
that the derivative of the wave-function vanish at $x=0$,
except for the higher excited states.
In section~\ref{classical}, we consider the classical limit of
the problem and find the classical trajectory of the negative mass particle as
well as the corresponding classical probability distributions of finding the particle at a particular point in $x$- and $p$-spaces.
It is found that the 1st and 2nd moments of these probability distributions
diverge, indicating the divergence of $\Delta x$ and $\Delta p$ 
in the classical limit also.
Section~\ref{conclusions} concludes with 
a summary of our results and some discussion on what they could mean.

\section{Expansion in the Eigenstates of $\bm{\hat{x}^2}$}
\label{sec2}

\subsection{Representations of $\bm{\hat{x}}$ and $\bm{\hat{p}}$}

The position and momentum operators obeying Eq.~(\ref{CommutationRelation})
can be represented in momentum space by \cite{Kempf:1994su}
\begin{eqnarray}
\hat{x} & = & i\hbar\,(1+\beta p^2)\,\frac{d}{dp}\;,  \cr
\hat{p} & = & p\;,
\label{Rep1D}
\end{eqnarray}
with the inner product between two states given by
\begin{equation}
\langle f | g \rangle
\;=\; \sqrt{\beta}\int_{-\infty}^{\infty} \frac{dp}{(1 + \beta p^2)}\;\tilde{f}^*(p)\,\tilde{g}(p)\;.
\label{Product1D}
\end{equation}
Here, the overall factor of $\sqrt{\beta}$ is introduced to render the wave-functions dimensionless,
while the weight $1/(1+\beta p^2)$ is necessary for the symmetricity of the operator $\hat{x}$.
The wave-functions are assumed to vanish as $p\rightarrow\pm\infty$.

It is useful to introduce the dimensionless variable
\begin{equation}
\theta \;\equiv\; \arctan(\sqrt{\beta}p) \;,
\label{thetadef}
\end{equation}
which maps the region $-\infty < p < \infty$ to
\begin{equation}
-\frac{\pi}{2} < \theta < \frac{\pi}{2} \;,
\label{rhoRange}
\end{equation}
and casts 
the $\hat{x}$ and $\hat{p}$ operators into the forms
\begin{eqnarray}
\hat{x} & = & i\hbar\sqrt{\beta}\,\dfrac{d}{d\theta} \;=\; \Delta x_{\min}\,i\dfrac{d}{d\theta}\;,\cr
\hat{p} & = & \dfrac{1}{\sqrt{\beta}}\tan\theta \;,
\label{xptheta}
\end{eqnarray}
with inner product given by
\begin{equation}
\langle f | g \rangle
\;=\; 
\int_{-\pi/2}^{\pi/2} d\theta\;\tilde{f}^*(\theta)\,\tilde{g}(\theta)\;.
\end{equation}
As in the $p$-representation,
we require the wave-functions to vanish at the domain boundaries
$\theta=\pm\pi/2$.

\subsection{The Eigenstates of $\bm{\hat{x}}$ and the Maximally Localized States}

Note that the necessary condition for the operator $\hat{x}$ 
in the $\theta$-representation to be symmetric is given by
\begin{equation}
\Bigl[ f^*(\theta) g(\theta) \Bigr]_{-\pi/2}^{\pi/2} \;=\; 0
\;.
\end{equation}
This would hold if all the wave-functions vanished at $\theta=\pm\pi/2$
as assumed above, or if the wave-functions that are non-zero at $\theta=\pm\pi/2$ satisfied
the boundary condition 
\begin{equation}
f(-\pi/2) \;=\; e^{i\delta} f(\pi/2)\;,
\label{PeriodicityCondition}
\end{equation}
where $\delta\in[-\pi,\pi)$ is a phase common to all such wave-functions.

If we allow for Eq.~(\ref{PeriodicityCondition}) with $\delta$ fixed, 
the operator $\hat{x}$ has eigenfunctions given by
\begin{equation}
\tilde{\phi}_{2z+\lambda}(\theta) 
\;=\; \dfrac{1}{\sqrt{\pi}}\;e^{-i(2z+\lambda)\theta}
\;=\; \dfrac{1}{\sqrt{\pi}}\;e^{-i\lambda\theta}e^{-i(2z)\theta}
\;,
\end{equation}
with eigenvalue $x=x_z(\lambda)\equiv(2z+\lambda)\Delta x_{\min}$, where $z\in\mathbb{Z}$
and $\lambda\equiv\delta/\pi\in[-1,1)$.
Since $\lambda=\delta/\pi$ is arbitrary, all values of $x$ are possible, except for
each choice of $\lambda$ the eigenvalues are discrete and separated by 
$2\Delta x_{\min}$ steps, reflecting the existence of the minimal length.\footnote{%
In the language of Kempf in Ref.~\cite{Kempf:1999xt}, the
eigenvalues $x_z(\lambda)$, $z\in\mathbb{Z}$ for each value of $\lambda$ provides a 
\textit{discretization} of the
$x$-axis, and the collection of all discretizations $\{x_z(\lambda)\}$,
$\lambda\in[-1,1)$ provides a \textit{partitioning} of the $x$-axis.}
For $\lambda=0$ the eigenvalues are even-integer multiples of $\Delta x_{\min}$,
while for $\lambda=-1$ the eigenvalues are odd-integer multiplies of $\Delta x_{\min}$.

A formal calculation of $\Delta x$ and $\Delta p$ for 
$\tilde{\phi}_{2z+\lambda}(\theta)$ yields
$\Delta x=0$ and $\Delta p=\infty$, which indicates that these
states do not satisfy Eq.~(\ref{MLUR}) and are thus `unphysical.'
Nevertheless, each set of these eigenfunctions sharing a common $\lambda$ are orthonormal, 
\begin{equation}
\braket{\phi_{2z+\lambda}}{\phi_{2z'+\lambda}}
\;=\; \delta_{zz'}\;,
\end{equation}
and complete.
That is, any well behaved wave-function $\tilde{f}(\theta)$ in the interval $\theta\in[-\pi/2,\pi/2]$ can be
expanded as
\begin{equation}
\tilde{f}(\theta)
\;=\; \sum_{z=-\infty}^{\infty} c_{2z+\lambda}\,\tilde{\phi}_{2z+\lambda}(\theta)\;,
\end{equation}
where
\begin{equation}
c_{2z+\lambda} 
\;=\; \braket{\phi_{2z+\lambda}}{f}
\;=\; \int_{-\pi/2}^{\pi/2} \tilde{\phi}_{2z+\lambda}^{*}(\theta)\,
\tilde{f}(\theta)\,d\theta
\;.
\end{equation}
Furthermore, it is straightforward to show that
\begin{equation}
\dfrac{\bra{f}\hat{x}^n\ket{f}}{\;\;(\Delta x_{\min})^n}
\;=\; \sum_{z\in\mathbb{Z}}
(2z+\lambda)^n |c_{2z+\lambda}|^2
\;,
\end{equation}
and that the sum on the right-hand-side is independent of the
choice of $\lambda$.
We can therefore interpret the coefficient $c_{2z+\lambda}=\braket{\phi_{2z+\lambda}}{f}$
as the probability amplitude for obtaining $2z+\lambda$ when $\hat{x}/\Delta x_{\min}$
is measured on the state $\ket{f}$.
Thus, the Fourier transform of the $\theta$-space wave-function
to $x/\Delta x_{\min}$-space
has physical meaning, despite the fact that the eigenstates of $\hat{x}$
are `unphysical.'

It has been suggested in Ref.~\cite{Kempf:1994su} 
that the `unphysical' eigenstates of $\hat{x}$ should be replaced by
the `maximally localized states,'
which in the $\theta$-representation are given by
\begin{eqnarray}
\tilde{\phi}_{2z+\lambda}^{ml}(\theta) 
& = & \sqrt{\dfrac{2}{\pi}}\;\cos\theta\;e^{-i(2z+\lambda)\theta}
\cr
& = & \dfrac{1}{\sqrt{2}}
\left[
 \tilde{\phi}_{2z+\lambda+1}(\theta)
+\tilde{\phi}_{2z+\lambda-1}(\theta)
\right]
\;.
\end{eqnarray}
Note that these functions vanish at $\theta=\pm\pi/2$.
For these states we have
$\Delta x=\Delta x_{\min}$ and $\Delta p=1/\sqrt{\beta}$,
so they are `physical' and `maximally localized.'
They can be used to expand the wave-function $\tilde{f}(\theta)$ as
\begin{equation}
\tilde{f}(\theta)
\;=\; \sum_{z=-\infty}^{\infty} c_{2z+\lambda}^{ml}\,\tilde{\phi}_{2z+\lambda}^{ml}(\theta)\;,
\end{equation}
provided that $\tilde{f}(\theta)/\cos\theta$ is well-behaved at $\theta=\pm\pi/2$.
However, the states with a common value of $\lambda$ are not orthonormal,
their inner product being given by
\begin{equation}
\braket{\phi_{2z+\lambda}^{ml}}{\phi_{2z'+\lambda}^{ml}}
\;=\; 
\delta_{zz'} + \dfrac{1}{2}\left(\delta_{z,z'+1}+\delta_{z,z'-1}\right)
\;.
\end{equation}
Consequently, the expansion coefficients $c_{2z+\lambda}^{ml}$ are not given by
$\braket{\phi_{2z+\lambda}^{ml}}{f}$ but by
\begin{equation}
c_{2z+\lambda}^{ml}
\;=\; \dfrac{1}{\sqrt{2\pi}}\int_{-\pi/2}^{\pi/2} 
\dfrac{d\theta}{\cos\theta}\;
e^{i(2z+\lambda)\theta}\,
\tilde{f}(\theta)
\;.
\end{equation}
Furthermore,
neither $c_{2z+\lambda}^{ml}$ nor $\braket{\phi_{2z+\lambda}^{ml}}{f}$
have any simple interpretation as a probability amplitude.
Indeed, the simplest way to utilize these coefficients would be
to recover the usual coefficients for the
expansion in $\tilde{\phi}_{2z+\lambda}(\theta)$ via
\begin{equation}
c_{2z+\lambda} \;=\;
\dfrac{1}{\sqrt{2}}
\left(
 c_{2z+\lambda+1}^{ml}
+c_{2z+\lambda-1}^{ml}
\right)
\;.
\end{equation}
Due to these complications, we refrain from using
these maximal localized states.

\subsection{The Schr\"odinger Equation}

Using the representations of $\hat{x}$ and $\hat{p}$ in Eq.~(\ref{xptheta}), 
the Schr\"odinger equation for $\hat{H}_1$ in $\theta$-space is given by 
\begin{equation}
\left(\dfrac{1}{2m\beta}\tan^2\theta
+ F\hbar\sqrt{\beta}\sqrt{-\dfrac{d^2}{d\theta^2}}
\right)\tilde{\psi}(\theta) 
\;=\; E\,\tilde{\psi}(\theta)\;.
\label{schrodingertheta1}
\end{equation}
There exist two characteristic lengths scales in this equation, namely the minimal length
$\Delta x_{\min}=\hbar\sqrt{\beta}$, and
\begin{equation}
a \;\equiv\; \left[\dfrac{\hbar^2}{2|m|F}\right]^{1/3}\;.
\label{adef2}
\end{equation}
The length scale
$a$ survives in the limit $\beta\rightarrow 0$ in which the canonical
commutation relation between $\hat{x}$ and $\hat{p}$ is recovered.
On the other hand,
$\Delta x_{\min}$ survives in the limit $|m|\rightarrow\infty$ in which
$a\rightarrow 0$.
Let us call the ratio of the two
\begin{equation}
\kappa \;\equiv\; \dfrac{\hbar\sqrt{\beta}}{a}\;=\; \dfrac{\Delta x_{\min}}{a}\;.
\end{equation}
Using $\kappa$, Eq.~(\ref{schrodingertheta1}) can be rewritten as
\begin{equation}
\left( \pm\dfrac{1}{\kappa^3}\tan^2\theta 
+ \sqrt{-\dfrac{d^2}{d\theta^2}}
\right) \tilde{\psi}(\theta)
\;=\; \varepsilon_\beta\,\tilde{\psi}(\theta)
\;,
\label{schrodingertheta2}
\end{equation}
where the sign in front of the $\tan^2\theta$ term indicates
the sign of the mass $m$, and
\begin{equation}
\varepsilon_\beta 
\;\equiv\; \dfrac{E}{F\hbar\sqrt{\beta}} 
\;=\; \dfrac{E}{F\Delta x_{\min}}
\;,
\end{equation}
that is, $\varepsilon_\beta$ is $E$ in units of $F\Delta x_{\min}$.
Another normalization of the eigenvalue we will be using is
\begin{equation}
\varepsilon_a \;\equiv\;
\kappa\varepsilon_\beta
\;=\; \dfrac{E}{Fa}
\;,
\end{equation}
that is, $\varepsilon_a$ is $E$ in units of $Fa$.

\subsection{The Expansion}

We expand the solution to Eq.~(\ref{schrodingertheta2}) 
in terms of the eigenstates of the operator
\begin{equation}
\hat{x}^2 \;=\; -(\Delta x_{\min})^2\dfrac{d^2}{d\theta^2}\;.
\end{equation}
Demanding that the wave-functions vanish at $\theta=\pm\pi/2$, which
correspond to $p=\pm\infty$, we find that the eigenvalues of $\hat{x}^2$ are 
$(n\,\Delta x_{\min})^2$, $n\in\mathbb{N}$,
with the $n$-th eigenstate given by
\begin{equation}
\tilde\varphi_n(\theta) 
\;=\; 
\sqrt{\dfrac{2}{\pi}}\,(-1)^{\left\lfloor\!\frac{n+1}{2}\!\right\rfloor} \times
\begin{cases}
\cos n\theta & \mbox{if $n$ odd}\;, \\
\sin n\theta & \mbox{if $n$ even}\;. 
\end{cases}
\end{equation}
Note that in terms of the eigenfunctions of $\hat{x}$, 
$\tilde{\varphi}_n(\theta)$ is a superposition of
$\tilde{\phi}_{n}(\theta)$ and $\tilde{\phi}_{-n}(\theta)$ with equal amplitude.
The above choice of sign allows us to write both the odd and even cases together as
\begin{equation}
\tilde\varphi_n(\theta) 
\;=\; 
\sqrt{\dfrac{2}{\pi}}\,\cos\theta\,U_{n-1}(\sin\theta)\;,
\end{equation}
where $U_{n-1}$
is the Chebyshev polynomial of the second kind \cite{SpecialFunctions1,SpecialFunctions2}:
\begin{equation}
U_{n-1}(\cos \xi) \;=\; \dfrac{\sin n\xi}{\sin \xi}\;,\qquad
n\;\in\;\mathbb{N}\;.
\end{equation}
The recursion relation for the Chebyshev polynomials
\begin{equation}
U_{n+1}(s)-2s\,U_n(s)+U_{n-1}(s) \;=\; 0
\end{equation}
allows us to write
\begin{equation}
\tilde\varphi_n(\theta)\sin\theta
\;=\; \dfrac{1}{2}\Bigl[\tilde\varphi_{n+1}(\theta) + \tilde\varphi_{n-1}(\theta)\Bigr] \;,
\end{equation}
which upon iteration gives us
\begin{equation}
\tilde\varphi_n(\theta)\sin^2\theta
\;=\; \dfrac{1}{4}\Bigl[\tilde\varphi_{n+2}(\theta) + 2\tilde\varphi_{n}(\theta) + \tilde\varphi_{n-2}(\theta)\Bigr] \;.
\label{phiRecursion}
\end{equation}
This relation will prove useful below.

Since $\hat{x}^2\tilde\varphi_n \;=\; (n\,\Delta x_{\min})^2\tilde\varphi_n$,
the action of the operator
\begin{equation}
|\hat{x}|
\;=\;\sqrt{\hat{x}^2}
\;=\; (\Delta x_{\min})\sqrt{-\dfrac{d^2}{d\theta^2}}
\label{xabsdef}
\end{equation}
on these states is given by
\begin{equation}
|\hat{x}|\,\tilde\varphi_n \;=\; (n\,\Delta x_{\min})\,\tilde\varphi_n\;.
\label{xabsonphi}
\end{equation}
Let
\begin{equation}
\tilde\psi(\theta) \;=\; \sum_{k=1}^\infty c_k\,\tilde\varphi_k(\theta)\;.
\end{equation}
Substituting this expansion into Eq.~(\ref{schrodingertheta2}) and using 
Eq.~(\ref{xabsonphi}) and the
recursion relation Eq.~(\ref{phiRecursion}), 
we find the following relations among the expansion coefficients:
\begin{eqnarray}
0 & = & \left(\pm\kappa^{-3}-B_3\right)c_3 + \left(\pm\kappa^{-3}+3B_1\right)c_1\;,\cr
0 & = & \left(\pm\kappa^{-3}-B_4\right)c_4 + 2\left(\pm\kappa^{-3}+B_2\right)c_2\;,\cr
0 & = & \left(\pm\kappa^{-3}-B_{k+2}\right)c_{k+2}  \cr
& & +\,2\left(\pm\kappa^{-3}+B_k\right)c_k 
+\left(\pm\kappa^{-3}-B_{k-2}\right)c_{k-2}\;,\quad(k\ge 3)\cr
& &
\label{recursion1}
\end{eqnarray}
where
\begin{equation}
B_k
\;\equiv\;
k-\varepsilon_\beta
\;.
\end{equation}
As can be seen, the odd and even coefficients in the expansion decouple as they should
since the odd wave-functions being cosines correspond to a parity even solution in $x$-space, 
and the even wave-functions being sines correspond to a parity odd solution in $x$-space.
The eigenvalues, $\varepsilon_\beta$, are determined by the condition
\begin{equation}
\sum_{k=1}^\infty |c_{k}|^2 \;=\;\mathrm{finite}\;.
\label{finitesum}
\end{equation}
%

\subsection{Negative mass case}

We begin with the negative mass case for which we were able to find exact solutions.  
We will elaborate on how we found these solutions later.

The eigenvalues are all positive and discrete, and are given by
\begin{equation}
\varepsilon_{\beta,n}^{(-)} \;=\; n + \dfrac{1}{\kappa^3}\;,\qquad
n\in\mathbb{N}\;,
\label{NegativeMassGuess}
\end{equation}
with odd $n$ corresponding to the even parity solutions
and even $n$ corresponding to the odd parity solutions.
Note that these eigenvalues are evenly spaced.
Thus, this characteristic is not exclusive to the canonical harmonic oscillator.
We will see another parallel with the
canonical harmonic oscillator when we discuss the classical limit of
our model in section~\ref{classical}.

When $n=2s-1$, $s\in\mathbb{N}$, 
the recursion relations for the odd coefficients with $\varepsilon_\beta$
set to $\varepsilon_{\beta,2s-1}^{(-)}$ become
\begin{eqnarray}
0 & = & (2-s)\,c_3 +[\,2\kappa^{-3}-3(1-s)\,]\,c_1
\cr
0 & = & [\,(j-s)+1\,]\,c_{2j+1} \cr
& & 
+2\,[\,\kappa^{-3}-(j-s)\,]\,c_{2j-1}
+[\,(j-s)-1\,]\,c_{2j-3}
\;,
\cr
& &
\label{recursionNegativeMassOddCoefs}
\end{eqnarray}
while for the $n=2s$, $s\in\mathbb{N}$, case
the recursion relations for the even coefficients with $\varepsilon_\beta$
set to $\varepsilon_{\beta,2s}^{(-)}$ become
\begin{eqnarray}
0 & = & (2-s)\,c_4 + [\,2\kappa^{-3}-2(1-s)\,]\,c_2\;,\cr
0 & = & [\,(j-s)+1\,]\,c_{2(j+1)} \cr
& & 
+2\,[\,\kappa^{-3}-(j-s)\,]\,c_{2j}
+[\,(j-s)-1\,]\,c_{2(j-1)}
\;.
\cr
& &
\label{recursionNegativeMassEvenCoefs}
\end{eqnarray}
Except of the coefficient of $(1-s)$ in the first lines, the
recursion relations are identical for the odd and even numbered coefficients.

The solutions to the above recursion relations can be written in terms of the
Bateman function, which was defined in Ref.~\cite{Bateman:1931} as
\begin{equation}
k_\nu(\mu)
\;\equiv\; \dfrac{2}{\pi}
\int_0^{\pi/2} \cos\left(\mu\tan\theta - \nu\,\theta\right)\,d\theta
\;.
\label{BatemanFunction}
\end{equation}
Note that this function is real for real $\mu$ and $\nu$.
We will also denote $k_\nu(\mu)$ as $k(\mu,\nu)$ when convenient.
The Bateman function with negative even-integer indices 
are identically zero, 
\begin{equation}
k_{-2n}(\mu) \;=\; 0\;,\qquad n\in\mathbb{N}\;,
\label{NegativeEvenBateman}
\end{equation}
while those with non-negative even-integer indices
appear in the following Fourier series \cite{Bateman:1931}:
\begin{equation}
e^{i\mu\tan\theta} \;\equiv\; \sum_{t=0}^{\infty} k_{2t}(\mu) \,e^{2it\theta}\;.
\label{exptanFS}
\end{equation}
Using the Bateman function,
the solutions to Eqs.~(\ref{recursionNegativeMassOddCoefs}) and 
(\ref{recursionNegativeMassEvenCoefs}) are respectively given by
\begin{equation}
c_{2j-1} \;=\; 
(-1)^j k_{2(j-s)}(\kappa^{-3}) 
\;,
\label{c2jminusonecases}
\end{equation}
and
\begin{equation}
c_{2j} \;=\; 
(-1)^j k_{2(j-s)}(\kappa^{-3}) 
\;.
\label{c2jcases}
\end{equation}
Note that due to Eq.~(\ref{NegativeEvenBateman}), the non-zero
coefficients start from the $j=s$ terms.
The non-zero coefficients from $j=s$ onwards are the same for
both the $n=2s-1$ and $n=2s$ cases, depending only on $j-s$ and $\kappa$.
Note also, that using Eq.~(\ref{NegativeMassGuess}),
the two cases can be written as
\begin{eqnarray}
c_{2j-1} & = & (-1)^j\;
k\left(\dfrac{1}{\kappa^3},\dfrac{1}{\kappa^3}+
\Bigl[\,(2j-1)-\varepsilon_{\beta,2s-1}^{(-)}\,\Bigr]\right)
\;,\cr
c_{2j} & = & (-1)^j\;
k\left(\dfrac{1}{\kappa^3},\dfrac{1}{\kappa^3}+
\Bigl[\,2j-\varepsilon_{\beta,2s}^{(-)}\,\Bigr]\right)
\;,
\label{BatemanNegativeMass}
\end{eqnarray}
where we have used the $k(\mu,\nu)$ notation for the Bateman function.

\begin{figure}[t]
\includegraphics[width=7.5cm]{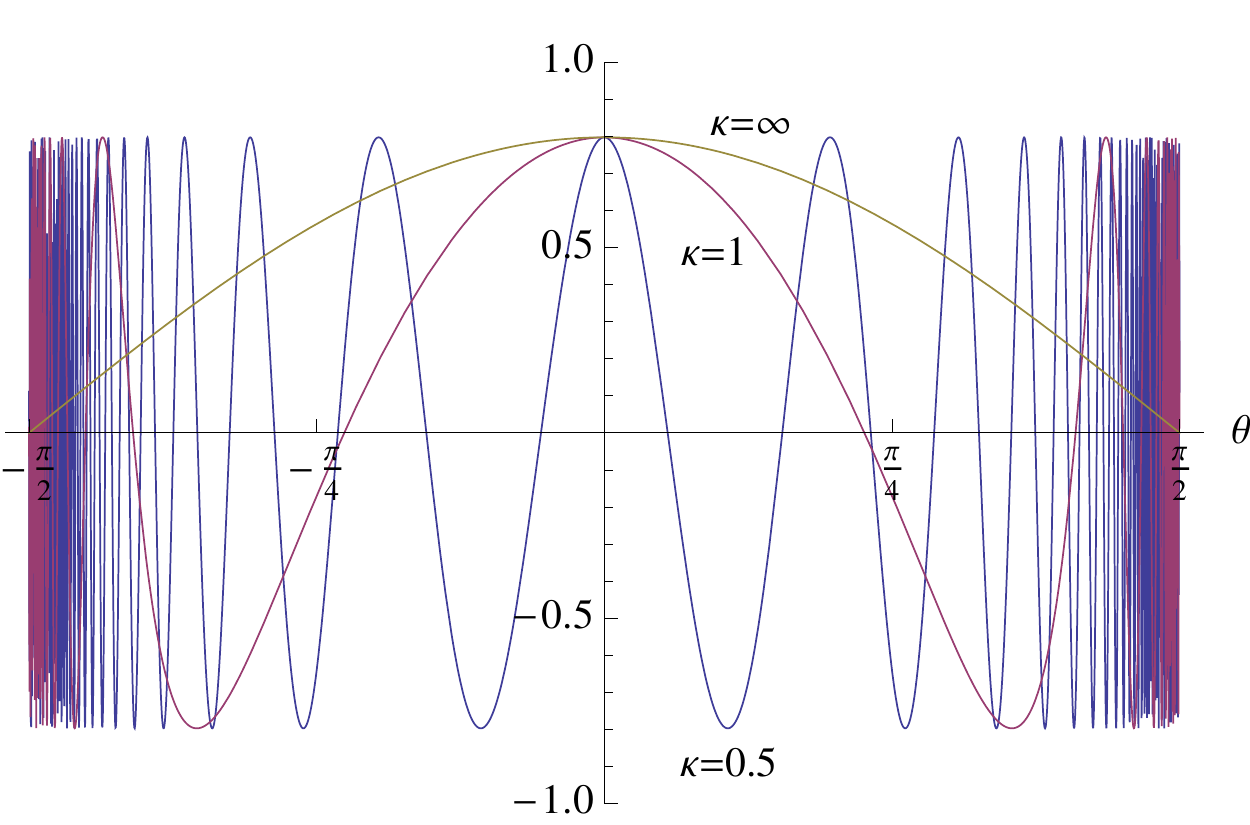}\\
(a) $n=1$\\
\includegraphics[width=7.5cm]{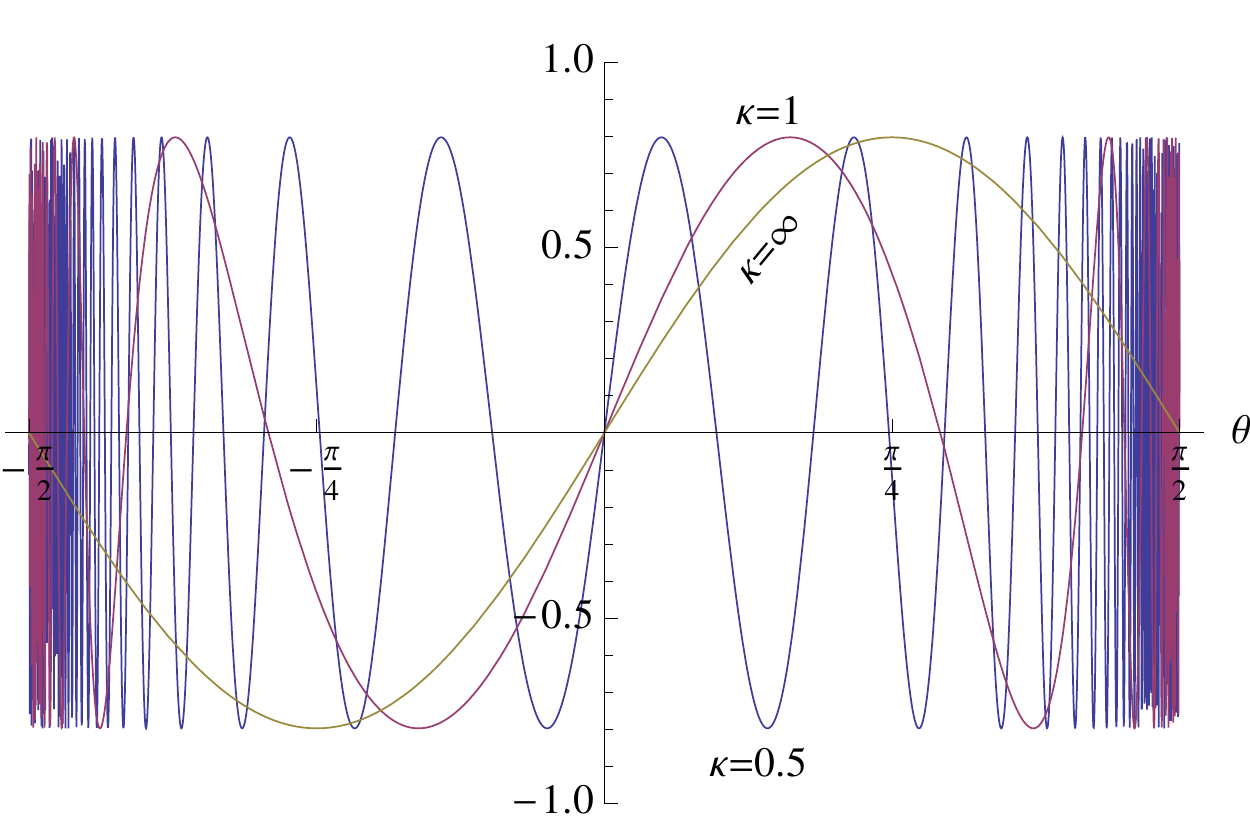}\\
(b) $n=2$\\
\includegraphics[width=7.5cm]{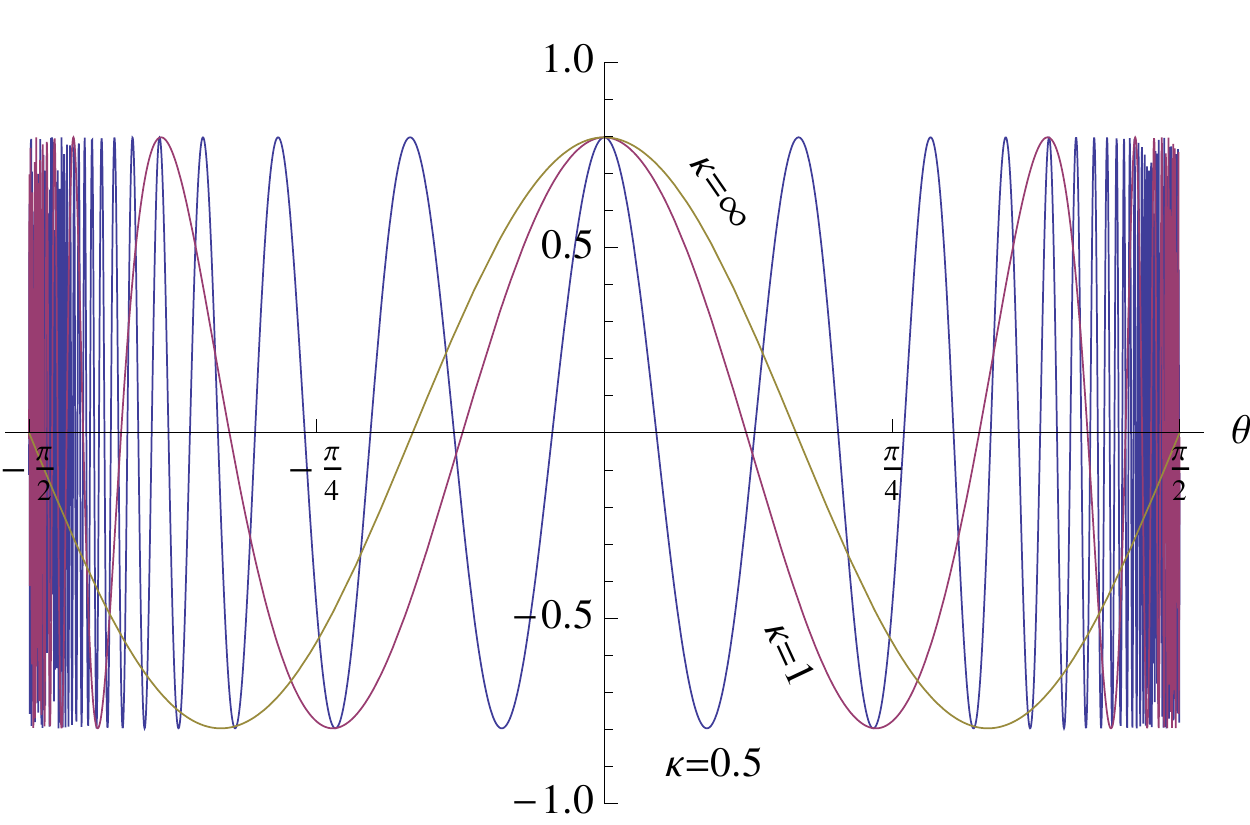}\\
(c) $n=3$
\caption{The eigenfunctions of $\hat{H}_1$ with negative mass in $\theta$-space
for the first three eigenvalues $\varepsilon_{\beta,n}^{(-)}$ ($n=1,2,3$), 
shown for the cases $\kappa=0.5$, $1$, and $\infty$.
}
\label{psinegativemass}
\end{figure}

In order to show that the above are indeed the solutions we seek,
we differentiate both sides of Eq.~(\ref{exptanFS}) by $\theta$ to find:
\begin{eqnarray}
\dfrac{\mu}{\cos^2\theta}\;\sum_{t=0}^{\infty} k_{2t}(\mu) \,e^{2it\theta}
& = & \sum_{t=0}^{\infty} (2t)\,k_{2t}(\mu) \,e^{2it\theta}
\;,
\label{exptanprime}
\end{eqnarray}
which can be rewritten using $\cos^2\theta=(1+\cos2\theta)/2=(2+e^{2i\theta}+e^{-2i\theta})/4$
as, 
\begin{eqnarray}
\lefteqn{\mu\sum_{t=0}^{\infty} k_{2t}(\mu) \,e^{2it\theta}}
\cr
& = & \left(1+\dfrac{e^{2i\theta}+e^{-2i\theta}}{2}\right)
\sum_{t=0}^{\infty} t\,k_{2t}(\mu) \,e^{2it\theta}
\;.
\end{eqnarray}
After some rearranging, this yields
\begin{eqnarray}
0 & = & k_2 - 2\mu k_0\;,\cr
0 & = & (t+1)k_{2(t+1)}-2(\kappa^{-3}-t)k_{2t}+(t-1)k_{2(t-1)}\;,\cr
&& 
\end{eqnarray}
for $t\ge 1$.
Setting $\mu=\kappa^{-3}$
and comparing with Eqs.~(\ref{recursionNegativeMassOddCoefs}) and 
(\ref{recursionNegativeMassEvenCoefs}),
we can check that 
Eq.~(\ref{c2jminusonecases}) satisfies Eq.(\ref{recursionNegativeMassOddCoefs}), while
Eq.~(\ref{c2jcases}) satisfies Eq.(\ref{recursionNegativeMassEvenCoefs}).
From Eq.~(\ref{exptanFS}), it is also straightforward to show 
(see Appendix~\ref{BatemanProperties}) that
\begin{equation}
\sum_{t=0}^{\infty}
\left[\,k_{2t}(\mu)\,\right]^2
\;=\; 1\;,
\end{equation}
for arbitrary $\mu$.
Thus, the states defined via Eqs.~(\ref{c2jminusonecases}) and (\ref{c2jcases})
are already normalized.
Eq.~(\ref{exptanFS}) also allows us to sum the series resulting from 
Eqs.~(\ref{c2jminusonecases}) and (\ref{c2jcases})
exactly, and we find
\begin{eqnarray}
\tilde{\psi}_{2s-1}(\theta)
& = & \sqrt{\dfrac{2}{\pi}}\;
\cos\left[\dfrac{1}{\kappa^3}\tan\theta + (2s-1)\theta\right]
\;,\cr
\tilde{\psi}_{2s}(\theta)
& = & \sqrt{\dfrac{2}{\pi}}\;
\sin\left[\dfrac{1}{\kappa^3}\tan\theta + 2s\theta\right]
\;.
\label{NegativeMassWaveFunctions}
\end{eqnarray}
In FIG.~\ref{psinegativemass}, we show the first three lowest energy eigenfunctions
for $\kappa=0.5$, $1$, and $\infty$.
In the $\kappa=\infty$ limit these functions respectively become 
$\sqrt{2/\pi}\cos[(2s-1)\theta]$ and $\sqrt{2/\pi}\sin[2s\theta]$, the
eigenfunctions of $\hat{x}^2$.

\subsection{Positive Mass Case}

\begin{figure}[t]
\includegraphics[width=8.5cm]{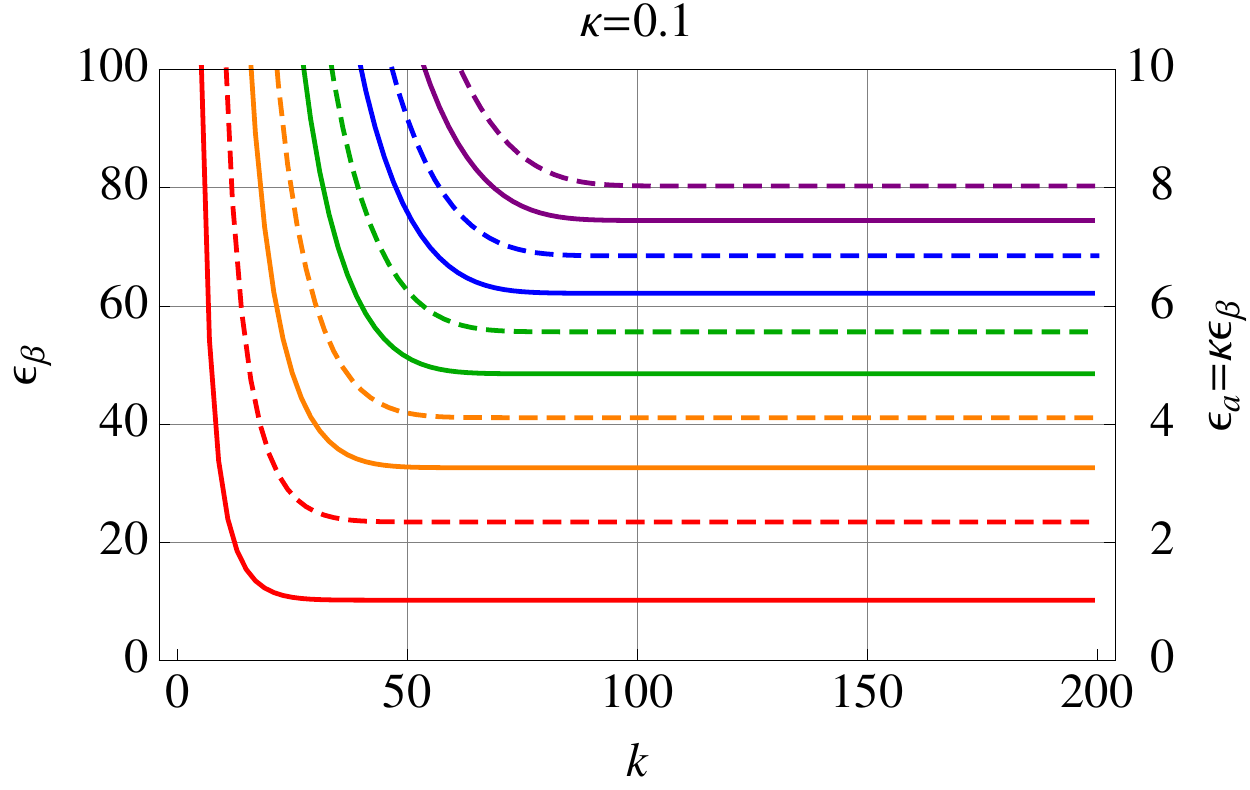}
\caption{The lowest five zeroes of $N_{k}(\varepsilon_\beta)$ for the case $\kappa=0.1$.
The solid lines connect the zeroes of $N_{k}(\varepsilon_\beta)$ with $k=\mathrm{odd}$,
and the dashed lines connect the zeroes of $N_{k}(\varepsilon_\beta)$ with $k=\mathrm{even}$.
These converge to the lowest ten eigenvalues of $\hat{H}_1$ with positive mass as 
$k\rightarrow\infty$.  
For larger values of $\kappa$ the convergence is faster.
}
\label{coeffzeroes}
\end{figure}

\begin{figure}[t]
\includegraphics[width=8cm]{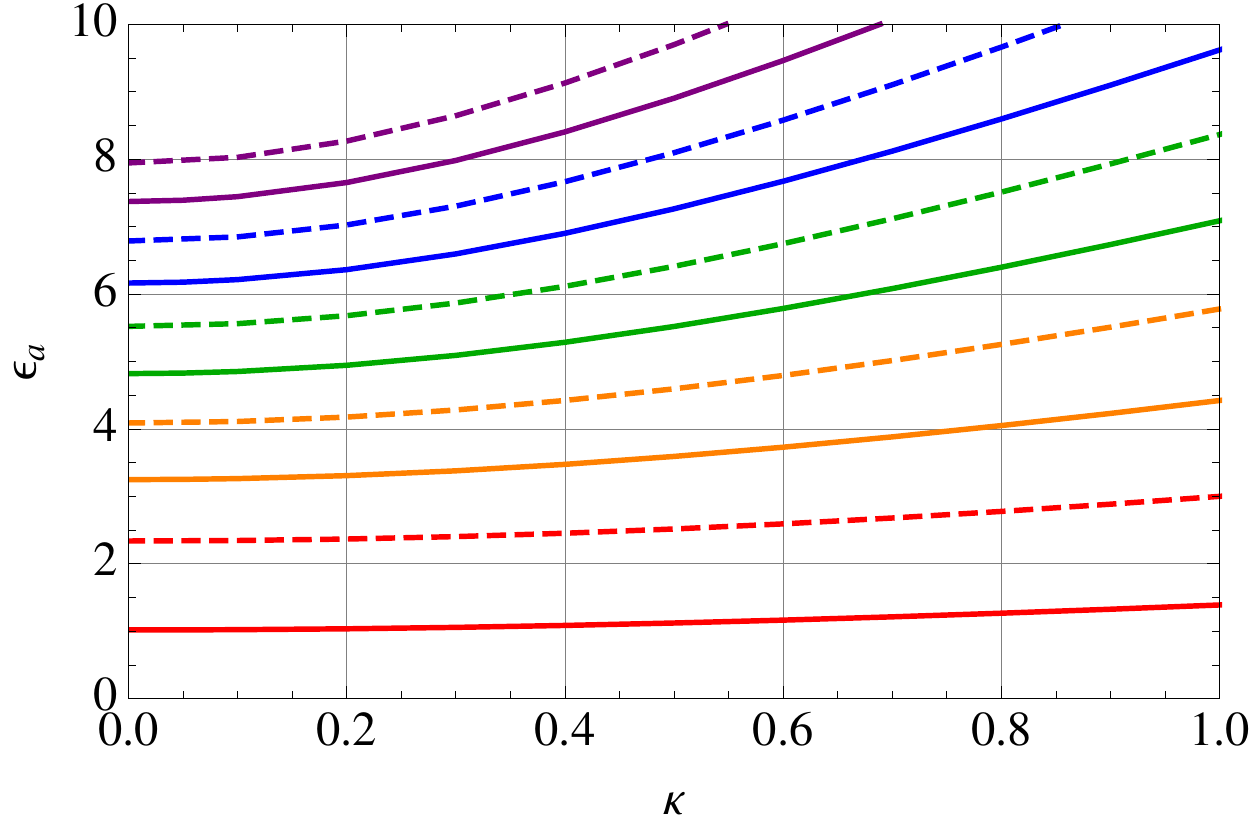}
\caption{The $\kappa$-dependence of the ten lowest eigenvalues of $\hat{H}_1$
for the positive mass case
scaled to $\varepsilon_a=\kappa\varepsilon_\beta=E/Fa$. 
The solid lines indicate the eigenvalues of even-parity states,
and the dashed lines indicate the eigenvalues of the odd-parity states.
They converge to the eigenvalues for the $\beta=0$ case as $\kappa\rightarrow 0$.
The odd-parity eigenvalues agree with those found by Benczik in Ref.~\cite{Benczik:2007we} for $\hat{H}_1'$ with an infinite potential wall at $x=0$.
}
\label{eigenvalues}
\end{figure}

For the positive mass case, we were unable to find exact analytical solutions to
Eq.~(\ref{recursion1}) and resorted to numerical techniques.
Using symbolic manipulation programs such as Mathematica, 
the recursion relation can be solved to express all the odd coefficients
in terms of $c_1$ and all the even coefficients in terms of $c_2$. 
For fixed $\kappa$, this will yield expressions with rational
functions of $\varepsilon_\beta$ multiplying the initial coefficients, that is:
\begin{eqnarray}
c_{2j-1} & = & \dfrac{N_{2j-1}(\varepsilon_\beta)}{D_{2j-1}(\varepsilon_\beta)}\;c_1\;,\cr
c_{2j}   & = & \dfrac{N_{2j}(\varepsilon_\beta)}{D_{2j}(\varepsilon_\beta)}\;c_2\;,
\label{coeffRational}
\end{eqnarray}
where $N_{2j-1}$, $D_{2j-1}$, $N_{2j}$, and $D_{2j}$ are all polynomials in $\varepsilon_\beta$.
In FIG.~\ref{coeffzeroes}, we plot the $k$-dependence of the zeroes of $N_{k}(\varepsilon_\beta)$
and find that they converge rapidly to fixed values indicating that
demanding $c_{k}$ to vanish for a large enough $k$ will let us find the value of $\varepsilon_\beta$
which would impose Eq.~(\ref{finitesum}).
Finding these zeroes for various values of $\kappa$ we obtain FIG.~\ref{eigenvalues}.

In the limit $\kappa\rightarrow 0$,
the even- and odd-parity eigenvalues found with this method 
converge to the eigenvalues for the $\beta=0$ case:
\begin{eqnarray}
\lim_{\kappa\rightarrow 0}\;\kappa\varepsilon_{\beta,2s-1}^{(+)} & = & 
\lim_{\kappa\rightarrow 0}\dfrac{E_{2s-1}^{(+)}}{Fa} \;=\; 
-\beta_s\;,
\cr
\lim_{\kappa\rightarrow 0}\;\kappa\varepsilon_{\beta,2s}^{(+)} & = & 
\lim_{\kappa\rightarrow 0}\dfrac{E_{2s}^{(+)}}{Fa} \;=\; 
-\alpha_s\;.
\label{betazerolimit}
\end{eqnarray}
Here, $\alpha_s < 0$ is the $s$-th zero of the Airy function $\Ai(\xi)$,
while $\beta_s < 0$ is the $s$-th zero of its derivative $\Ai'(\xi)$,
both numbered in descending order.
(See appendix~\ref{CanonicalReview}.)
In the opposite limit $\kappa\rightarrow\infty$, which corresponds to
$m\rightarrow\infty$, we find for both parities
\begin{eqnarray}
\lim_{\kappa\rightarrow\infty}\varepsilon_{\beta,n}^{(+)} & = & 
\lim_{\kappa\rightarrow\infty}\dfrac{E_{n}^{(+)}}{F\Delta x_{\min}} \;=\; 
n\;.
\label{minfinitelimit}
\end{eqnarray}
Thus, the eigenvalues for the positive mass case connect smoothly to
those for the negative mass case, Eq.~(\ref{NegativeMassGuess}), at
$1/m = 0$.

\begin{figure}[t]
\includegraphics[width=7.5cm]{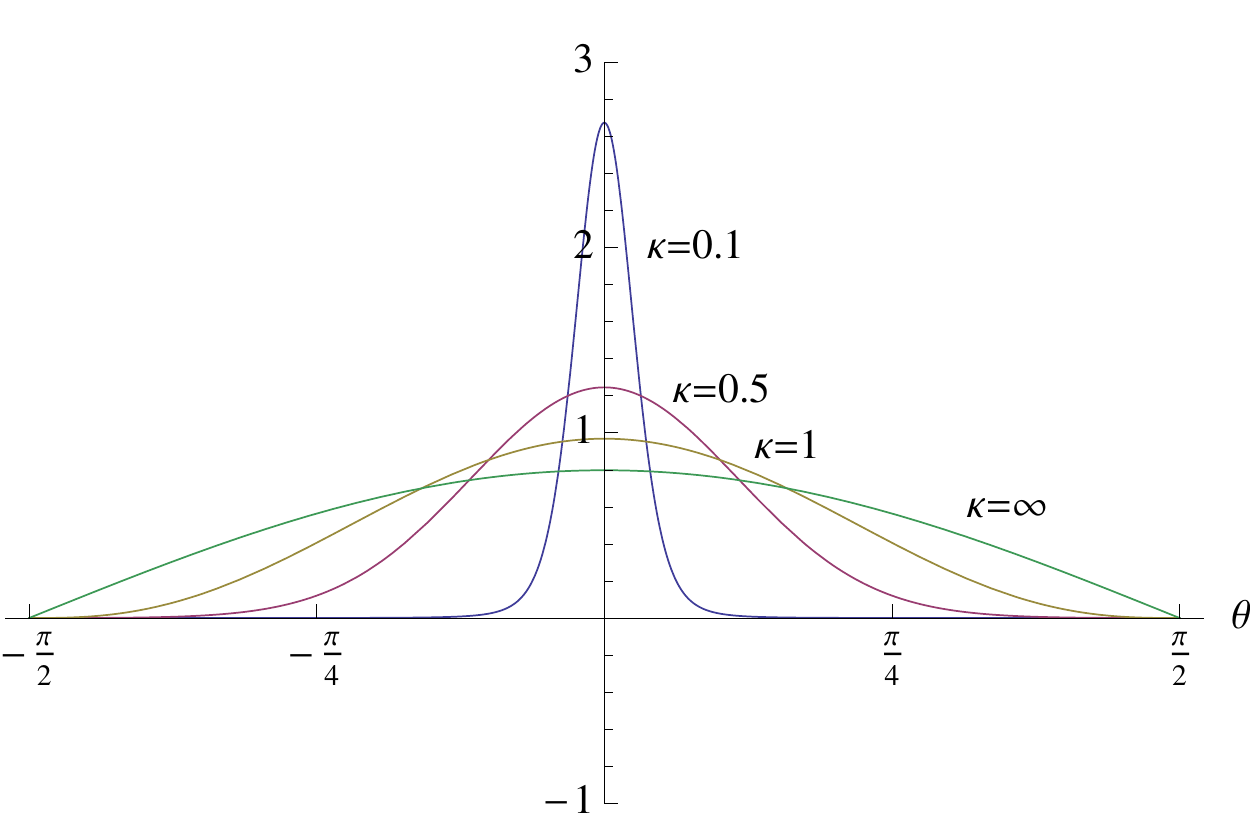}\\
(a) $n=1$\\
\includegraphics[width=7.5cm]{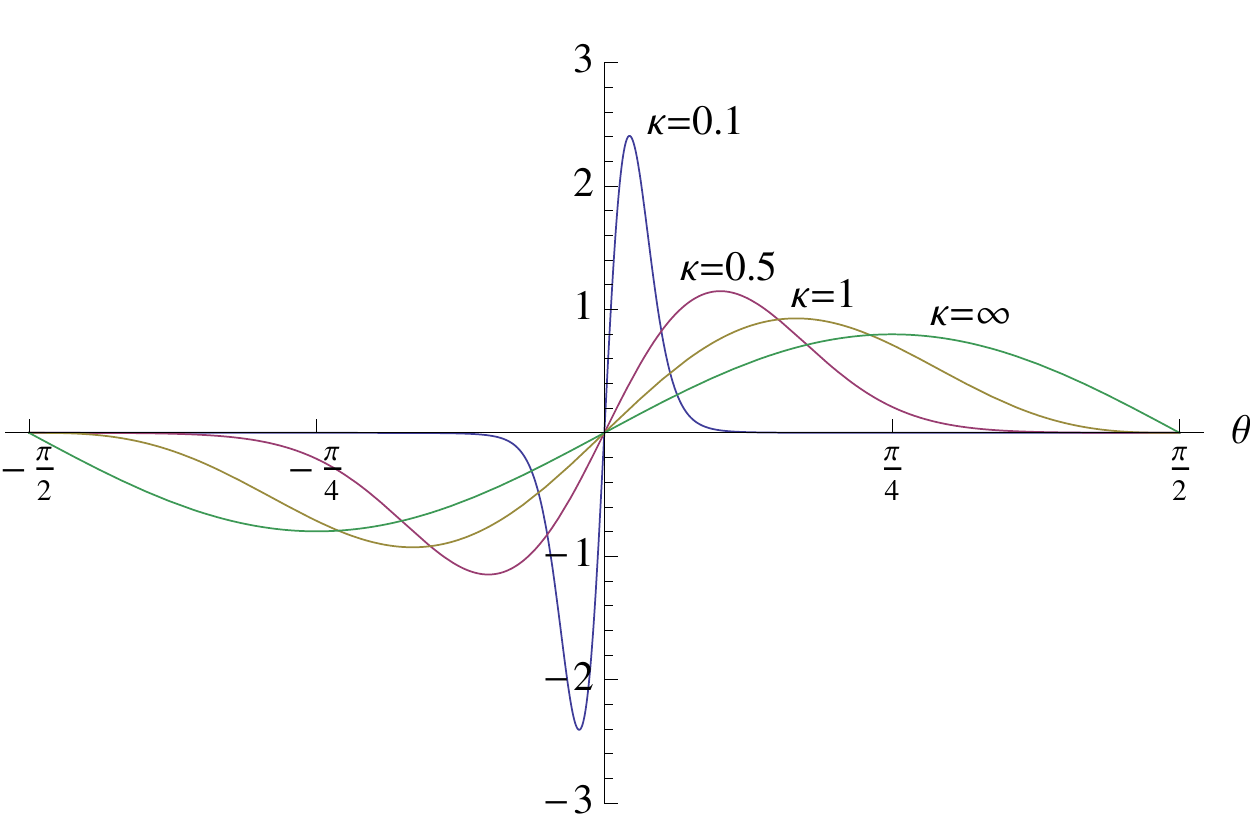}\\
(b) $n=2$\\
\includegraphics[width=7.5cm]{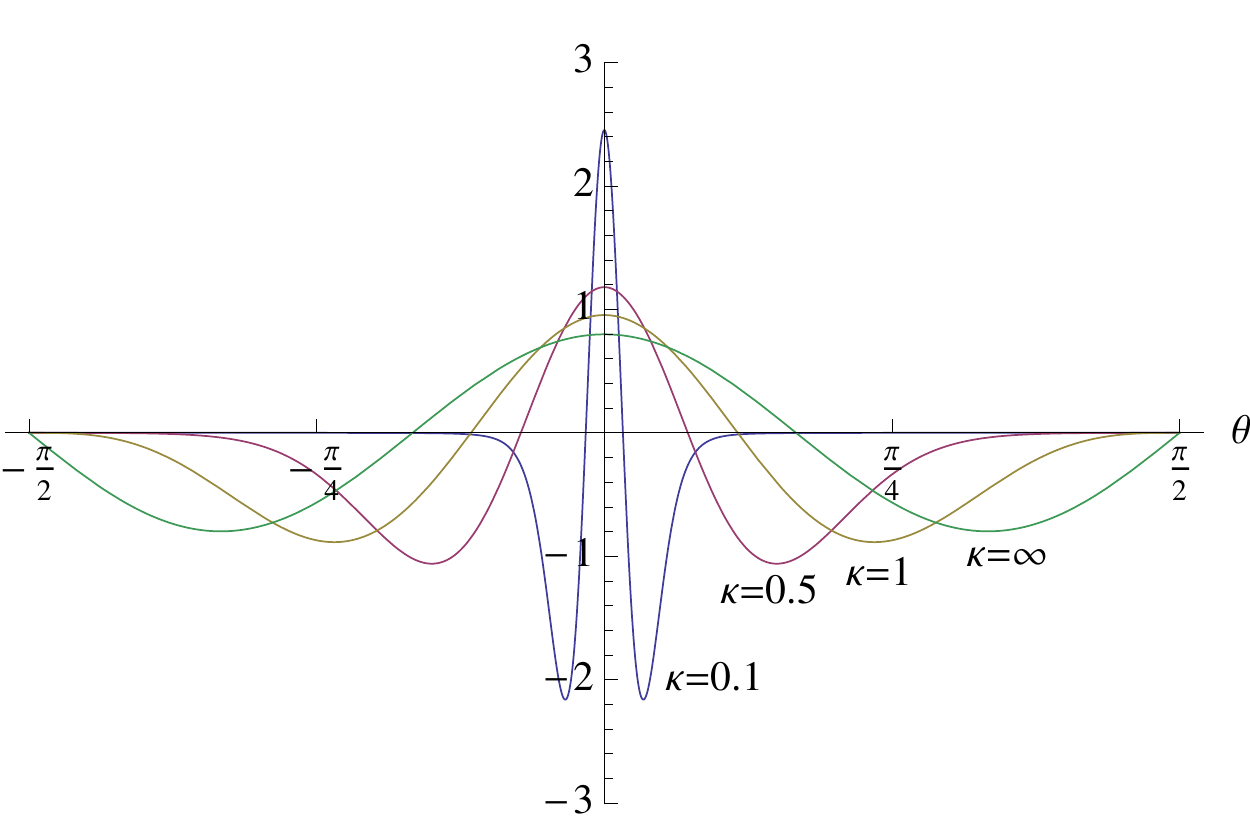}\\
(c) $n=3$
\caption{The eigenfunctions of $\hat{H}_1$ with positive mass in $\theta$-space
for the three lowest eigenvalues
shown for $\kappa=0.1$, $0.5$, $1$, and $\infty$.
}
\label{PsiPositiveMass}
\end{figure}

Once the eigenvalues are obtained, Eq.~(\ref{coeffRational}) can be used
to calculate the expansion coefficients of the eigenstates.
A complication arises when a zero of the numerator function is also a zero of the
denominator function.
This happens, for instance, to the even coefficients when $\kappa=1$, the lowest
eigenvalue being $\varepsilon_\beta=3$.
This is also a zero of the denominator function $D_{2j}(\varepsilon_\beta)$ for
$j\ge 3$. 
What this tells us is that the sequence of coefficients terminates
after $c_4$ for this set of parameters, 
that is, all the coefficients including and beyond $c_6$ are all zero.
The two non-zero coefficients in this case must be fixed from the recursion relation 
so that $c_6$ will be zero as required:
\begin{equation}
c_2 \;=\; \dfrac{2}{\sqrt{5}}\;,\qquad
c_4 \;=\; -\dfrac{1}{\sqrt{5}}\;.
\end{equation}
Proceeding in this way, we can determine the expansion coefficients of the
eigenstate for each eigenvalue $\varepsilon_{\beta,n}^{(+)}$.

Numerically, these coefficients are found to satisfy the following
relations, up to normalizations, analogous to Eq.~(\ref{BatemanNegativeMass}) for the negative mass case:
\begin{eqnarray}
c_{2j-1} & = & 
k\left(\dfrac{1}{\kappa^3},\dfrac{1}{\kappa^3}-
\Bigl[\,(2j-1)-\varepsilon_{\beta,2s-1}^{(+)}\,\Bigr]\right)
\;,\cr
c_{2j} & = & 
k\left(\dfrac{1}{\kappa^3},\dfrac{1}{\kappa^3}-
\Bigl[\,2j-\varepsilon_{\beta,2s}^{(+)}\,\Bigr]\right)
\;.
\label{BatemanPositiveMass}
\end{eqnarray}
Furthermore, the odd-parity energy eigenvalues satisfy
\begin{equation}
k\left(\dfrac{1}{\kappa^3},\dfrac{1}{\kappa^3}+\varepsilon_{\beta,2s}^{(+)}\right)\;=\;0\;.
\label{czero}
\end{equation}
We will elaborate on why this is the case in section~\ref{alternative}.
The $\theta$-space eigenfunctions constructed from these coefficients 
for the three lowest eigenvalues for several representative values of $\kappa$
are shown in
FIG.~\ref{PsiPositiveMass}.

\subsection{Uncertainties}

\begin{figure}[t]
\includegraphics[width=7.5cm]{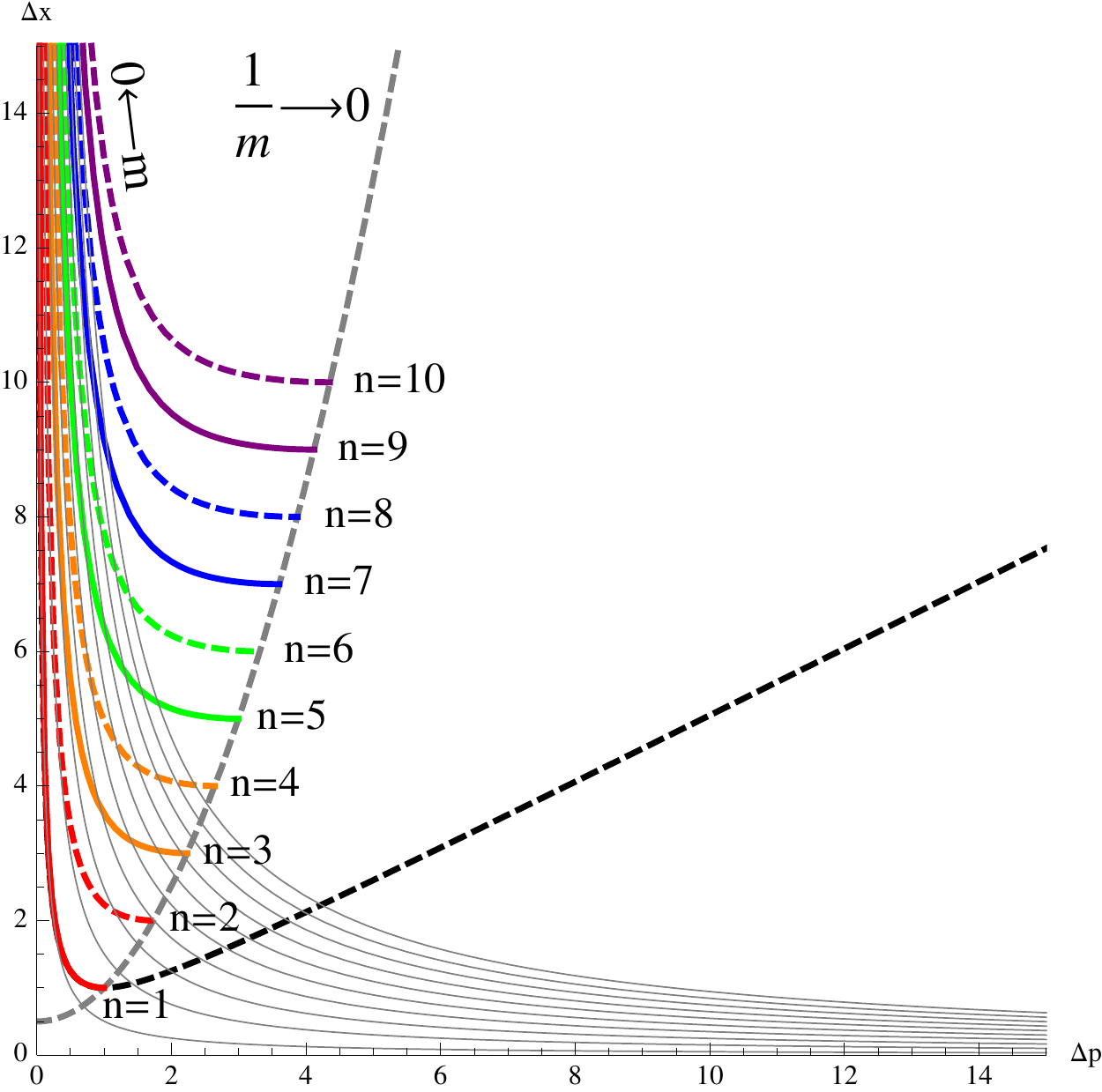}
\caption{The uncertainties in $x$ and $p$ for the first ten eigenstates of $\hat{H}_1$.
$\Delta p$ is in units of $1/\sqrt{\beta}$, while $\Delta x$ is in units of
$\Delta x_{\min}=\hbar\sqrt{\beta}$.
When the mass is positive, $(\Delta p,\Delta x)$ follows the
solid curves for the parity even states, and the dashed curves for parity odd states,
approaching the curve $(\Delta x/\Delta x_{\min})=(1+\beta\Delta p^2)/2$ 
(shown in dashed gray) as $1/m\rightarrow +0$.
When $1/m$ crosses over into the negative, both $\Delta p$ and $\Delta x$ diverge.
}
\label{DelPDelX}
\end{figure}

The expansion coefficients found in the previous subsections can be utilized to calculate 
the uncertainties in $x$ and $p$ for each state.
Let us write
\begin{equation}
\ket{\psi} \;=\;\sum_{k=1}^{\infty} c_k\ket{k}\;,\qquad
\sum_{k=1}^{\infty} |c_k|^2 \;=\; 1\;,
\end{equation}
where $(\hat{x}/\Delta x_{\min})^2\ket{k} = k^2\ket{k}$, $\braket{k}{\ell}=\delta_{k\ell}$.
Since all the eigenstates of $\hat{H}_1$ are also eigenstates of parity, and also since
particles do not go anywhere when they are bound, it is clear that
\begin{equation}
\vev{\hat{x}} \,=\, \bra{\psi}\hat{x}\ket{\psi} \,=\, 0\,,\qquad
\vev{\hat{p}} \,=\, \bra{\psi}\hat{p}\ket{\psi} \,=\, 0\,,
\end{equation}
for both positive and negative mass cases.
It is also straightforward to show that
\begin{equation}
\vev{\hat{x}^2}
\;=\; \bra{\psi}\hat{x}^2\ket{\psi}
\;=\;(\Delta x_{\min})^2\sum_{k=1}^{\infty}|c_k|^2 k^2\;.
\label{vevxx}
\end{equation}
To calculate $\hat{p}^2$, we need
\begin{eqnarray}
\lefteqn{\bra{k}\hat{p}^2\ket{\ell}}
\cr
& = & \dfrac{1}{\beta}\int_{-\pi/2}^{\pi/2}
d\theta\;\tilde{\varphi}_k^*(\theta)\,\tan^2\theta \;\tilde{\varphi}_\ell(\theta)
\cr
& = & 
\begin{cases}
0  & \mbox{if $k-\ell=\mathrm{odd}$} \\
\dfrac{1}{\beta}\bigl[\,2\min(k,\ell) - \delta_{k\ell}\,\bigr] & \mbox{if $k-\ell=\mathrm{even}$}
\end{cases}
\end{eqnarray}
the proof of which can be found in the appendix of Ref.~\cite{Lewis:2011fg}.
Using this, we can calculate $\vev{\hat{p}^2}$ via
\begin{equation}
\vev{\hat{p}^2}
\;=\; \bra{\psi}\hat{p}^2\ket{\psi} 
\;=\; \sum_{k=1}^{\infty}\sum_{\ell=1}^{\infty}
c_k^* c_{\ell} \bra{k}\hat{p}^2\ket{\ell}
\;.
\label{vevpp}
\end{equation}
%

\subsubsection{Positive Mass Case}

The results of our numerical calculations are shown in FIG.~\ref{DelPDelX} for the
positive mass states.
In the infinite mass limit, the energy eigenstates will simply be the
eigenstates of $\hat{x}^2$ with uncertainties given by
\begin{eqnarray}
\Delta x_n & = & \sqrt{\bra{n}\hat{x}^2\ket{n}} \;=\; n\,\Delta x_{\min} \;=\; n\,\hbar\sqrt{\beta}\;,\cr
\Delta p_n & = & \sqrt{\bra{n}\hat{p}^2\ket{n}} \;=\; \sqrt{\dfrac{2n-1}{\beta}}\;.
\end{eqnarray}
So as $m\rightarrow\infty$, the points $(\Delta p_n,\Delta x_n)$ will terminate on the curve
\begin{equation}
\dfrac{\Delta x}{\Delta x_{\min}}
\;=\; \dfrac{1+\beta\Delta p^2}{2}\;,
\label{InfiniteMassCurve}
\end{equation}
which is shown in dashed gray.

\subsubsection{Negative Mass Case}

For the negative mass states, it turns out that both $\Delta p$ and $\Delta x$ diverge.
This can be seen either by using the expansion coefficients listed in Eqs.~(\ref{c2jminusonecases}) and (\ref{c2jcases}) with Eqs.~(\ref{vevxx}) and (\ref{vevpp}), or by using the wave-functions given in Eq.~(\ref{NegativeMassWaveFunctions}).
For instance, using the expansion coefficients we find 
\begin{eqnarray}
\dfrac{\vev{\hat{x}^2}_{2s-1}}{\Delta x_{\min}^2}
& = & \sum_{\ell=0}^{\infty} (2\ell+2s-1)^2\left[k_{2\ell}(\kappa^{-3})\right]^2
\;,
\cr
\dfrac{\vev{\hat{x}^2}_{2s}}{\Delta x_{\min}^2}
& = & \sum_{\ell=0}^{\infty} (2\ell+2s)^2\left[k_{2\ell}(\kappa^{-3})\right]^2
\;,
\end{eqnarray}
and both these sums are divergent since
\begin{eqnarray}
\sum_{\ell=0}^{\infty} (2\ell)   \left[k_{2\ell}(\mu)\right]^2 & = & \infty\;,\cr
\sum_{\ell=0}^{\infty} (2\ell)^2 \left[k_{2\ell}(\mu)\right]^2 & = & \infty\;,
\label{BatemanDivergences}
\end{eqnarray}
for arbitrary $\mu$ as shown in Appendix~\ref{BatemanProperties}.
To see the divergence of $\Delta p$, 
the simplest way would be to use the relation
\begin{equation}
\vev{\hat{p}^2}
\;=\; 2|m|
\left(
F\vev{|\hat{x}|}
-\vev{\hat{H}_1}
\right)
\;,
\end{equation}
and note that
\begin{eqnarray}
\dfrac{\vev{|\hat{x}|}_{2s-1}}{\Delta x_{\min}}
& = & \sum_{\ell=0}^{\infty} (2\ell+2s-1)\left[k_{2\ell}(\kappa^{-3})\right]^2
\;,
\cr
\dfrac{\vev{|\hat{x}|}_{2s}}{\Delta x_{\min}}
& = & \sum_{\ell=0}^{\infty} (2\ell+2s)\left[k_{2\ell}(\kappa^{-3})\right]^2
\;,
\end{eqnarray}
which are again both divergent.
Thus, unlike the harmonic oscillator case studied in Ref.~\cite{Lewis:2011fg}, 
the uncertainties $\Delta x$ and $\Delta p$ of the negative mass states do not
inhabit the $\Delta x\sim \Delta p$ branch of the MLUR curve.

In section~\ref{classical}, we will see that the divergence of $\Delta x$ and $\Delta p$
for the negative mass states can be understood classically
by taking the $\hbar\rightarrow 0$ limit of Eq.~(\ref{CommutationRelation}) and 
looking at the behavior of the classical particle whose Hamiltonian is given by
$H_1$. But before that, let us look at an alternative approach in 
deriving the results of this section, which will
clarify how the Bateman function solution was discovered.

\section{Alternative Approach}
\label{alternative}

\subsection{The Schr\"odinger Equation}

Recall that the standard procedure in solving for the eigenvalues and
eigenstates of $\hat{H}_1$, Eq.~(\ref{H1}), in the canonical $\beta=0$ case is to solve the
Schr\"odinger equation for $\hat{H}'_1$, Eq.~(\ref{H1prime}), and then
impose the boundary condition $\psi'(0)=0$ or $\psi(0)=0$, respectively, to obtain the
parity even or odd eigenvalues.
In this section, we will explore whether an analogous technique works when $\beta\neq 0$.

Using the same representation of $\hat{x}$ and $\hat{p}$ as above, namely
Eq.~(\ref{xptheta}), the Schr\"odinger equation for $\hat{H}'_1$ 
is obtained from Eq.~(\ref{schrodingertheta2}) by making the replacement
\begin{equation}
\sqrt{-\dfrac{d^2}{d\theta^2}} \quad\rightarrow\quad
i\dfrac{d}{d\theta}\;,
\end{equation}
to yield
\begin{equation}
\left( \pm\dfrac{1}{\kappa^3}\tan^2\theta 
+ i\dfrac{d}{d\theta}
\right) \tilde{\psi}(\theta)
\;=\; \varepsilon_\beta\,\tilde{\psi}(\theta)
\;.
\label{H1primeSchrodinger}
\end{equation}
The solution to this equation is easily seen to be
\begin{equation}
\tilde{\psi}^{(\pm)}(\theta,\varepsilon_\beta)
\;=\; 
\exp\left[i
\left\{
\pm\dfrac{1}{\kappa^3}\bigl(\tan\theta - \theta\bigr)
-\varepsilon_\beta\,\theta
\right\}
\right]
\;.
\end{equation}
Fourier transforming to $\chi_\beta\equiv x/\Delta x_{\min}$ space, we find:
\begin{eqnarray}
\lefteqn{
\psi^{(\pm)}(\chi_\beta-\varepsilon_\beta)
\;=\;
\dfrac{1}{\pi}\int_{-\pi/2}^{\pi/2}
d\theta\;e^{i\chi_\beta\theta}\;
\tilde{\psi}^{(\pm)}(\theta,\varepsilon_\beta)
}
\cr
& = & \dfrac{1}{\pi}
\int_{-\pi/2}^{\pi/2}\!d\theta
\,\exp\biggl[i
\biggl\{
\pm\dfrac{1}{\kappa^3}\bigl(\tan\theta - \theta\bigr)
+\bigl(\chi_\beta-\varepsilon_\beta
\bigr)\theta
\biggr\}
\biggr]
\cr
& = & \dfrac{2}{\pi}\int_{0}^{\pi/2}
\cos\biggl[
\dfrac{1}{\kappa^3}\tan\theta
-\left\{\dfrac{1}{\kappa^3}\mp \bigl(\chi_\beta-\varepsilon_\beta\bigr)
\right\}\theta
\biggr]\,d\theta
\cr
& = &
k\left(\dfrac{1}{\kappa^3}\,,\,
\dfrac{1}{\kappa^3}\mp \bigl(\chi_\beta-\varepsilon_\beta\bigr)
\right)
\;,
\label{NaiveFourier}
\end{eqnarray}
where in the last line, we have made use of the Bateman function introduced
in Eq.~(\ref{BatemanFunction}).
As discussed in section~\ref{sec2}B, 
$\tilde{\psi}^{(\pm)}(\theta)$ can be recovered from the values of
$\psi^{(\pm)}(\chi_\beta-\varepsilon_\beta)$ sampled at the discrete points 
$\chi_\beta=2z+\lambda$, $z\in\mathbb{Z}$,
for arbitrary $\lambda$.
Rescaling variables to $\varepsilon_a=\kappa\varepsilon_\beta$ and
$\chi_a=\kappa\chi_\beta$, 
it is straightforward to show that
in the positive mass case, we have
\begin{equation}
\lim_{\kappa\rightarrow 0}\;\dfrac{1}{2\kappa}\;
\psi^{(+)}(\chi_a/\kappa-\varepsilon_a/\kappa)
\;=\;
\Ai\left(\chi_a-\varepsilon_a\right)\;.
\end{equation}
That is, our solution converges to the $\beta=0$ case 
in this limit as it should.

\subsection{Odd Parity Solutions}

\begin{figure}[t]
\includegraphics[width=8cm]{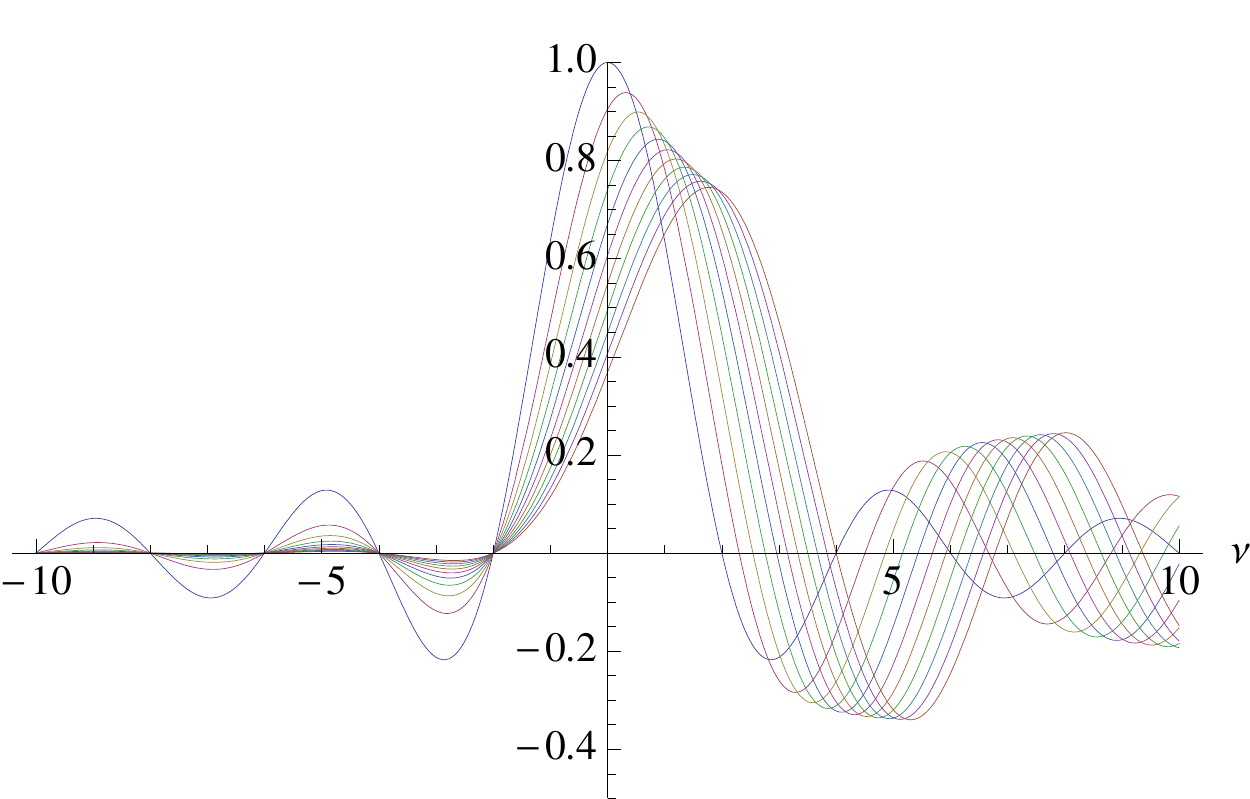}
\includegraphics[width=8cm]{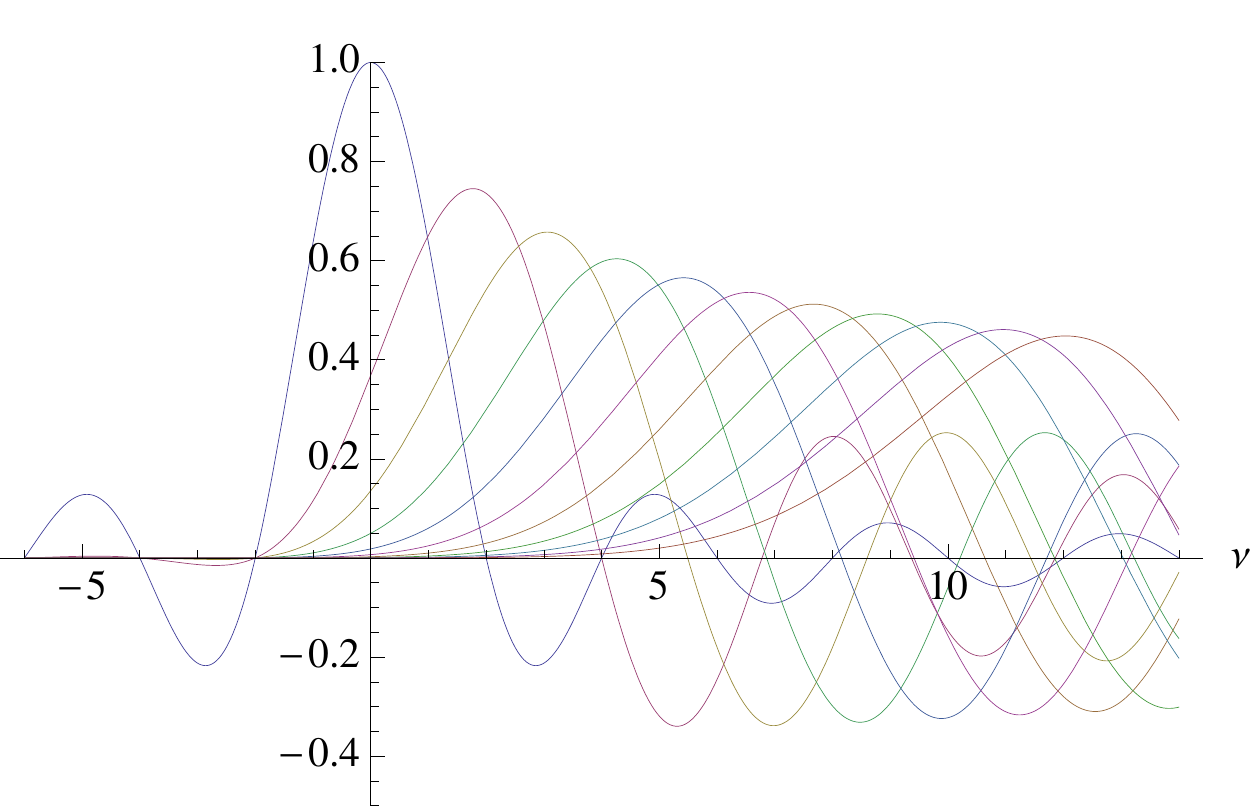}
\caption{The $\nu$-dependence of the Bateman function $k(\mu,\nu)$ for fixed $\mu$.
The top figure shows the graphs from $\mu=0$ to $\mu=1$ at $\Delta\mu=0.1$ intervals,
while the bottom figures shows those from $\mu=0$ to $\mu=10$ at $\Delta\mu=1$ intervals.
The $\mu=0$ case is left-right symmetric.
Note that $k(-\mu,\nu)=k(\mu,-\nu)$, so the graphs for the negative $\mu$ cases can be
obtained by simply flipping the direction of the $\nu$-axis.
}
\label{BatemanPlot}
\end{figure}

\begin{figure}[t]
\includegraphics[width=8cm]{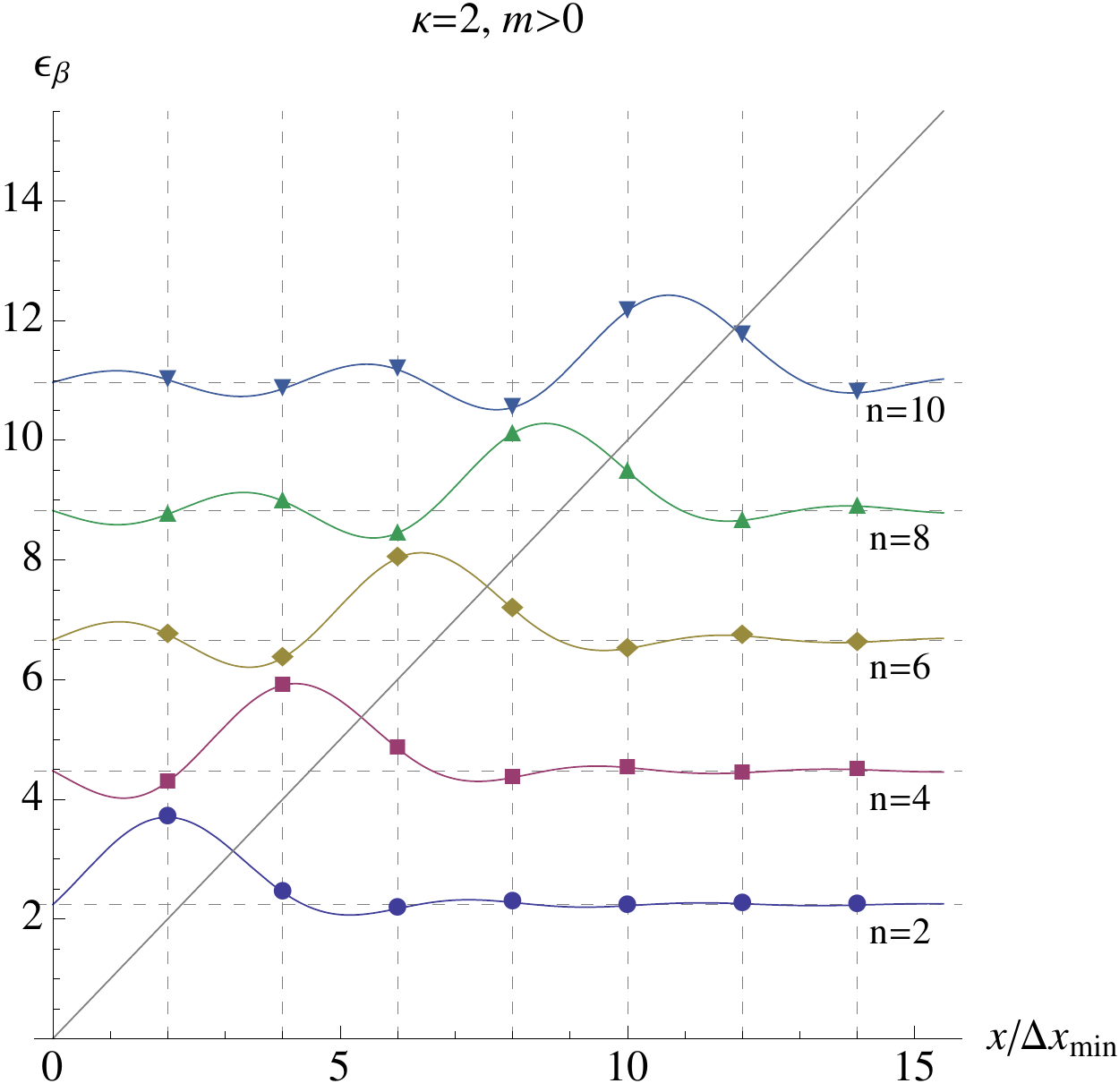}\\
\medskip
\includegraphics[width=8cm]{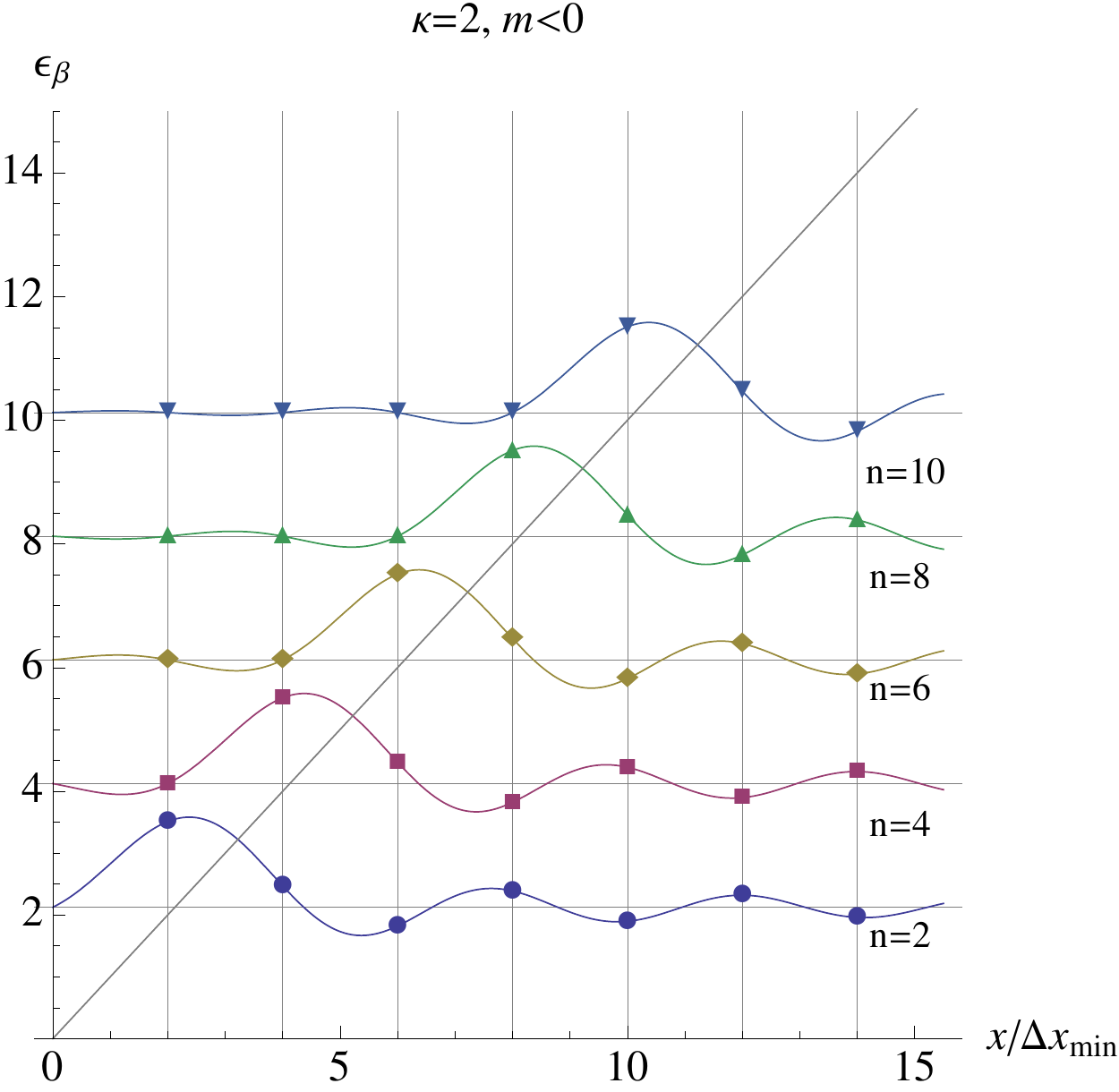}
\caption{The wave-functions for the first five odd-parity energy eigenstates
in $x/\Delta x_{\min}$-space for the positive (top) and negative (bottom) mass cases when
$\kappa=\Delta x_{\min}/a=2$. 
In both cases, the wave-functions vanish at $x=0$.
The values at even-integer multiples of $\Delta x_{\min}$
correspond to the expansion coefficients discussed in section~\ref{sec2}.
In the positive mass case, the wave-function in the 
physically forbidden region ($x/\Delta x_{\min}>\varepsilon_\beta$) oscillates instead of damping exponentially
as in the $\beta=0$ limit.
In the negative mass case, the wave-function is zero at 
even-integer multiples of $\Delta x_{\min}$ such that
$x<n$.
}
\label{PsiOdd2}
\end{figure}

\begin{figure}[t]
\includegraphics[width=8cm]{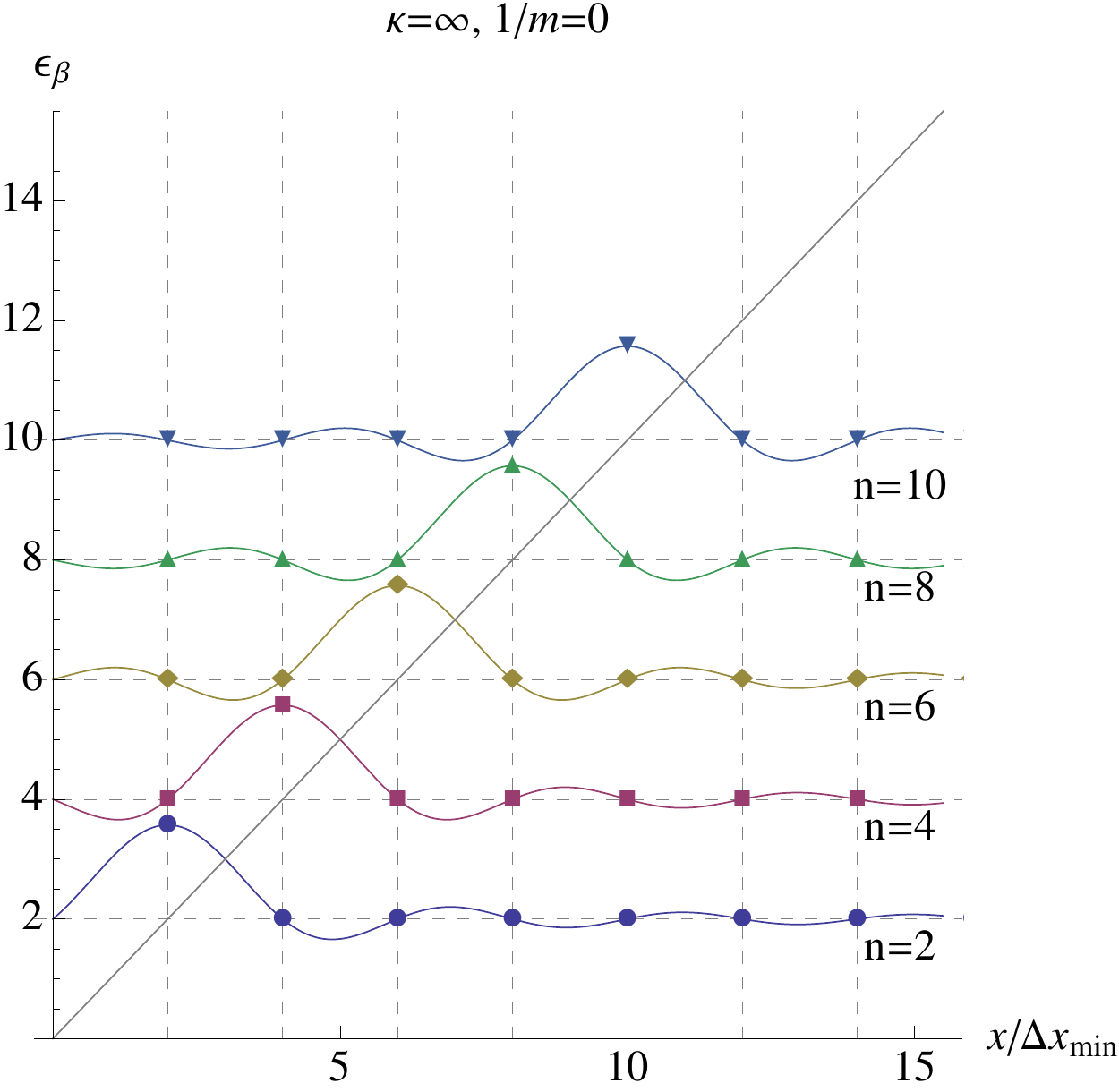}
\caption{The wave-functions for the first five odd-parity energy eigenstates
in $x/\Delta x_{\min}$-space in the limit
$\kappa=\Delta x_{\min}/a\rightarrow \infty$. 
Both the positive and negative mass cases converge to the same function.
Note that the functions are now left-right symmetric with respect to the
turning points.  Their values are zero at all even-integer multiples
of $\Delta x_{\min}$ except at the turning points where
$x/\Delta x_{\min}=n$.}
\label{PsiOddInf}
\end{figure}

We first consider the odd-parity solutions.
Though it is not clear that the concept of the wave-function vanishing
at $x=0$ makes sense in the presence of a minimal length, let us 
nevertheless impose the boundary condition $\psi^{(\pm)}(-\varepsilon_\beta)=0$ as in 
the $\beta=0$ case:
\begin{equation}
\psi^{(\pm)}(-\varepsilon_\beta) \;=\;
k\left(\dfrac{1}{\kappa^3}\,,\,
\dfrac{1}{\kappa^3}\pm \varepsilon_\beta
\right)
\;=\; 0\;.
\label{kcondition}
\end{equation}
Here, the sign in front of $\varepsilon_\beta$ is that of the mass.  
The $\nu$-dependence of the Bateman function $k(\mu,\nu)$ for fixed $\mu$ is shown for
several values of $\mu$ in FIG.~\ref{BatemanPlot}.
For fixed $\mu=\kappa^{-3}$, the Bateman function $k(\mu,\nu)$ 
has countable-infinite number of zeroes along both the positive and negative $\nu$ axes.

Let the $s$-th positive zero of $k(\kappa^{-3},\nu)$ be $\nu_s(\kappa)$.
If imposing Eq.~(\ref{kcondition}) is correct, then 
in the positive mass case
these zeroes should correspond to energy eigenvalues
given by
\begin{equation}
\varepsilon_{\beta,2s}^{(+)}
\;=\; \nu_s(\kappa) - \dfrac{1}{\kappa^3}\;,
\label{Eplus}
\end{equation}
and indeed these match preciously the values we obtained in 
section~\ref{sec2}F where they were found to satisfy Eq.~(\ref{czero}).
These energies are all positive, as they should be, and
agree with the energies derived by Benczik in Ref.~\cite{Benczik:2007we}.
In Benczik's approach, the boundary condition was given by 
\begin{equation}
U\left(
-\dfrac{\kappa^{-3}+\varepsilon_{\beta}}{2};\,0\;;\dfrac{2}{\kappa^3}
\right)\;=\;0\;,
\label{Ucondition}
\end{equation}
where $U(\alpha;\gamma;z)$ is Kummer's function of the second kind (see appendix~\ref{Sandor}),
which is related to the Bateman function via \cite{Wolfram}
\begin{equation}
k(\mu,\nu)\;=\; \dfrac{e^{-\mu}}{\Gamma\left(1+\frac{\nu}{2}\right)}\;
U\left(-\dfrac{\nu}{2};0;2\mu\right)\;,
\label{kUGamma}
\end{equation}
provided that $\mu$ is positive.
Clearly, the condition given in Eq.~(\ref{kcondition}) for the positive mass case is the same as the condition given in Eq.~(\ref{Ucondition}).

The negative zeroes $k(\mu,\nu)$ for fixed $\mu$ are 
independent of $\mu$, and thus of $\kappa$, 
and are given by the even negative integers:
\begin{equation}
\nu_{-s} \;=\; -2s\;,\qquad s\,=\,1,\,2,\,3,\cdots\,.
\label{NegativeZeroes}
\end{equation}
In the expression of Eq.~(\ref{kUGamma}), they are the
poles of the $\Gamma$-function in the denominator. 
(These do not appear in the approach of Benczik \cite{Benczik:2007we}.)
For the positive mass case, these will lead to negative energies,
corresponding to the particle in the negative $x$ region with
an infinite potential wall at $x=0$.
For the negative mass case, however, these correspond to positive energies:
\begin{equation}
\varepsilon_{\beta,2s}^{(-)}\;=\;2s+\dfrac{1}{\kappa^3}\;,
\label{Eminus}
\end{equation}
in agreement with our results of section~\ref{sec2}E.

Thus, imposing the boundary condition $\psi^{(\pm)}(-\varepsilon_\beta)=0$ to determine the eigenvalues $\varepsilon_\beta$ 
leads to results that are consistent with our previous approach.
The (un-normalized) parity odd wave-functions in $\chi_\beta$-space are therefore
\begin{equation}
\psi_{2s}^{(\pm)}(\chi_\beta)
\;=\; k\left(
\dfrac{1}{\kappa^3},
\dfrac{1}{\kappa^3}\mp
\left[
\chi_\beta
-\varepsilon_{\beta,2s}^{(\pm)}
\right]
\right)
\;.
\end{equation}
Comparing to the second lines of Eqs.~(\ref{BatemanNegativeMass}) and (\ref{BatemanPositiveMass}),
we can see that the expansion coefficients $c_{2j}$ found in section~\ref{sec2}
are equal to the values of $\psi_{2s}^{(\pm)}(\chi_\beta)$ sampled at the discrete points 
$\chi_\beta=2j$, $j\in\mathbb{N}$ up to phases.
Indeed, it was via this approach that we first found 
the solution 
Eq.~(\ref{c2jcases}) to Eq.~(\ref{recursionNegativeMassEvenCoefs}).
In Figs.~\ref{PsiOdd2} we plot the 
first five odd-parity eigenfunctions with the lowest 
eigenvalues for both positive and negative masses
using $\kappa=2$ as a representative case.
The limiting case $\kappa\rightarrow\infty$ ($1/m\rightarrow 0$) is 
shown in FIG.~\ref{PsiOddInf}.

\subsection{Even Parity Solutions}

\begin{figure}[t]
\includegraphics[width=8cm]{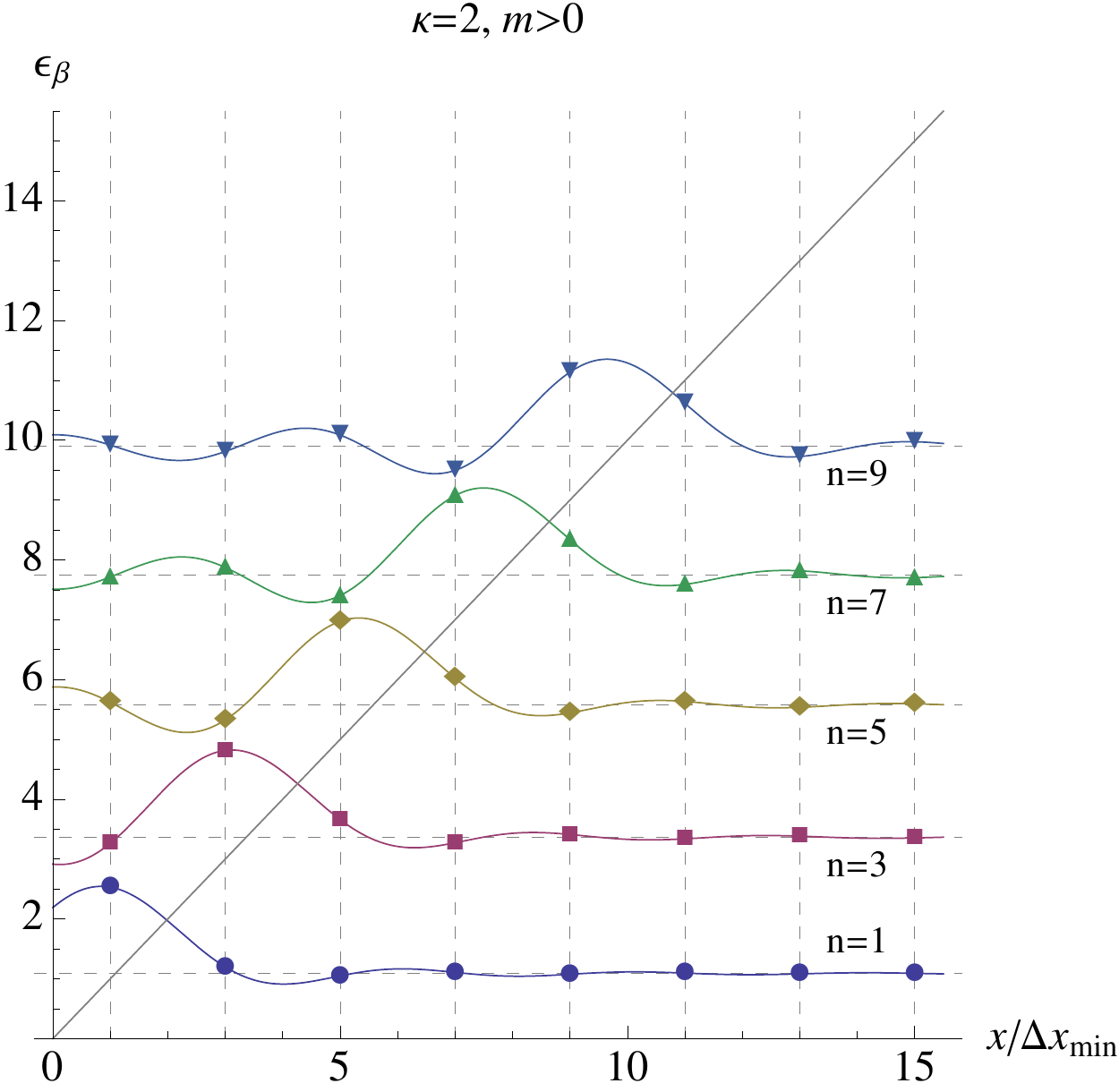}\\
\bigskip
\includegraphics[width=8cm]{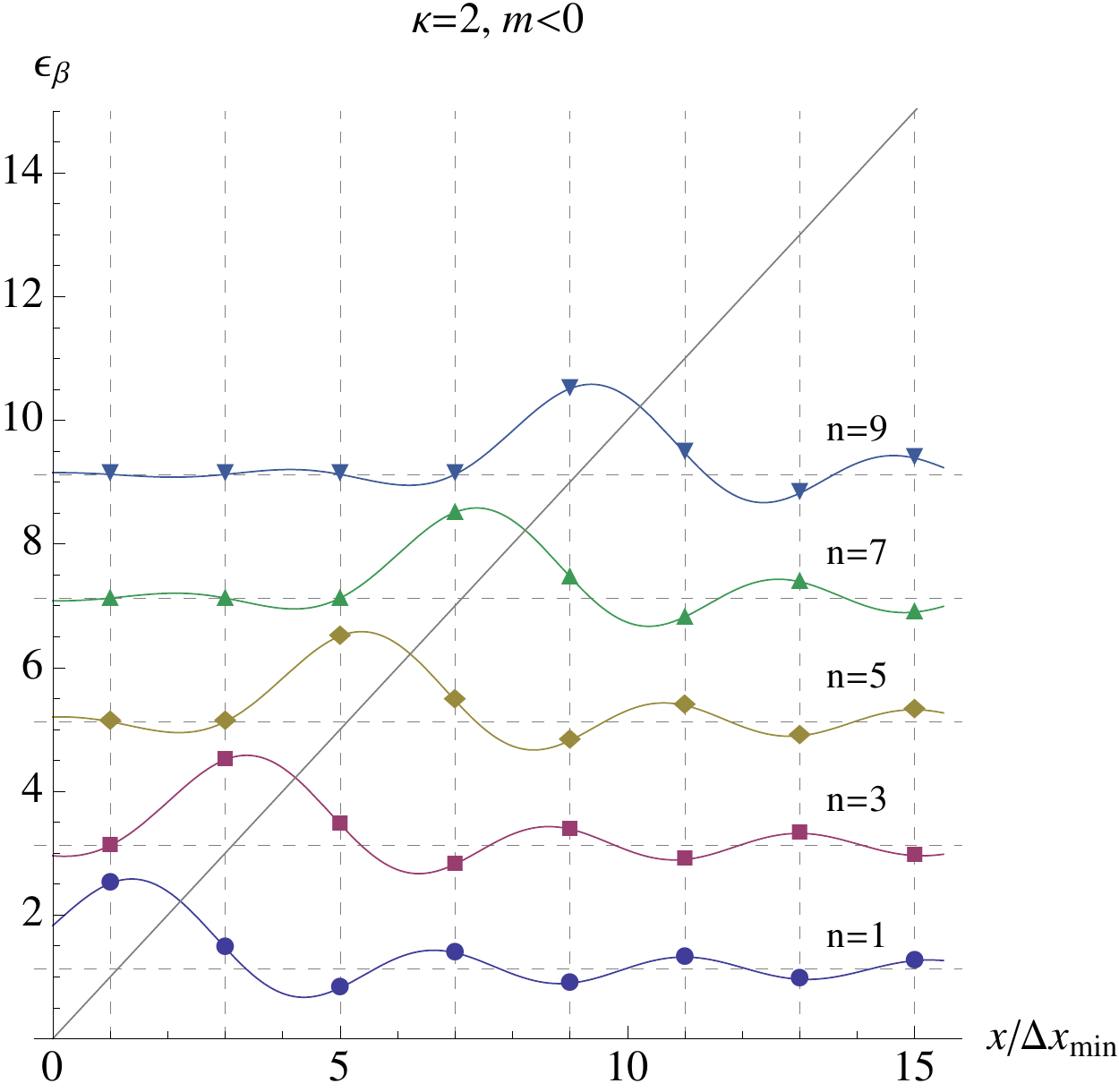}
\caption{The wave-functions for the first five energy eigenstates
in $x$-space for the positive (top) and negative (bottom) mass cases when
$\kappa=\Delta x_{\min}/a=2$. 
The values at odd integer multiples of $\Delta x_{\min}$
correspond to the expansion coefficients discussed in section~\ref{sec2}.
}
\label{PsiEven2}
\end{figure}

\begin{figure}[t]
\includegraphics[width=8cm]{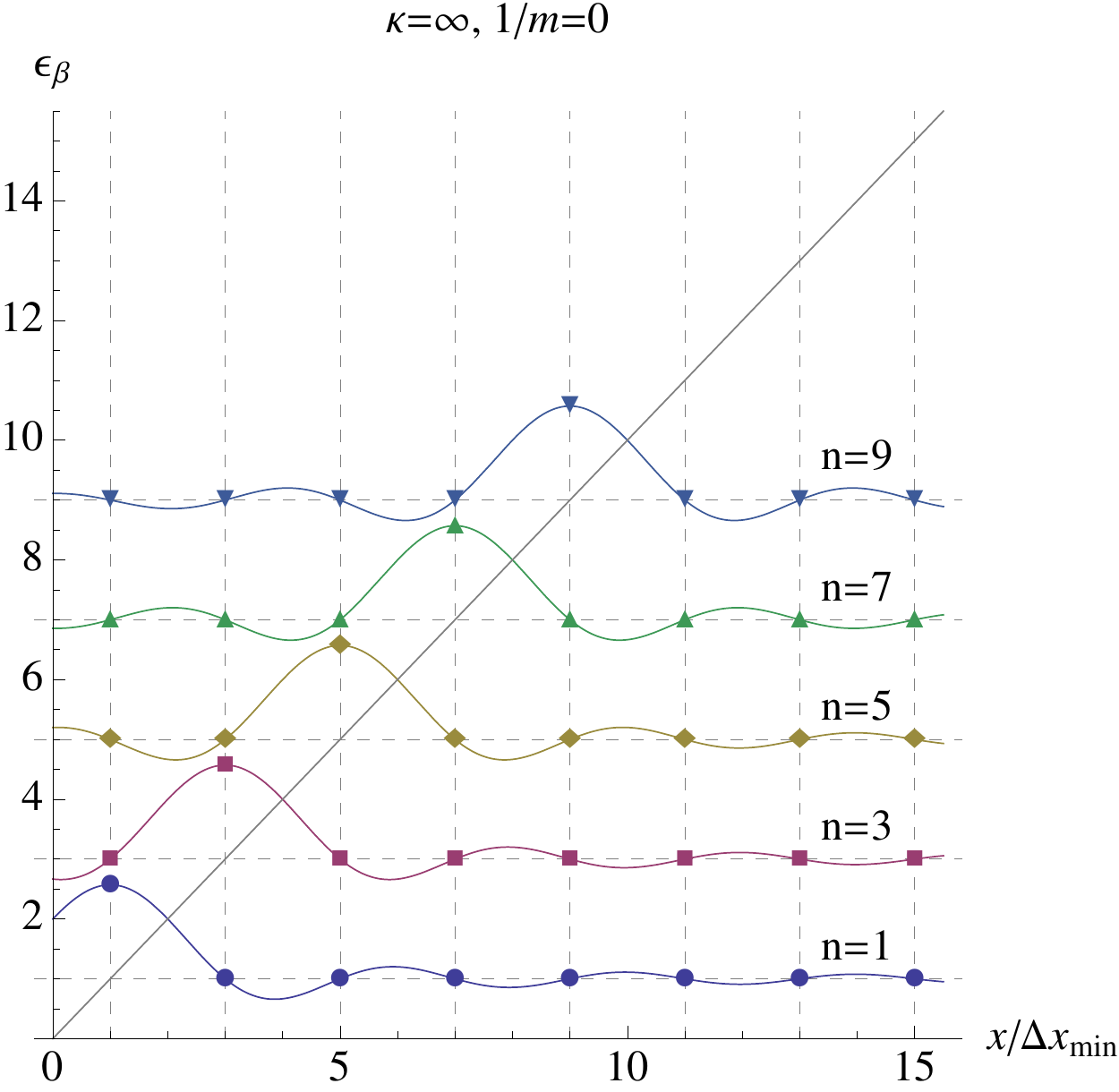}
\caption{The wave-functions for the first five energy eigenstates
in $x$-space in the limit
$\kappa=\Delta x_{\min}/a\rightarrow \infty$. 
Both the positive and negative mass cases converge to the same functions.
Note that the functions are now left-right symmetric with respect to the
turning points.  Their values is zero at all odd integer multiples
of $\Delta x_{\min}$ except at the turning points.}
\label{PsiEvenInf}
\end{figure}

\begin{figure}[t]
\includegraphics[width=8cm]{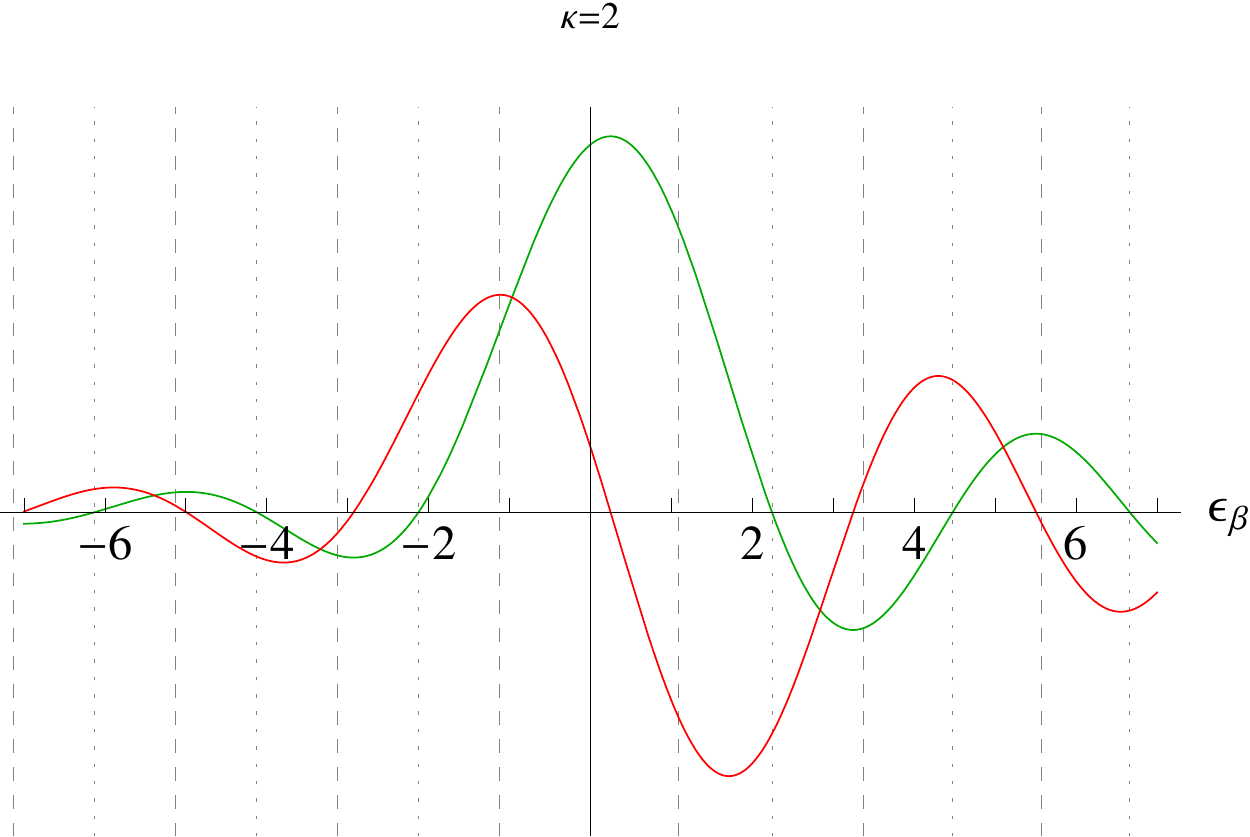}
\caption{Plots of $\psi^{(+)}(-\varepsilon_\beta)$ (green)
and $\psi^{\prime(+)}(-\varepsilon_\beta)$ (red)
compared to the even-parity (dashed vertical lines) and
odd-parity (dotted vertical lines) energy eigenvalues.
The positive eigenvalues are $\varepsilon_{\beta,n}^{(+)}$,
while the negative eigenvalues are $-\varepsilon_{\beta,n}^{(-)}$. 
The zeroes of $\psi^{(+)}(-\varepsilon_\beta)$ match the
odd-parity eigenvalues $\pm\varepsilon_{\beta,2s}^{(\pm)}$ exactly.
For the even-parity case, there is a mismatch between
the number zeros of $\psi^{\prime(+)}(-\varepsilon_\beta)$
and the number of eigenvalues between $-\varepsilon_{\beta,2}^{(-)}$
and $\varepsilon_{\beta,2}^{(+)}$.
However, the zeroes of $\psi^{\prime(+)}(-\varepsilon_\beta)$
do agree with the even-parity eigenvalues 
$\pm\varepsilon_{\beta,2s-1}^{(\pm)}$ for large $s$.
}
\label{BatemanPrimeZero2}
\end{figure}

In the $\beta=0$ limit, the energy eigenvalues for the even-parity
states are obtained by demanding that the derivative of the wave-function
vanish at $x=0$.
Again, it is not clear whether this notion can be extended to the
$\beta\neq 0$ case with non-zero minimal length.
Granted, we do have a wave-function with a continuous variable
$\chi_\beta=x/\Delta x_{\min}$.
However, as discussed in section~\ref{sec2}B, only its values at the discrete points 
$\chi_\beta=2n+\lambda$, $n\in\mathbb{N}$, have physical meaning
for each choice of $\lambda\in[-1,1)$, so taking the derivative with respect to
$\chi_\beta$ may be problematic.

Indeed,
if we naively impose the condition
\begin{eqnarray}
\psi^{\prime(\pm)}(-\varepsilon_\beta)
& = &
\dfrac{d}{d\chi_\beta}\psi^{(\pm)}(\chi_\beta-\varepsilon_\beta)
\bigg|_{\chi_\beta=0}
\cr
& = &
\mp\dfrac{\partial}{\partial\nu}k
\left(
\dfrac{1}{\kappa^3},
\dfrac{1}{\kappa^3}\pm\varepsilon_\beta
\right)
\;=\; 0
\label{PsiPrimeZero}
\end{eqnarray}
and solve for $\varepsilon_\beta$, 
then the eigenvalues for the even-parity states that we found in the previous section will not be reproduced.
This can be checked numerically, but can also be seen graphically.
In FIG.~\ref{PsiEven2} we plot the wave-functions
\begin{equation}
\psi_{2s-1}^{(\pm)}(\chi_\beta)
\;=\; k\left(
\dfrac{1}{\kappa^3},
\dfrac{1}{\kappa^3}\mp
\left[
\chi_\beta
-\varepsilon_{\beta,2s-1}^{(\pm)}
\right]
\right)
\;,
\end{equation}
for the $\kappa=2$ case using the eigenvalues found in section~\ref{sec2}.
FIG.~\ref{PsiEvenInf} shows the wave-functions in the $\kappa=\infty$ limit.
As is evident from these figures, the derivative of the ground state
wave-function is non-zero at $x=0$.
Despite this, the values of this wave-function sampled at
$\chi_\beta=2j-1$, $j\in\mathbb{N}$, agree with the coefficients
$c_{2j-1}$ derived in section~\ref{sec2}.

The disagreement can also be deduced from the fact that the zeroes of 
$\partial_\nu k(\kappa^{-3},\nu)$ are separated by the zeroes of $k(\kappa^{-3},\nu)$, 
so there is only one zero of $\partial_\nu k(\kappa^{-3},\nu)$
between $\nu_{-1}=-2$ and $\nu_1(\kappa)$, whereas there are two 
energy eigenvalues between $-\varepsilon_{\beta,2}^{(-)}$ 
and $\varepsilon_{\beta,2}^{(+)}$, namely $-\varepsilon_{\beta,1}^{(-)}$
and $\varepsilon_{\beta,1}^{(+)}$.  
Thus, there is a mismatch between the number of zeroes and the number of states.
This situation is shown graphically in FIG.~\ref{BatemanPrimeZero2} for the
$\kappa=2$ case.

Thus, for the even-parity case, imposing Eq.~(\ref{PsiPrimeZero})
will not give us the energy eigenvalues $\varepsilon^{(\pm)}_{\beta,2s-1}$.
However, if we look at the excited state wave-functions in Figs.~(\ref{PsiEven2}) and (\ref{PsiEvenInf}), we note that the derivative of the wave-functions at $x=0$ approaches zero as $n$ is increased. 
This can also be seen in FIG.~\ref{BatemanPrimeZero2} 
where the zeroes of $\Psi^{\prime(+)}(-\varepsilon_\beta)$ farther away from the
origin agree better with the energy eigenvalues.
So the derivative of the wave-function at $x=0$ deviates most from zero
for the ground state, and the deviation is reduced as one looks at
higher and higher excited states.

Physically, this can be understood as due to the existence of the minimal
length ``blurring out'' the position of the origin $x=0$.
This leads to ``phase shifts'' in the wave-functions, the most affected
being the ground-state which has the largest probability amplitude at the origin.
The higher excited states with smaller amplitudes at $x=0$ are less
affected.  
And the odd-parity states, with zero probability amplitude at the origin, are
not affected at all.
This situation is similar to the Coulomb potential problem discussed in
Ref.~\cite{Benczik:2005bh}.
There, the energy eigenvalues of the $s$-wave states were affected non-perturbatively
by the existence of the minimal length, while those for the $\ell\ge 1$ states were not.

\section{Classical Limit}
\label{classical}

\subsection{The classical equation of motion and its solution in the range $\bm{x\ge 0}$}

\begin{figure}[t]
\includegraphics[width=8cm]{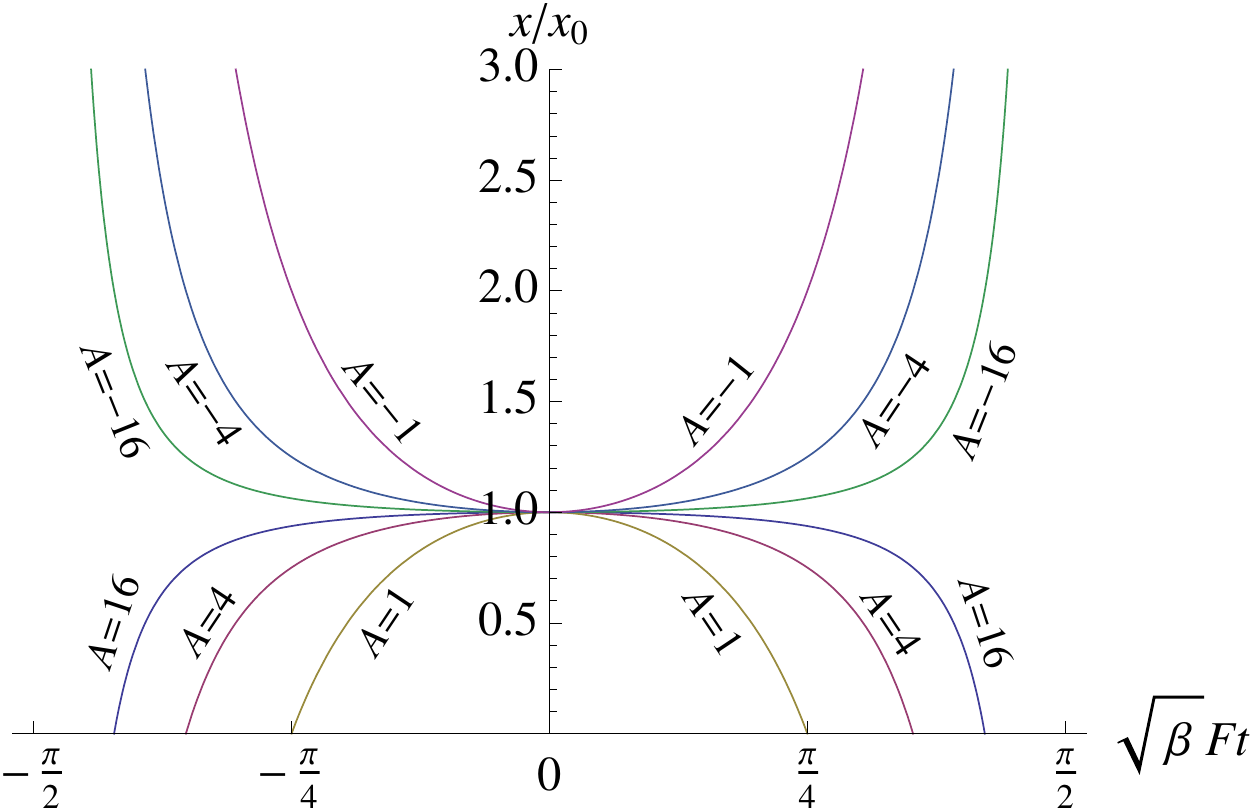}
\caption{The solution to the classical equations of motion for
various values of the parameter $A=2m\beta E$.}
\label{ClassicalX}
\end{figure}

Let us now look at the classical limit of our problem to 
obtain a better understanding of our results.
We assume that the classical limit of the deformed commutation relation, Eq.~(\ref{CommutationRelation}),
is obtained by the usual correspondence between commutators and Poisson brackets:
\begin{equation}
\dfrac{1}{i\hbar}\bigl[\,\hat{A},\,\hat{B}\,\bigr]
\quad\rightarrow\quad\bigl\{\,A,\,B\,\bigr\}\;.
\end{equation}
Therefore, we have
\begin{eqnarray}
\{\,x,\,x\,\} & = & 0\;,\cr
\{\,p,\,p\,\} & = & 0\;,\cr
\{\,x,\,p\,\} & = & (1+\beta p^2)\;.
\label{ModifiedPoisson}
\end{eqnarray}
Our Hamiltonian was
\begin{equation}
H_1 \;=\; \frac{p^2}{2m} + F|x|\;,\qquad F\,>\,0\;,
\label{LinearHamiltonian}
\end{equation}
but if we restrict our attention to motion in the range $x\ge 0$
we can use
\begin{equation}
H'_1 \;=\; \frac{p^2}{2m} + Fx\;,\qquad F\,>\,0\;.
\label{LinearHamiltonian2}
\end{equation}
Then, our equations of motion will be
\begin{eqnarray}
\dot{x} 
& = & \{\,x,\,H'_1 \} \;=\; \dfrac{1}{m}(1+\beta p^2)\,p\;,\cr
\dot{p} 
& = & \{\,p,\,H'_1 \} \;=\; -F(1+\beta p^2)\;.
\label{EOM}
\end{eqnarray}
Note that $\dot{x}$ and $p$ have opposite sign when the mass $m$ is negative.
$\dot{p}$ is also always negative, due to our restriction of attention to the
$x\ge 0$ region.
Changing the variable from $p$ to $\theta$, Eq.~(\ref{thetadef}), these equations become
\begin{eqnarray}
\dot{x} 
& = & \dfrac{1}{m\sqrt{\beta}}
\left[\dfrac{\tan\theta}{\cos^2\theta}\right]
\,=\, \dfrac{1}{2m\sqrt{\beta}}\,\dfrac{d}{d\theta}
\left[\tan^2\theta\right]\;, \cr
\dot{\theta} & = &  -\sqrt{\beta}F\;.
\label{ClassicalEOM}
\end{eqnarray}
The equation for $\theta$ is trivially solved to yield
\begin{equation}
\theta(t) \;=\; -\sqrt{\beta}Ft\;,
\end{equation}
where we have set the clock so that $\theta(0)=0$
($p(0)=0$), that is,
the particle is at the turning point at $t=0$.
The corresponding $t$-dependence of the momentum $p$ is 
\begin{equation}
p(t) 
\;=\; \dfrac{1}{\sqrt{\beta}}\tan[\,\theta(t)\,]
\;=\; -\frac{1}{\sqrt{\beta}}\tan(\sqrt{\beta}Ft)
\;.
\label{pClassical}
\end{equation}
Taking the ratio of $\dot{x}$ to $\dot{\theta}$ we find
\begin{equation}
\dfrac{\dot{x}}{\dot{\theta}}
\;=\; \dfrac{dx}{d\theta} 
\;=\; -\,\dfrac{1}{2m\beta F}\;
\dfrac{d}{d\theta}
\left[\tan^2\theta\right]
\;,
\end{equation}
which can be integrated to yield
\begin{equation}
x(\theta) \;=\; x_0 -\frac{1}{2m\beta F}\tan^2\theta\;,
\label{xtheta}
\end{equation}
where $x_0$ is the turning point at which
$Fx_0=E$, the particle's total mechanical energy.
Since $\theta(t)=-\sqrt{\beta}Ft$ we obtain
\begin{equation}
x(t) 
\;=\; x_0 -\dfrac{1}{2m\beta F}\tan^2(\sqrt{\beta}Ft) \;.
\label{xClassical}
\end{equation}
It is straightforward to show that in the limit $\beta\rightarrow 0$, 
this solution reduces to
\begin{eqnarray}
\lim_{\beta\rightarrow 0} x(t) & = & x_0 -\dfrac{1}{2}\left(\dfrac{F}{m}\right)t^2\;,\cr
\lim_{\beta\rightarrow 0} p(t) & = & -Ft\;.
\end{eqnarray}
Note that $F/m$ is the acceleration, which can be either positive or
negative depending on the sign of $m$.
Rewriting Eq.~(\ref{xClassical}) as
\begin{equation}
\dfrac{x(t)}{x_0} 
\;=\; 1
- \dfrac{1}{A}\tan^2(\sqrt{\beta}Ft)\;,\quad A=2m\beta E\;,
\end{equation}
we plot this solution for various values of 
the dimensionless parameter $A=2m\beta E$ in FIG.~\ref{ClassicalX}.
Negative values of $A$ correspond to the negative mass case.

\subsection{Continuation into the $\bm{x\le 0}$ range}

\begin{figure}[t]
\hspace{0.1cm}\includegraphics[width=8cm]{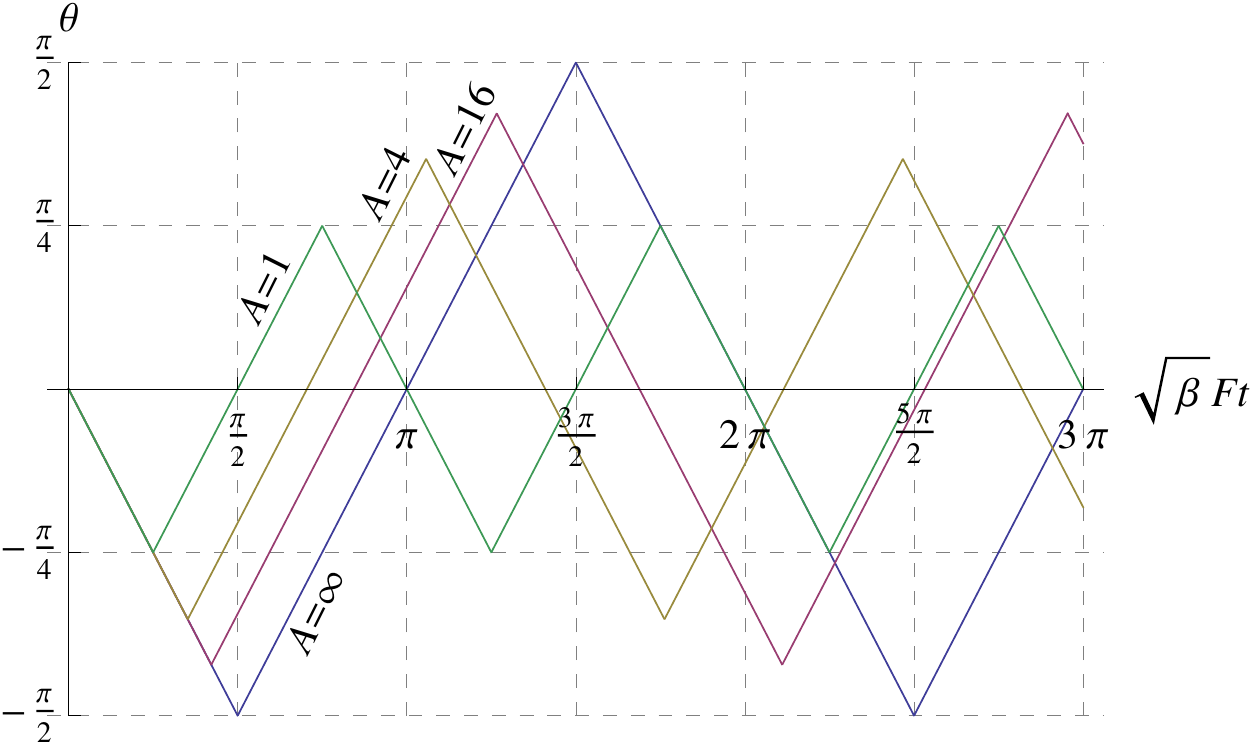}\\
\hspace{0.2cm}\includegraphics[width=8cm]{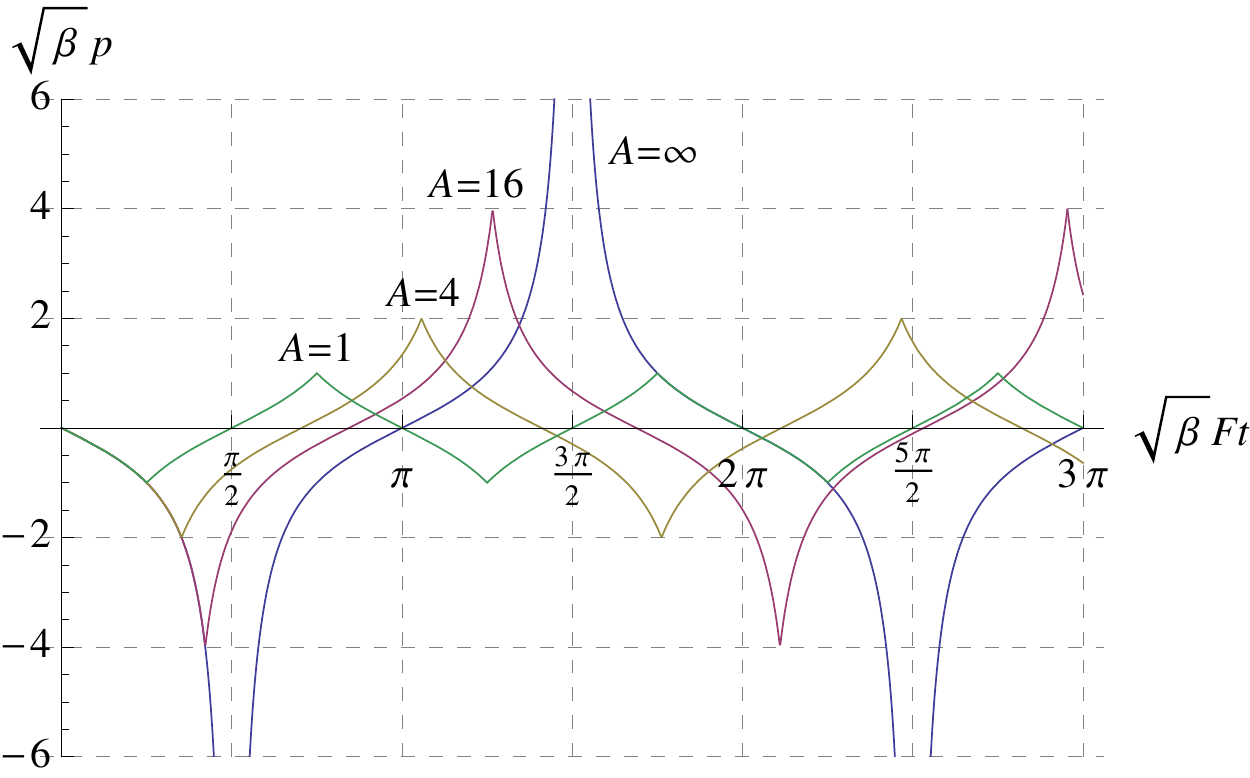}\\
\includegraphics[width=8.2cm]{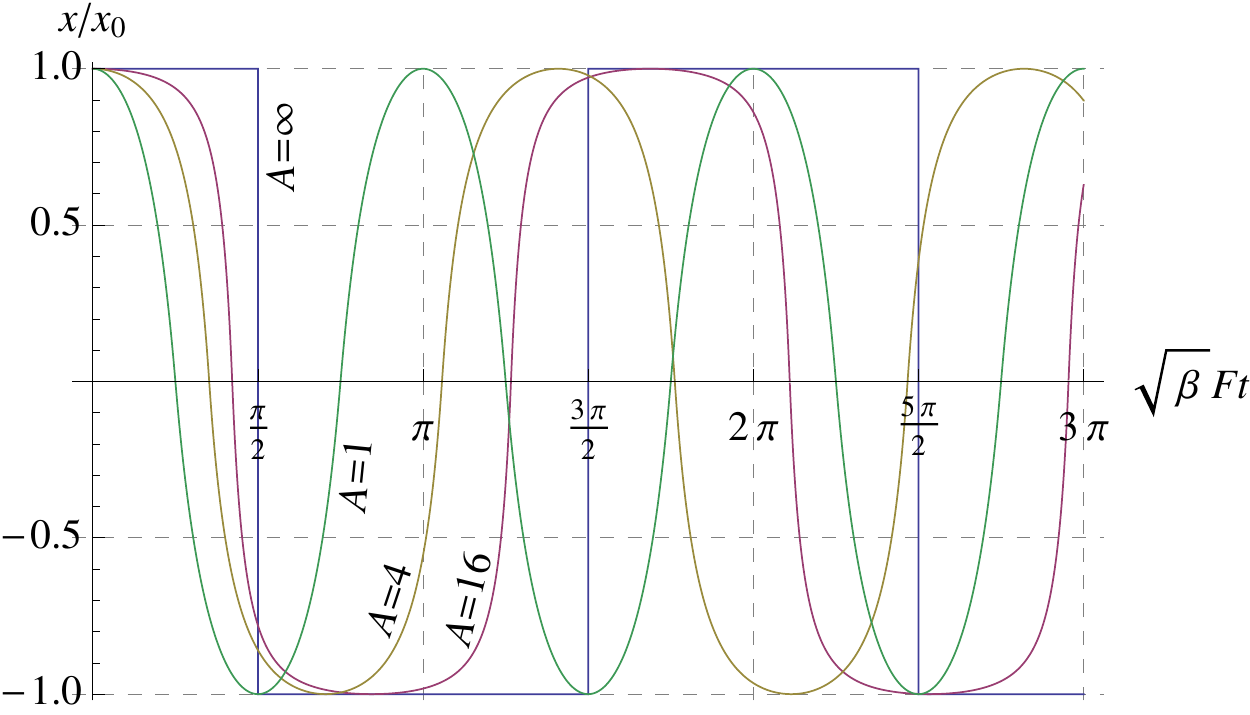}
\caption{The solution to the classical equations of motion for
various values of the parameter $A=2m\beta E>0$.}
\label{TdependenceMplus}
\end{figure}

\begin{figure}[t]
\includegraphics[width=8cm]{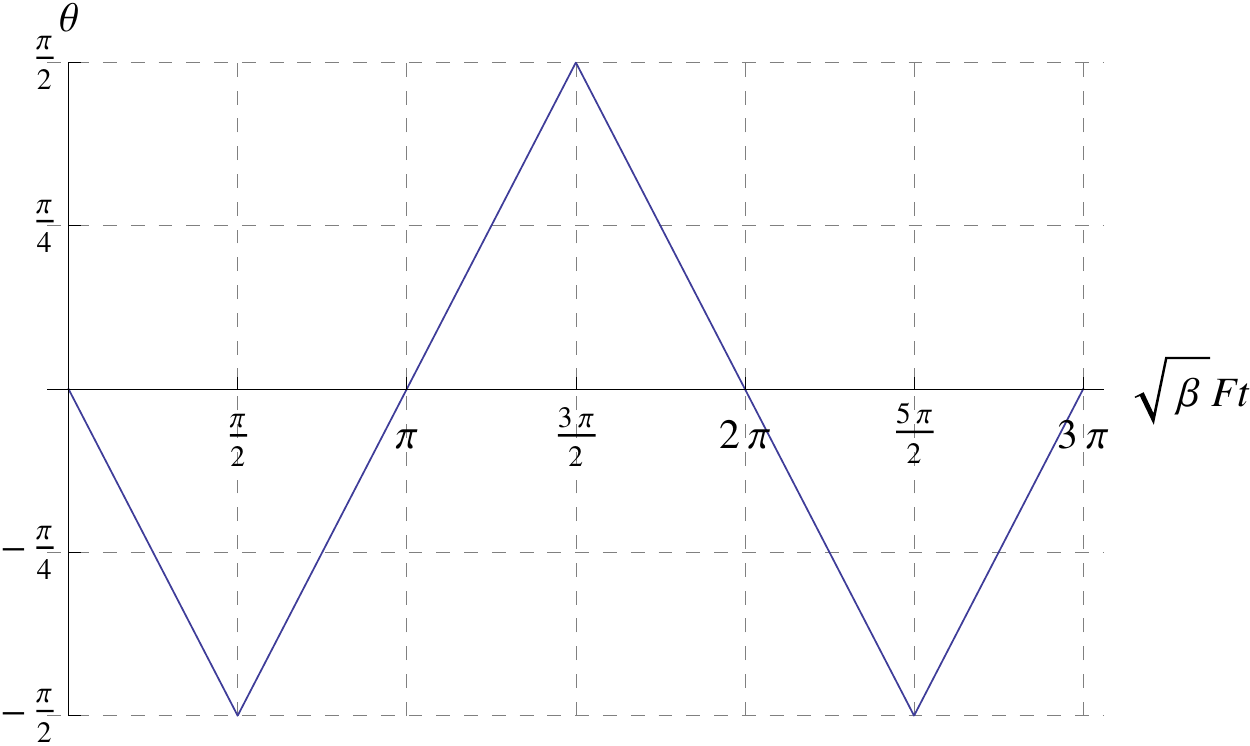}\\
\includegraphics[width=8cm]{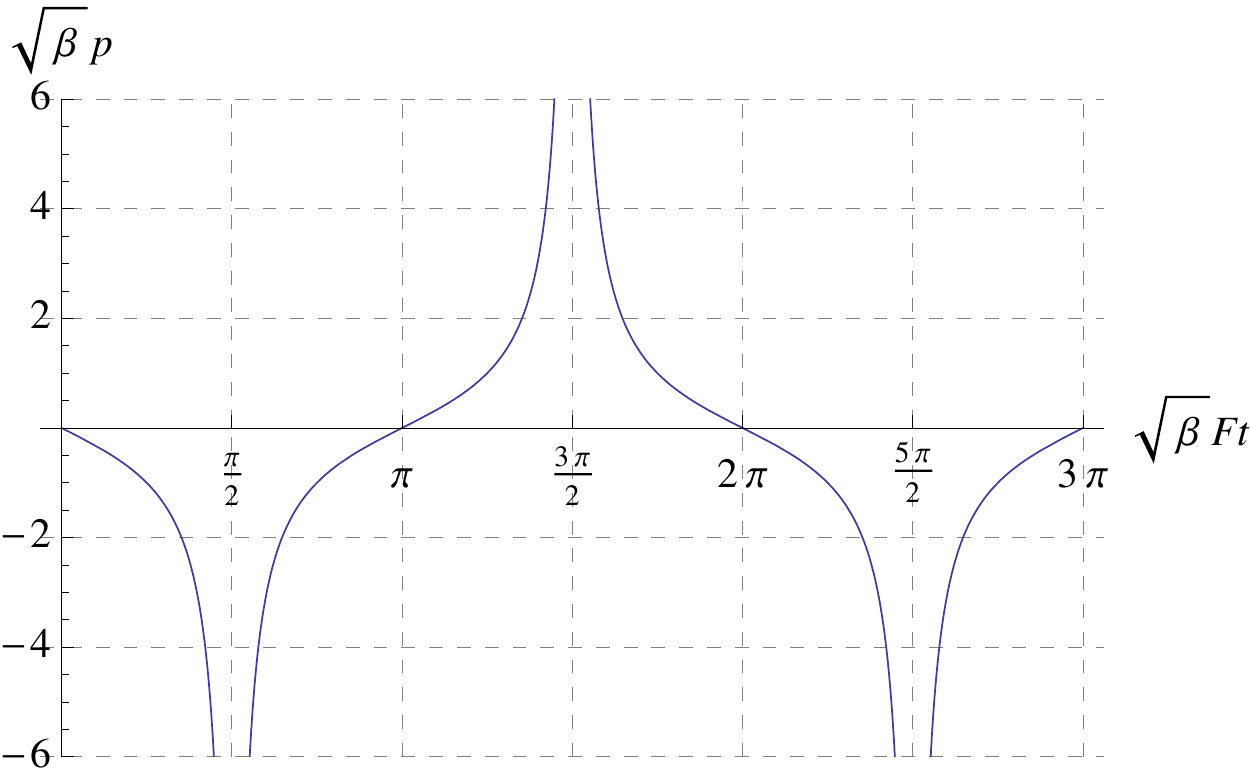}\\
\includegraphics[width=8cm]{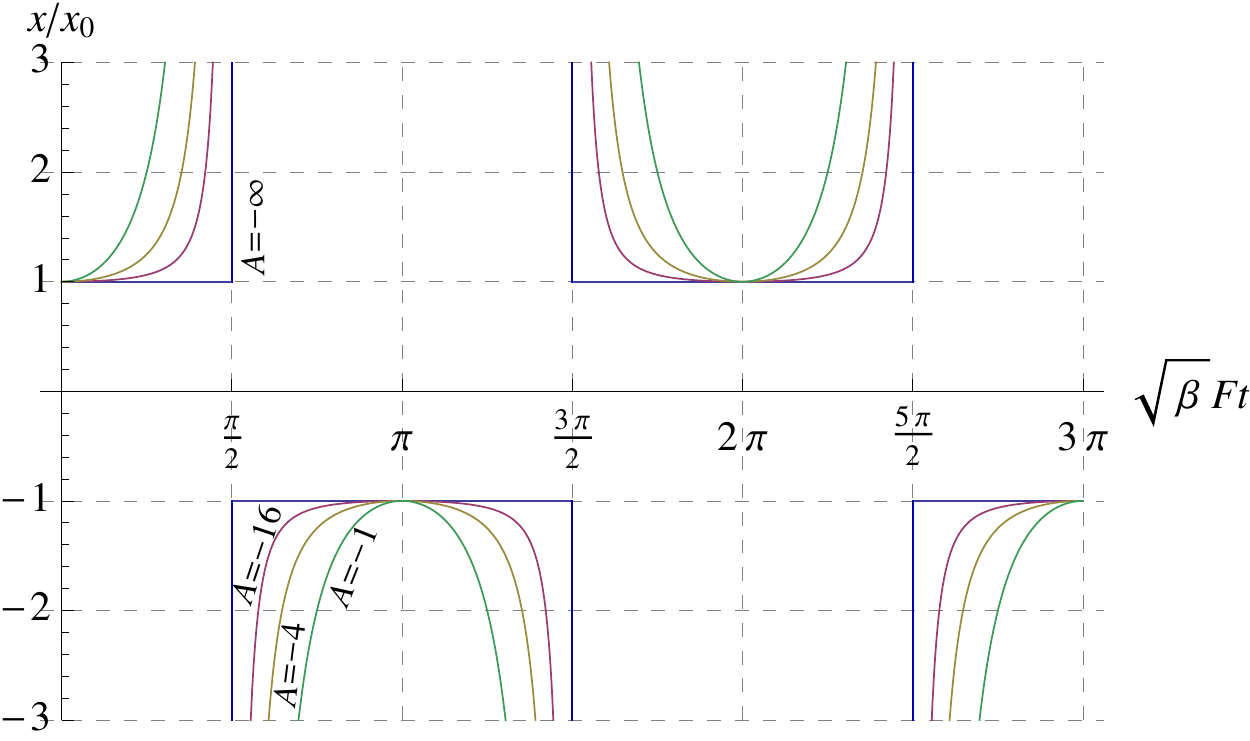}
\caption{The solution to the classical equations of motion for
various values of the parameter $A=2m\beta E<0$.
It is assumed that $x$-space is compactified at $x=\infty$.
The behaviors of $\theta(t)$ and $p(t)$, and the
oscillation period are independent of the value of $A$.
}
\label{TdependenceMminus}
\end{figure}

\subsubsection{Positive Mass Case}

In the positive mass case, the particle starts out from $x=x_0$ at time $t=0$,
and will reach $x=0$ at time
\begin{equation}
t 
\;=\; \frac{1}{\sqrt{\beta}F}\tan^{-1}\left(\sqrt{A}\,\right)
\;\equiv\; \dfrac{T_+}{4} \;,
\label{Tplus}
\end{equation}
at which point the motion will connect smoothly to the parity flipped solution for the
range $x\le 0$.
The particle will oscillate back and forth between the positive and negative turning points
with period $T_+$ as shown in FIG.~\ref{TdependenceMplus}.

\subsubsection{Negative Mass Case}

When the mass is negative, the particle starts out from $x=x_0$ at time $t=0$,
and will reach $x=+\infty$ at time
\begin{equation}
t \;=\; \frac{\pi}{2\sqrt{\beta}F} \;=\; \dfrac{T_-}{4}\;.
\label{Tminus}
\end{equation}
As in the harmonic oscillator case discussed in Ref.~\cite{Lewis:2011fg}, 
we assume that $x$-space is compactified at $x=\infty$ so that
the particle will oscillate between the positive and negative turning points
via $x=\infty$ with period $T_-$
as shown in FIG.~\ref{TdependenceMminus}.

\begin{figure}[t]
\includegraphics[width=7cm]{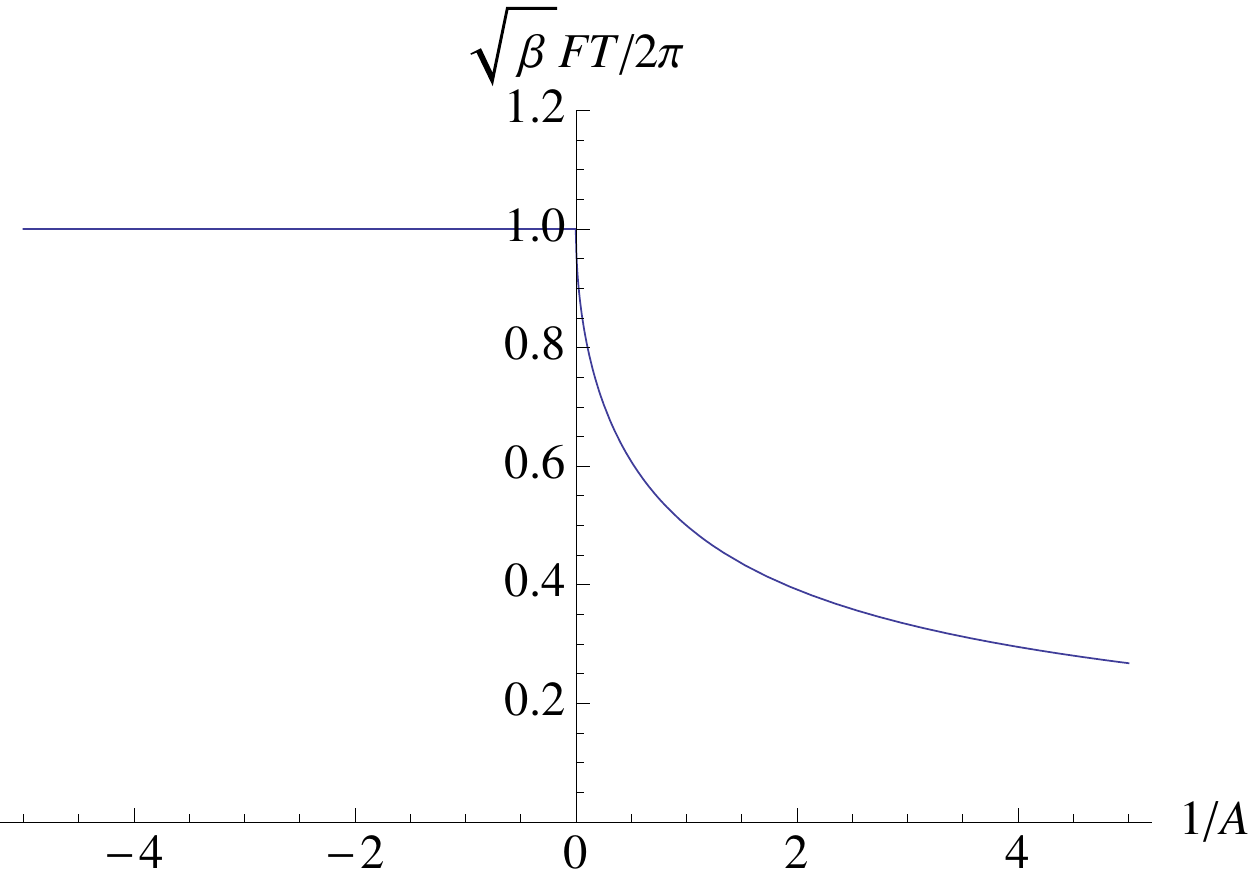}\\
\caption{The dependence of the oscillation period $T$ 
on the value of $A=2m\beta E$.
Note that $T$ is independent of $A$ when $A$ is negative.
}
\label{AdependenceofT}
\end{figure}

Note that while the oscillation period depends on
$A=2m\beta E$ when the mass is positive,  
it is independent of $A=2m\beta E$ when the mass is negative.
See FIG.~\ref{AdependenceofT}.
We note that this independence of the period on the energy $E$ is a feature
shared with the canonical harmonic oscillator.
Given that the quantum mechanical versions of both have equally spaced
energy eigenvalues, perhaps the two characteristics are connected.
Note also that due to the finiteness (non-infiniteness) of $T_-$, the particle has 
non-zero probability of being at finite $x$. 
This can be understood as what makes bound states with discrete positive
energy eigenvalues possible even when the mass is negative, effectively inverting the potential.

\subsection{Classical Probabilities}

As we have just seen, in the classical limit the 
particle bounces back and forth between the two turning points
via $x=0$ when the mass is positive, and via $x=\infty$ when the mass is negative.
Here, we calculate the classical probability densities in $|x|$- and $|p|$-spaces
from which we calculate the classical uncertainties.

\subsubsection{Positive Mass Case}

For the positive mass case, we note that
\begin{equation}
\dfrac{T_+}{4}
\;=\; \int_0^{T_+/4}dt 
\;=\; \int_{x_0}^{0}\dfrac{dx}{\dot{x}}
\;=\; \int_{0}^{-\sqrt{A/\beta}}\dfrac{dp}{\dot{p}}
\;,
\end{equation}
for the first one-quarter period of oscillation from $t=0$ to $t=T_+/4$.
Thus, we can identify
\begin{equation}
P(x)\;=\; -\dfrac{4}{T_+\dot{x}}\;,\quad\mbox{and}\quad
\tilde{P}(p)\;=\; -\dfrac{4}{T_+\dot{p}}\;,
\end{equation}
as the probability densities 
for the ranges $0<x<x_0$ and $-\sqrt{A/\beta}<p<0$, respectively.
Simply replacing $x$ and $p$ with their absolute values in
the final expressions will give us the probability densities
in $|x|$- and $|p|$-spaces.
This yields
\begin{eqnarray}
\dfrac{\tilde{P}\left(|p|\right)}{\sqrt{\beta}}
& = &
\dfrac{1}{\tan^{-1}(A)}\left(\dfrac{1}{1+\beta |p|^2}\right)
\;,\cr
x_0 P\left(|x|\right)
& = & \dfrac{1}{2\tan^{-1}(A)}
\dfrac{\sqrt{A}}{\sqrt{1-\dfrac{|x|}{x_0}}
\left[1+A\left(1-\dfrac{|x|}{x_0}\right)\right]}
\;.
\cr
& &
\end{eqnarray}
Note that $|x|$ and $|p|$ are restricted to the ranges
$0\le |x|\le x_0$ and $0\le|p|\le\sqrt{A/\beta}$, respectively.

\subsubsection{Negative Mass Case}

For the negative mass case, we note
\begin{equation}
\dfrac{T_-}{4}
\;=\; \int_0^{T_-/4}dt 
\;=\; \int_{x_0}^{\infty}\dfrac{dx}{\dot{x}}
\;=\; \int_{0}^{-\infty}\dfrac{dp}{\dot{p}}
\;,
\end{equation}
and make the following identifications:
\begin{equation}
P(x)\;=\; \dfrac{4}{T_-\dot{x}}\;,\quad\mbox{and}\quad
\tilde{P}(p)\;=\; -\dfrac{4}{T_-\dot{p}}\;.
\end{equation}
This yields
\begin{eqnarray}
\dfrac{\tilde{P}\left(|p|\right)}{\sqrt{\beta}}
& = &
\dfrac{2}{\pi}\left(\dfrac{1}{1+\beta |p|^2}\right)
\;,\cr
x_0 P\left(|x|\right)
& = & \dfrac{1}{\pi}
\dfrac{\sqrt{|A|}}{\sqrt{\dfrac{|x|}{x_0}-1}
\left[1+|A|\left(\dfrac{|x|}{x_0}-1\right)\right]}
\;.
\cr
& &
\end{eqnarray}
In this case, the ranges of $|x|$ and $|p|$ are
$x_0\le |x|$ and  $0\le |p|$, respectively.
That is, both ranges are infinite.

\begin{figure}[t]
\includegraphics[width=8cm]{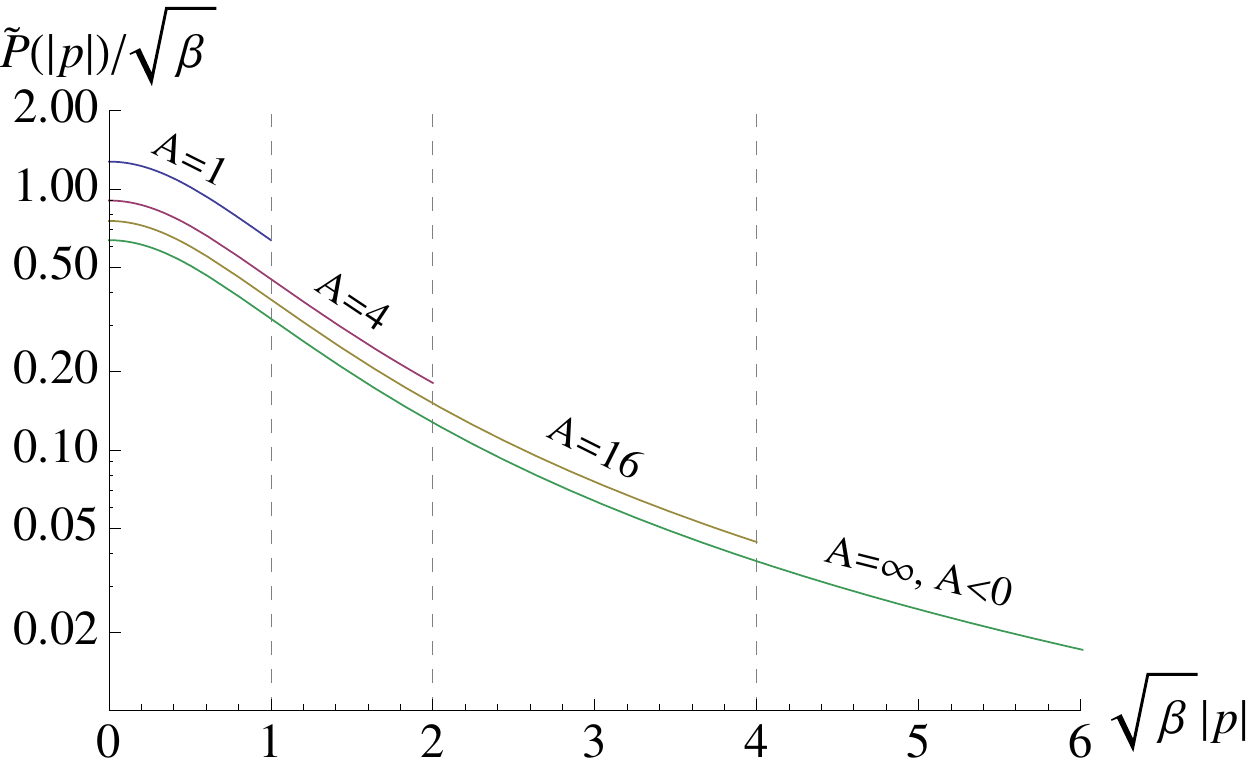}
\includegraphics[width=8cm]{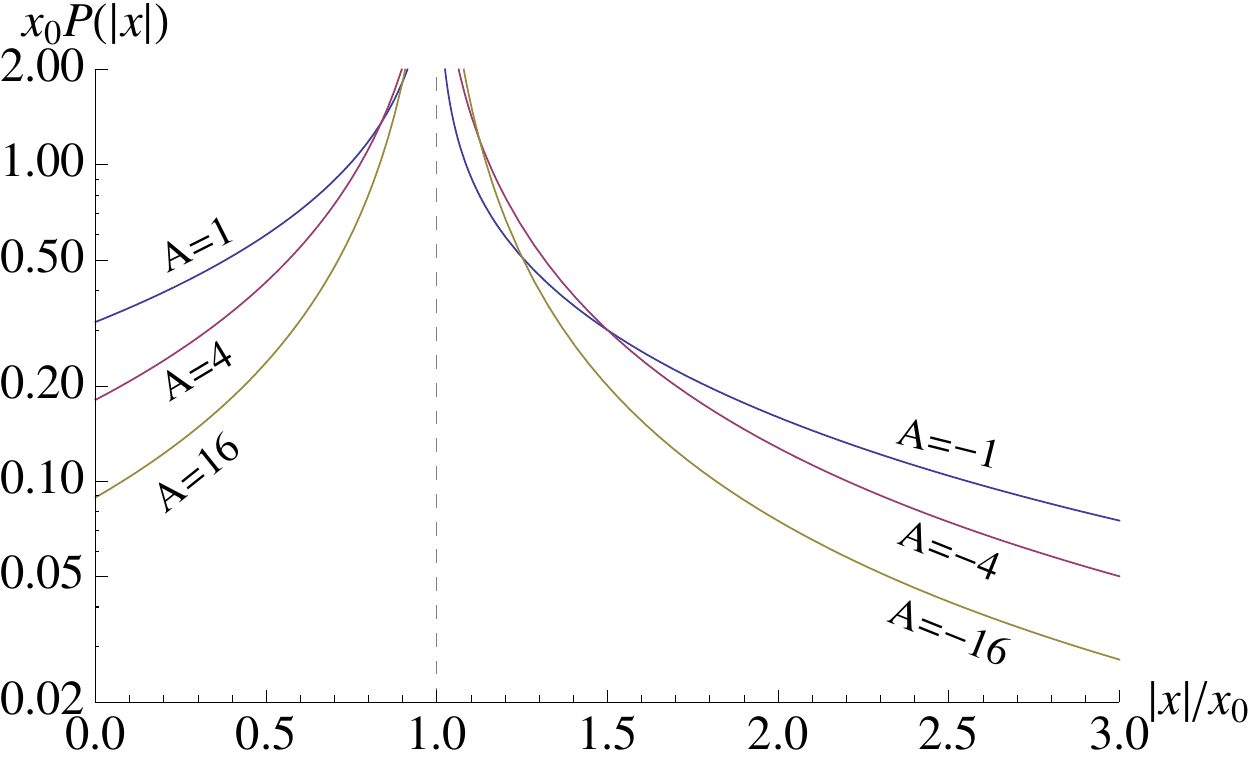}
\caption{The classical probability distributions in 
$|p|$- and $|x|$-spaces for
several values of the parameter $A=2m\beta E$.}
\label{ProbabilityDistribution}
\end{figure}

\subsubsection{Uncertainties}

The above probability distributions are
plotted for several values of $A$, both positive and negative,
in FIG.~\ref{ProbabilityDistribution}.
Looking at the $|x|$-space distribution, 
we see the probability density diverges at $|x|=x_0$
so the particle has the highest probability of being found in the vicinity of the turning points.
In the infinite (both positive and negative) mass limit, 
the distribution will become a $\delta$-function located there.

For the positive mass case ($A>0$), the ranges of both $x$ and $p$ are finite, so
the expectations values of $x^2$ and $p^2$ are also finite.
However, for the negative mass case ($A<0$) both ranges are infinite,
and for large $x$ and $p$ we find
\begin{equation}
P\left(|x|\right)\;\sim\;|x|^{-3/2}\;,\qquad
\tilde{P}\left(|p|\right)\;\sim\;|p|^{-2}\;,
\end{equation}
that is
\begin{equation}
x^2 P\left(|x|\right)\;\sim\;|x|^{1/2}\;,\qquad
p^2 \tilde{P}\left(|p|\right)\;\sim\;\mbox{constant}\;.
\end{equation}
and it is clear that the expectation values of $x^2$ and $p^2$ will both diverge.
Thus the divergences of $\Delta x$ and $\Delta p$ can be seen in the classical limit also.

\subsubsection{Comparison with Quantum Probablities}

Let us compare the classical probability distributions derived above
with the quantum ones to confirm that the distributions
follow each other.  
Here, we only consider the negative mass case where $\Delta x$ and $\Delta p$ diverge.

\begin{figure}[t]
\includegraphics[width=8cm]{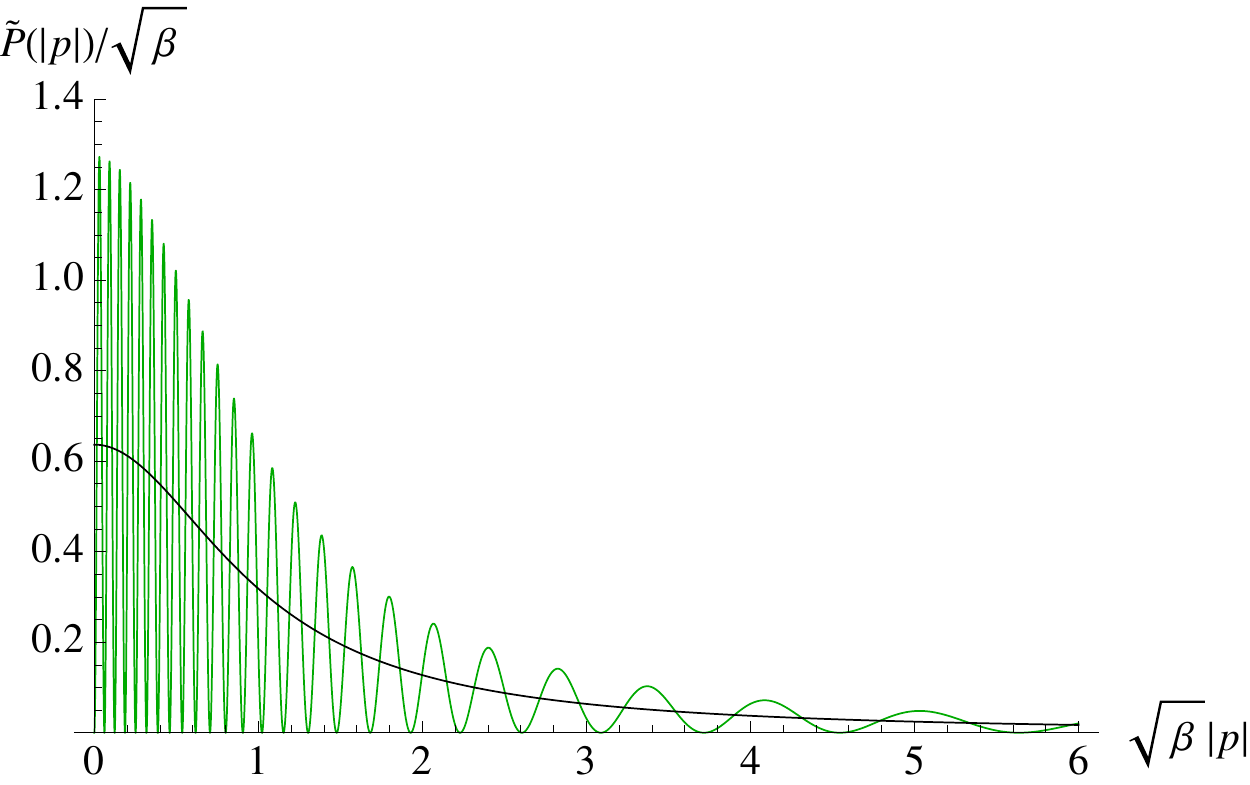}
\includegraphics[width=8cm]{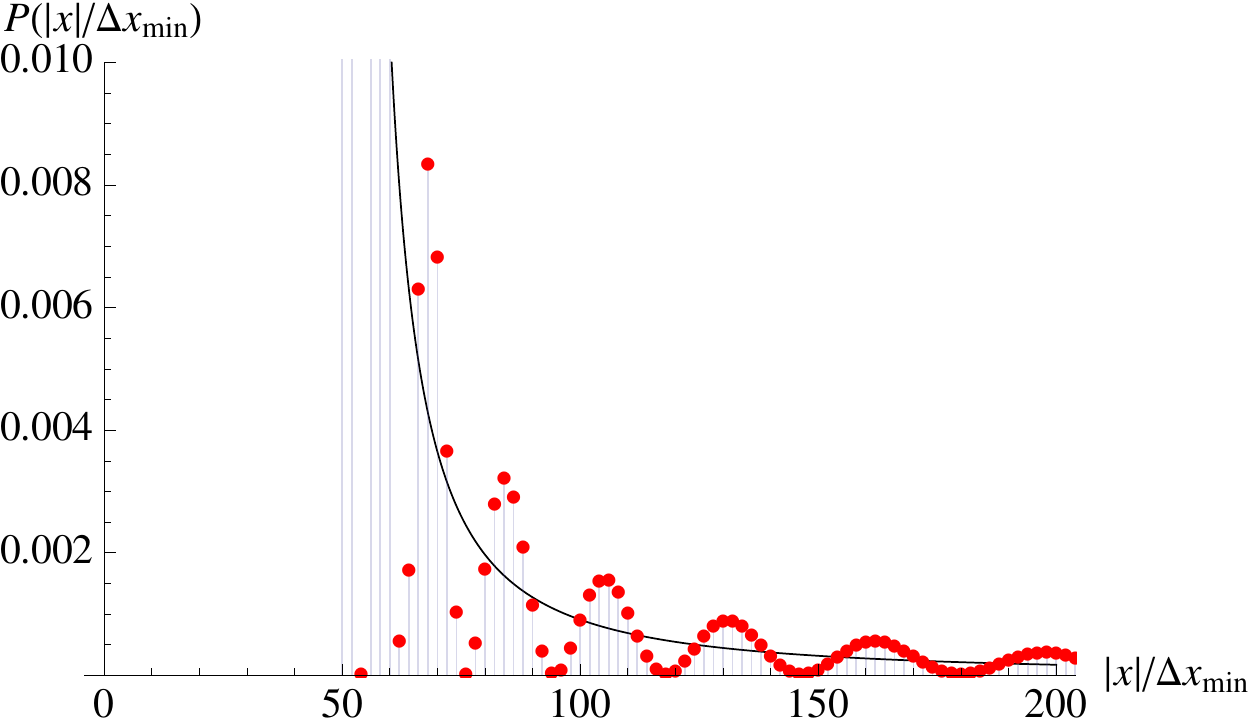}
\caption{
Comparison of the quantum and classical probabilities for the case $\kappa=1$, $n=50$.
Top: quantum (green) and classical (black)
probability densities in $|p|$-space.
Bottom: the discrete quantum probabilities (red dots) and the
properly normalized classical distribution function (black line) 
in $|x|$-space.
}
\label{CQcomparison}
\end{figure}

To see the classical$\leftrightarrow$quantum correspondence,
we note that
\begin{eqnarray}
|A| 
& = & 2|m|\beta E  
\;=\; \varepsilon_\beta\kappa^3
\;,\cr
x_0
& = & \dfrac{E}{F} 
\;=\; \varepsilon_\beta \Delta x_{\min}
\;,\cr
\dfrac{x}{x_0}
& = & \left(\dfrac{x}{\Delta x_{\min}}\right)\dfrac{1}{\varepsilon_\beta}
\;=\; \left(\dfrac{x}{\Delta x_{\min}}\right)\dfrac{\kappa^3}{|A|}
\;.
\end{eqnarray}
The value of $|A|$ that corresponds to the $n$-th negative-mass quantum state is
\begin{equation}
|A_{n}^{(-)}|
\;=\; \varepsilon_{\beta,n}^{(-)}\kappa^3
\;=\; 1 + n\kappa^3\;,
\end{equation}
cf. Eq.~(\ref{NegativeMassGuess}).
Since the eigenvalues of $|\hat{x}|$ are discretized to be integer multiples of $\Delta x_{\min}$,
the ratio $x/x_0$ should be replaced by
\begin{equation}
\dfrac{x}{x_0} 
\;=\;
\left(\dfrac{x}{\Delta x_{\min}}\right) \dfrac{\kappa^3}{|A_{n}^{(-)}|}
\;\rightarrow\; \dfrac{\kappa^3 k}{|A_{n}^{(-)}|}
\;,\qquad
k\in\mathbb{N}\;,
\end{equation}
with
\begin{equation}
\int\dfrac{dx}{x_0} \;\rightarrow\; \sum_{k}\dfrac{\kappa^3}{|A_{n}^{(-)}|}\;,
\end{equation}
and the discretized classical probabilities which correspond to
the $n$-th quantum state become
\begin{equation}
x_0 P(|x|) 
\;\rightarrow\; \dfrac{\kappa^3}{|A_{n}^{(-)}|}\,P(k\Delta x_{\min}) 
\;=\; \dfrac{1}{n+\dfrac{1}{\kappa^3}}\,P(k\Delta x_{\min})\;.
\end{equation}
This function should be compared to 
$|c_k|^2$, where
$c_k$ are the expansion coefficients of section~\ref{sec2}.
However, since every other $c_k$ is zero, we will instead plot
$|c_k|^2/2$ only when it is non-zero for the ease of comparison.
This is shown 
in FIG.~\ref{CQcomparison}
for the $\kappa=1$, $n=50$ case.
Comparison between the classical and quantum probability densities is also shown
in $p$-space where
the quantum probability distribution is obtained
by using the $\theta$-space wave-functions from Eq.~(\ref{NegativeMassWaveFunctions}),
and changing the variable to $p=\tan\theta/\sqrt{\beta}$.
As can be seen, smoothing out the quantum distributions will give us the classical ones.

\section{Summary and Conclusions}
\label{conclusions}

In this paper, we work out the eigenvalues and eigenfunctions of the
Hamiltonian $\hat{H}_1$, Eq.~(\ref{H1}), when the position and momentum operators
are assumed to obey the deformed commutation relation Eq.~(\ref{CommutationRelation}).
As in the harmonic oscillator case discussed in a previous publication \cite{Lewis:2011fg},
we find that $\hat{H}_1$ allows for an infinite ladder of discrete positive 
eigenvalues, not just when the mass is positive, but also when the mass is negative.
The energy eigenvalues for the negative mass case are evenly spaced.
Calculating the uncertainties $\Delta x$ and $\Delta p$ for the corresponding eigenstates,
we find that for the positive mass case $\Delta x\sim 1/\Delta p$, 
and as $1/m\rightarrow +0$ the uncertainties approach the curve given in Eq.~(\ref{InfiniteMassCurve}).
The same curve separated the positive- and negative-mass regions in $\Delta p$-$\Delta x$ space
for the harmonic oscillator \cite{Lewis:2011fg}.
However, if $1/m$ is decreased through zero so that the mass turns negative,
we find that, instead of the points $(\Delta p,\Delta x)$ continuing on to 
smooth curves with a $\Delta x\sim\Delta p$ behavior as in the harmonic oscillator case, both
$\Delta x$ and $\Delta p$ diverge.
Thus, the $\Delta x\sim\Delta p$ behavior cannot be seen in the eigenstates of
$\hat{H}_1$ with negative mass.\footnote{%
We have explored the possibility of `regularizing' these divergences by deforming the
$\theta$-space integration into the upper complex plane.
Unfortunately, this attempt leads to finite but negative values for $\Delta x^2$.
}

Taking the classical limit, we find that the time it takes for the
negative-mass particle to reach infinity from the turning points is finite (non-infinite).
Consequently, the particle has a finite (non-zero) probability to be found
at finite $x$. 
It also means that we must compactify $x$-space with an addition of the infinity point $x=\infty$
and allow the particle to oscillate between the two turing points via $x=\infty$.
These points can be understood as what 
make the existence of bound states with discrete energy eigenvalues possible
even when the mass is negative.

Furthermore, calculating the moments of the classical probability densities in $x$- and $p$-spaces, 
we find that the uncertainties $\Delta x$ and $\Delta p$ diverge in the classical limit also. 
This is due to the tails of the probability
distributions not falling fast enough as $|x|\rightarrow\infty$ and $|p|\rightarrow\infty$.
Thus, though the negative-mass particle spends enough time near the turning points
to allow for bound states, it does not spend enough time there to allow for finite $\Delta x$.

A curious fact is that the classical period of oscillation of the negative-mass particle via $x=\infty$ is 
independent of the particle energy.
Together with the fact that the quantum energy eigenvalues are evenly spaced,
this suggests either a direct connection between the negative-mass case and the canonical harmonic oscillator, or a common property shared between the two that would lead to this result.
A related question would be whether `coherent states' exist for the
negative mass V-shaped potential where $\vev{x}$ obeys the classical equation of motion, just as in the canonical harmonic oscillator case.  Would such a state have finite $\Delta x$ which would `localize' it 
in some fashion?
These points will be further explored in future publications.

Answering the question posed in the introduction,
we find that the $\Delta x\sim\Delta p$ behavior seen in 
the eigenstates of the negative-mass harmonic oscillator Hamiltonian \cite{Lewis:2011fg} is not universal.
Simply making the mass negative for other potentials will not necessarily lead to 
a similar behavior.
Assuming the deformed commutation relation, Eq.~(\ref{CommutationRelation}),
between $\hat{x}$ and $\hat{p}$ does not guarantee that
$\Delta x\sim\Delta p$ can be realized.

What if we looked at particles in other potentials?
Consider the Hamiltonian
\begin{equation}
\hat{H}_q \;=\; \dfrac{\hat{p}^2}{2m} + F|\hat{x}|^q\;,
\end{equation}
where the action of the operator $|\hat{x}|^q$ on an eigenstate
of $\hat{x}^2$ with eigenvalue $\sigma^2$ ($\sigma>0$) is assumed to be
\begin{equation}
|\hat{x}|^q\ket{\sigma^2}\;=\;\sigma^q\ket{\sigma^2}\;.
\end{equation}
Since this class of Hamiltonians reduce to $F|\hat{x}|^q$ when the
mass is taken to infinity, in that limit the eigenstates of $\hat{H}_q$ will 
reduce to those of $\hat{x}^2$ with uncertainties on the curve of Eq.~(\ref{InfiniteMassCurve}).
Thus an educated guess would be that $\Delta x\sim 1/\Delta p$ for these
Hamiltonians as well, as long as the mass is kept positive, and that the points
$(\Delta p,\Delta x)$ will follow trajectories similar to those shown in FIG.~\ref{DelPDelX} which 
terminate on the Eq.~(\ref{InfiniteMassCurve}) curve.

For the negative mass case, consider
the classical limit in which the equation of motion in the range $x\ge 0$ will be given by
\begin{eqnarray}
\dot{x} 
& = & -\dfrac{1}{|m|}(1+\beta p^2)\,p\;,\cr
\dot{p} 
& = & -qF(1+\beta p^2)\,x^{q-1}\;.
\label{EOMq}
\end{eqnarray}
Using the conservation of energy, 
\begin{equation}
E \;=\; -\dfrac{p^2}{2|m|}+F x^q\;,
\end{equation}
we can write $x$ in terms of $p$ and vice versa:
\begin{eqnarray}
x & = & \left[\dfrac{1}{F}\left(E+\dfrac{p^2}{2|m|}\right)\right]^{1/q}\;\sim\;p^{2/q} \;,\cr
p & = & \sqrt{2|m|(Fx^q-E)}\;\sim\;x^{q/2}\;.\vphantom{\Bigg|}
\end{eqnarray}
Thus
\begin{eqnarray}
P(x) & \;\sim\; & \dot{x}^{-1} \;\sim\; p^{-3} \;\sim\; x^{-3q/2}\;,\cr
\tilde{P}(p) & \;\sim\; & \;\dot{p}^{-1} \;\sim\; p^{-2}x^{1-q} \;\sim\; p^{-2(2-1/q)}\;,
\end{eqnarray}
and
\begin{eqnarray}
x^2 P(x)         & \;\sim\; & x^{2-3q/2}\;,\cr
p^2 \tilde{P}(p) & \;\sim\; & p^{-2(1-1/q)}\;. 
\end{eqnarray}
So for the classical expectation values of $x^2$ and $p^2$ to be finite,
we need
\begin{equation}
2\;<\;q\;.
\end{equation}
Since all potentials with $q<2$ will have divergent 
$\Delta x$ and $\Delta p$ classically,
this suggests that their quantum counterparts are also divergent,
just as in the $q=1$ case.
The borderline $q=2$ case seems exceptional.
This corresponds to the harmonic oscillator discussed in Ref.~\cite{Lewis:2011fg}
where we had $x^2 P(x)\sim x^{-1}$ and $p^2\tilde{P}(p)\sim p^{-1}$.
So the classical uncertainties for this case 
are log-divergent even though the quantum uncertainties are not.
It turns out that for the harmonic oscillator, 
the asymptotic quantum probabilities in $x$- and $p$-spaces
decay faster than their classical counterparts
provided that the condition of Eq.~(\ref{HOcondition}) is satisfied, which was necessary for 
normalizable bound states to exist in the first place.
These considerations imply that to look for the $\Delta x\sim\Delta p$ behavior
one should look at potentials with $q$ larger than 2.

One final problem we would like to point out is the question of how
the uncertainties $\Delta x$ and $\Delta p$ should be defined for 
a particle in the potential of  Eq.~(\ref{HalfPotential}),
which corresponds to a particle bouncing in a uniform gravitational field
with a rigid floor at $x=0$ \cite{Brau:2006ca,Benczik:2007we,Nozari:2010qy}.
The energy eigenvalues for this problem are the same as the odd-parity states of $\hat{H}_1$,
and the $x/\Delta x_{\min}$-space eigenfunctions corresponding to the lowest five
will be the same as those shown in FIG.~\ref{PsiOdd2}.
The eigenfunctions in the infinite mass limit will be as those shown in
FIG.~\ref{PsiOddInf}, with the particle localized at $x/\Delta x_{\min}=n$ for
the $n$-th state.
Unlike the V-shaped potential case, however, the wave-function 
does not continue into the $x<0$ region with a negative turning point at 
$x/\Delta x_{\min}=-n$ which shares in the localization.
So a naive calculation of $\Delta x$ for the infinite mass states would yield zero, 
in clear violation of Eq.~(\ref{MLUR}).
The crux of this problem seem to lie in how one should take into account the
fuzziness of the location of the infinite potential wall at $x=0$ in the presence of the
minimal length.
We have considered a variety of ways in which this problem may be avoided
but are yet to find a satisfactory resolution.
Here, we only allude to the existence of this problem and 
refrain from discussing it further.

\section*{Acknowledgments}

We wish to acknowledge the work of Sandor Benczik in his Ph.D. thesis \cite{Benczik:2007we}
which provided a starting point for this paper.
We are also indebted to Lay Nam Chang, George Hagedorn, Will Loinaz, Sourav Mandal, Djordje Minic,
Eric Sharpe, and Uwe T\"auber for
helpful discussions.  
This work was supported in part by the U.S. Department of Energy, 
grant number DE-FG05-92ER40677, task A (ZL and TT) and by
the World Premier International Research Center Initiative, MEXT, Japan (TT).

\newpage
\appendix

\section{Review of the Canonical $\bm{\beta=0}$ Case}
\label{CanonicalReview}

\subsection{Solution in Coordinate Space}

\begin{figure}[t]
\includegraphics[width=8cm]{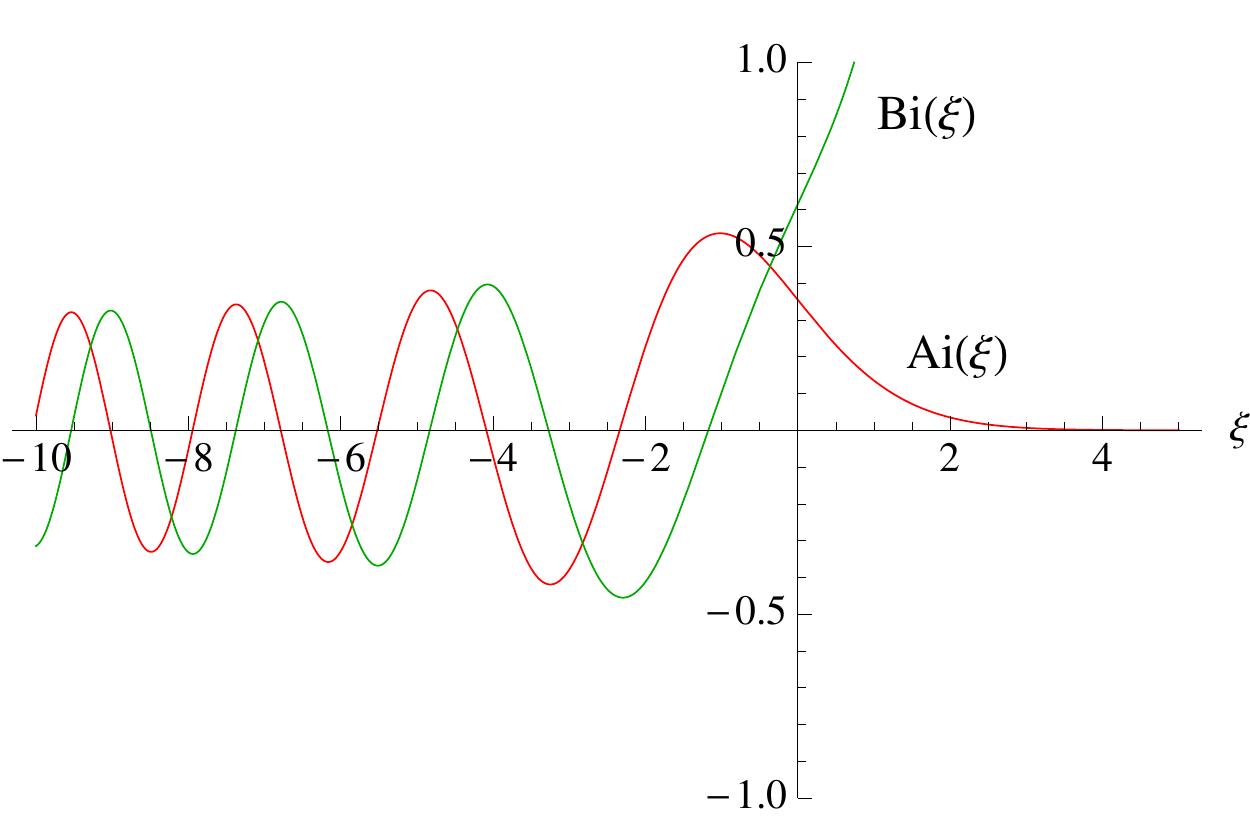}
\caption{The Airy functions $\Ai(\xi)$ and $\Bi(\xi)$.}
\label{AiryAiBiGraphs}
\end{figure}

As mentioned in the main text,
the eigenvalues and eigenstates of
$\hat{H}_1$, Eq.~(\ref{H1}), in the canonical $\beta=0$, $m>0$ case are obtained by solving the
Schr\"odinger equation for $\hat{H}'_1$, Eq.~(\ref{H1prime}),
in the region $x\ge 0$,
and then imposing the boundary condition
$\psi'(0)=0$ or $\psi(0)=0$ at $x=0$
to obtain the parity even and odd states, respectively.
The said Schr\"odinger equation is
\begin{equation}
\left(-\dfrac{\hbar^2}{2m}\,\dfrac{d^2}{dx^2} + Fx\right)\psi(x) \;=\; E\,\psi(x)\;,
\label{SchrodingerPlus}
\end{equation}
which possesses a characteristic length scale 
given by
\begin{equation}
a \;=\; \left[\dfrac{\hbar^2}{2mF}\right]^{1/3}\;.
\label{adefappendix}
\end{equation}
Defining dimensionless variable and eigenvalue by
\begin{equation}
\chi \;\equiv\; \dfrac{x}{a}\;,\qquad
\varepsilon_a \;\equiv\; \left(\dfrac{2mE}{\hbar^2}\right)a^2 \;=\; \dfrac{E}{Fa}\;,
\label{chidef}
\end{equation}
the above Schr\"odinger equation can be written as
\begin{equation}
\dfrac{d^2}{d\chi^2}\,\psi - \bigl(\chi-\varepsilon_a\bigr)\psi \;=\; 0\;.
\label{SchrodingerPositiveChi}
\end{equation}
Further shifting the variable to $\xi=\chi-\varepsilon_a$, we obtain 
\begin{equation}
\dfrac{d^2}{d\xi^2}\,\psi - \xi\psi \;=\; 0\;,
\label{AiryDiffEq}
\end{equation}
the solution to which is the Airy function \cite{AiryFunctionBook}:
\begin{eqnarray}
\psi(\xi) 
\;=\; \Ai(\xi) 
& = & \dfrac{1}{\pi}\int_0^\infty \cos\left(\dfrac{1}{3}t^3+\xi t\right) dt \cr
& = & \dfrac{1}{2\pi}\int_{-\infty}^{\infty} e^{i\left(\frac{1}{3}t^3+\xi t\right)}\,dt
\;.
\label{AiDef}
\end{eqnarray}
The other solution linearly independent from $\Ai(\xi)$ 
is the Airy function of the second kind
\begin{equation}
\Bi(\xi) \;=\; 
\dfrac{1}{\pi}\int_0^\infty
\left[\,
 \exp\left(-\dfrac{1}{3}t^3+\xi t\right)
+\sin\left(\dfrac{1}{3}t^3+\xi t\right)
\right] dt
\;,
\end{equation}
which diverges as $\xi\rightarrow\infty$, as shown in FIG.~\ref{AiryAiBiGraphs}, so is not normalizable.

\begin{figure}[t]
\includegraphics[width=6cm]{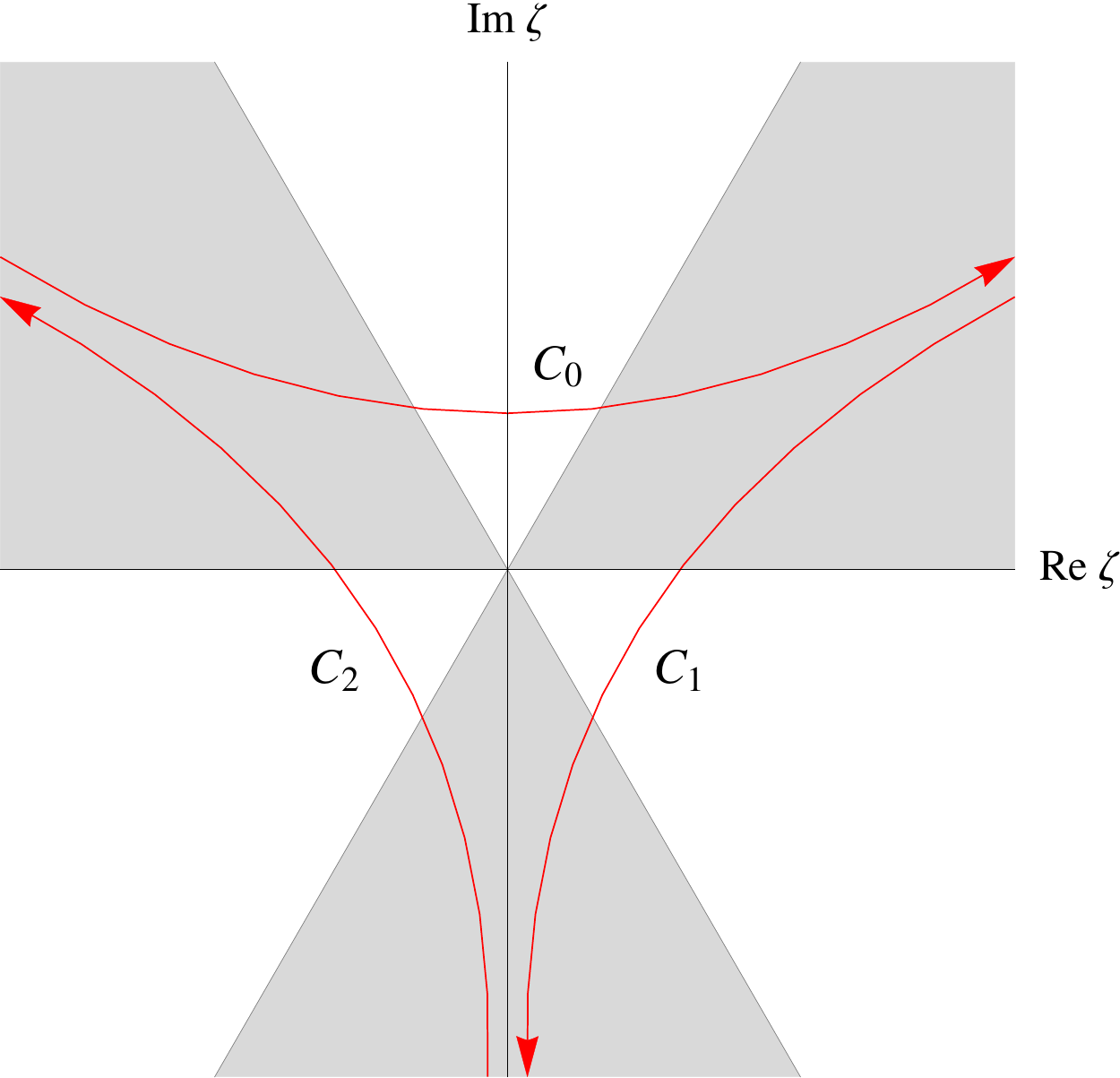}
\caption{The contours for the Fourier-Laplace transform of the Airy function
from momentum to coordinate space.
The contours continue inside the shaded areas all the way to infinity.}
\label{Fig02}
\end{figure}

\subsection{Solution in Momentum Space}

Before continuing on to determine the eigenvalues $\varepsilon_a$, 
it is instructive to see how the Schr\"odinger equation for $\hat{H}'_1$ can be solved in 
momentum space \cite{AiryFunctionBook}.


The Schr\"odinger equation for $\hat{H}'_1$ in momentum space is
\begin{equation}
\left(\dfrac{p^2}{2m}+i\hbar F\dfrac{d}{dp}\right)\tilde{\psi}(p) \;=\; E\,\tilde{\psi}(p)\;.
\end{equation}
Using the length scale $a$ defined in Eq.~(\ref{adefappendix}), 
we define the dimensionless variable
\begin{equation}
\zeta \;\equiv\; \dfrac{a\,p}{\hbar}\;.
\end{equation}
The momentum space Schr\"odinger equation in the variable $\zeta$ is
\begin{equation}
\left(\zeta^2 + i\dfrac{d}{d\zeta}\right)\tilde{\psi}(\zeta) \;=\; \varepsilon_a\,\tilde{\psi}(\zeta)\;,
\end{equation}
where $\varepsilon_a$ was defined in Eq.~(\ref{chidef}).
Being a first order differential equation, this can be solved easily to yield
\begin{equation}
\tilde{\psi}(\zeta) \;=\;
\exp\left[i\left(\dfrac{\zeta^3}{3} - \varepsilon_a\,\zeta\right)\right]\;,
\label{PsiZetaBetaZero}
\end{equation}
up to a normalization constant.
To obtain the coordinate space wave-function in $\chi$, we must Fourier transform $\tilde{\psi}(\zeta)$.


The Fourier-Laplace transform of the above function requires integration along
a contour which would render the integral finite and well-defined.
For this, the real part of the argument of the exponential must go to 
negative infinity at the end-points of the contour,
and there exist three possible contours which are shown in
FIG.~\ref{Fig02}.
Since there are three contours, 
integration along these contours gives us three functions, 
but only two of them are linearly independent since
the integral along $C_0+C_1+C_2$ is clearly zero.
Thus, though the momentum space Schr\"odinger equation 
is a first-order differential equation with only one solution, two linearly
independent solutions in coordinate space are still obtained by Fourier-Laplace 
integrals along linearly independent contours.

The integration along $C_0$ can be deformed to lie along the real $\zeta$ axis
and leads to the Airy function of the first kind:
\begin{eqnarray}
\lefteqn{\dfrac{1}{2\pi}\int_{C_0}
d\zeta\;\tilde{\psi}(\zeta)\,e^{i\chi\zeta}}
\cr
& = & \dfrac{1}{\pi}\int_{0}^{\infty}
d\zeta\;\cos\left[\dfrac{\zeta^3}{3} + (\chi-\varepsilon_a) \zeta\right]
\cr
& = & 
\Ai(\chi-\varepsilon_a)
\;.\phantom{\bigg|}
\end{eqnarray}
It should be kept in mind that we are taking the limit in which the contour $C_0$
approaches the real $\zeta$ axis from above, and the integration should be understood as such.
The Airy function of the second kind is a linear combination of the $C_1$ and $C_2$
integrals which can be deformed to lie along the real $\zeta$ and negative imaginary $\zeta$
axes:
\begin{eqnarray}
\lefteqn{\dfrac{i}{2\pi}
\left(\int_{-C_2}-\int_{-C_1}\right)
d\zeta\;\tilde{\psi}(\zeta)\,e^{i\chi\zeta}
}
\cr
& = &
\dfrac{1}{\pi}\int_0^{\infty}
d(i\zeta)\;
\exp\left[-\dfrac{(i\zeta)^3}{3} + (\chi-\varepsilon_a) (i\zeta)\right]
\cr
& & 
+\dfrac{1}{\pi}
\int_0^\infty
d\zeta\;
\sin\left[\dfrac{\zeta^3}{3} + (\chi-\varepsilon_a)\zeta\right]
\cr
& = & 
\Bi(\chi-\varepsilon_a)
\;.\phantom{\bigg|}
\end{eqnarray}
%

%
\begin{figure}[t]
\includegraphics[width=8.5cm]{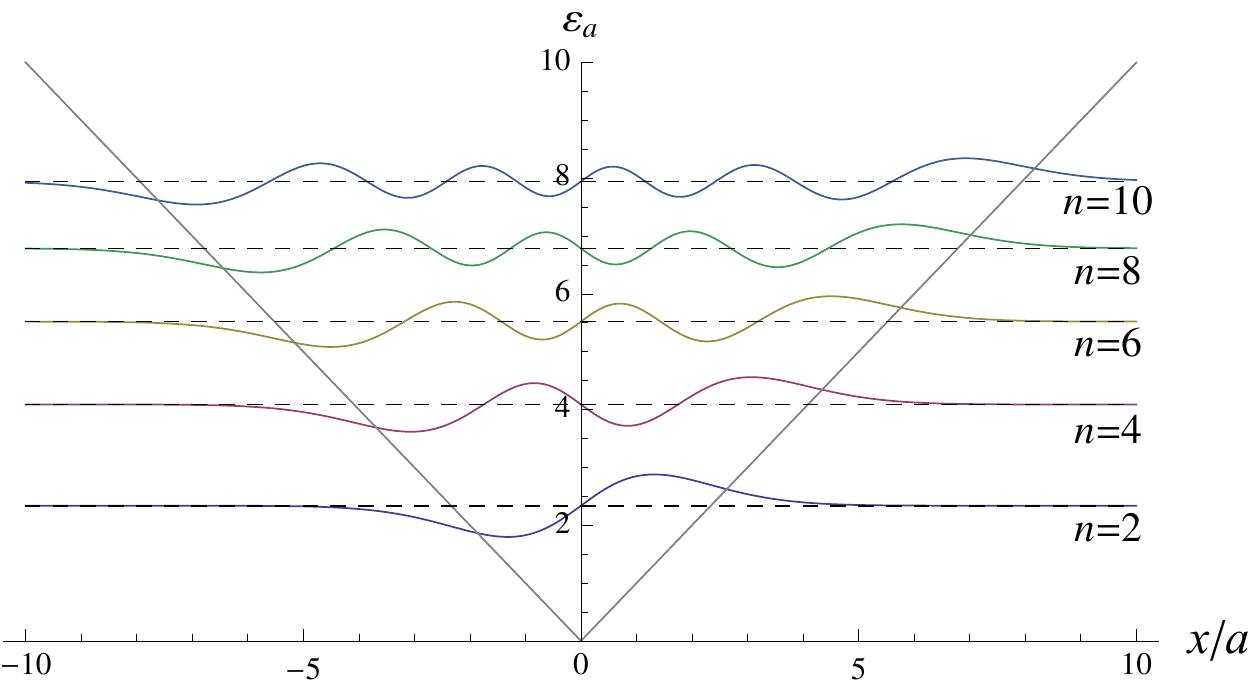}
\caption{The first five parity-odd eigenfunctions of $\hat{H}_1$.}
\label{CanonicalOddEigenfunctions}
\end{figure}

\subsection{Parity Even States}

The boundary condition $\psi(\chi=0)=0$ translates to
$\psi(\xi=-\varepsilon_a)=0$, so $-\varepsilon_a$ must be a zero point of $\Ai(\xi)$.
Let $\alpha_n$, $n=1,2,3,\cdots$, be the zero-points of $\Ai(\xi)$ arranged in descending order,
that is:
\begin{equation}
\cdots\;<\;\alpha_3\;<\;\alpha_2\;<\;\alpha_1\;<\;0\;.
\end{equation}
They are all negative, and $\alpha_n$ is encoded in Mathematica as \texttt{AiryAiZero[n]}.
Thus,
\begin{equation}
\varepsilon_{a,2n}\,=\,-\alpha_n
\quad\rightarrow\quad
E_{2n} 
\,=\, -\dfrac{\hbar^2}{2ma^2}\alpha_n
\,=\, -\alpha_n Fa
\;.
\end{equation}
Note that, despite the minus sign,
these energies are positive since the $\alpha_n$'s are negative.
These also give the energy eigenvalues for a particle in the 
half-potential Eq.~(\ref{HalfPotential}).
The corresponding eigenfunctions are
\begin{equation}
\psi_{2n}(\chi)\;\propto\;
\begin{cases}
\phantom{-}\Ai(\phantom{-}\chi+\alpha_n) & \mbox{for $\chi>0$} \\
-\Ai(-\chi+\alpha_n) & \mbox{for $\chi<0$}
\end{cases}
\end{equation}
Using the fact that $\Ai(\xi)$ is the solution to Eq.~(\ref{AiryDiffEq}),
it is straightforward to show that
\begin{equation}
\int_{-\infty}^{\infty} \bigl[\,\Ai(|\chi|+\alpha_n)\,\bigr]^2\,d\chi 
\;=\;
2\bigl[\,\Ai'(\alpha_n)\,\bigr]^2
\;,
\end{equation}
where the prime denotes differentiation.
The normalized eigenfunctions are therefore:
\begin{equation}
\psi_{2n}(\chi) \;=\; \dfrac{\mathrm{sign}(\chi)}{\sqrt{2}|\Ai'(\alpha_n)|}\;\Ai(|\chi|+\alpha_n)\;.
\end{equation}
The first five of these eigenfunctions are plotted in Fig~\ref{CanonicalOddEigenfunctions}.

\noindent
If we change the variable back to $x = a\chi$, then the wavefunction is
\begin{eqnarray}
\psi_{2n}(x) 
& = & \dfrac{\mathrm{sign}(x)}{\sqrt{2a}\,|\Ai'(\alpha_n)|}\;\Ai\left(\dfrac{|x|}{a}+\alpha_n\right)
\;.
\end{eqnarray}
%

\subsection{Parity Odd States}

\begin{figure}[t]
\includegraphics[width=8cm]{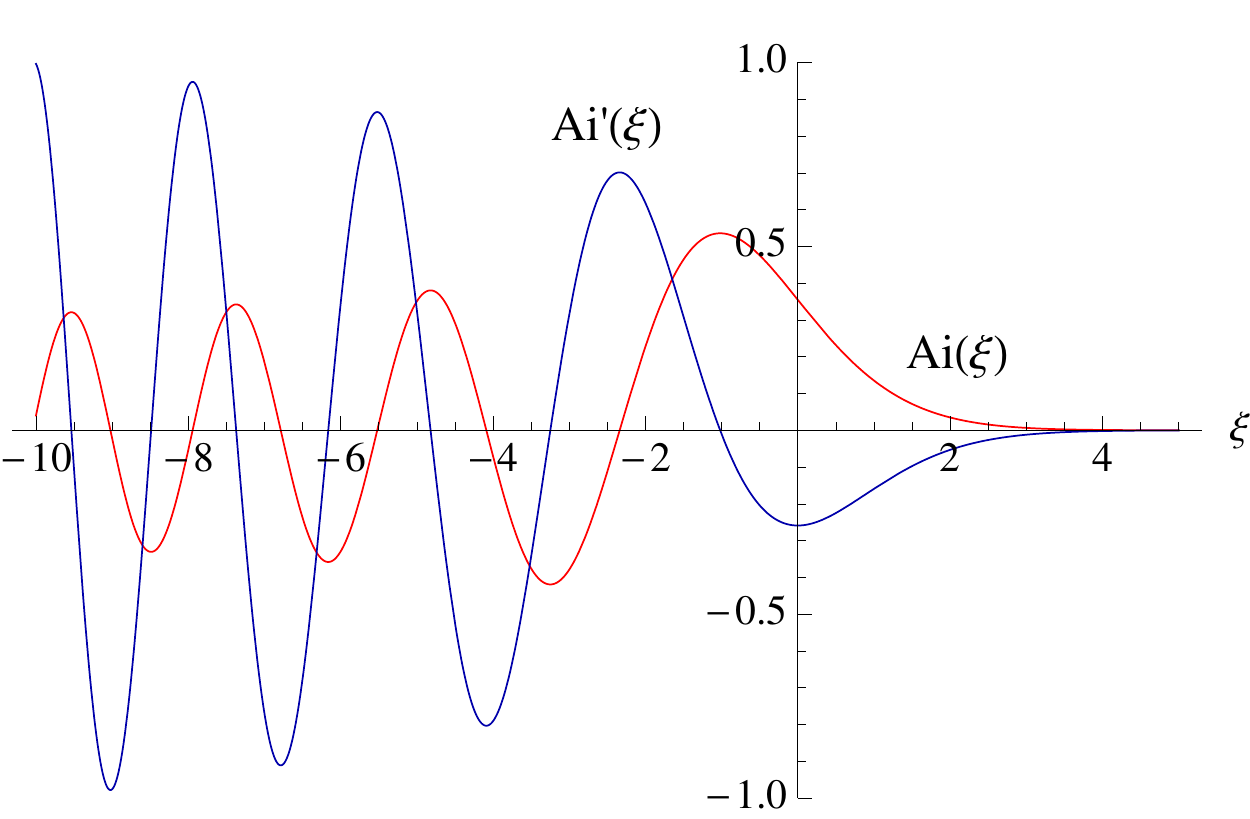}
\caption{The Airy function $\Ai(\xi)$ and its derivative $\Ai'(\xi)$.
Note that the zeroes of $\Ai(\xi)$ and $\Ai'(\xi)$ are all negative and
separate each other.
}
\label{AiryAiAiPrimeGraphs}
\end{figure}

\begin{figure}[t]
\includegraphics[width=8.5cm]{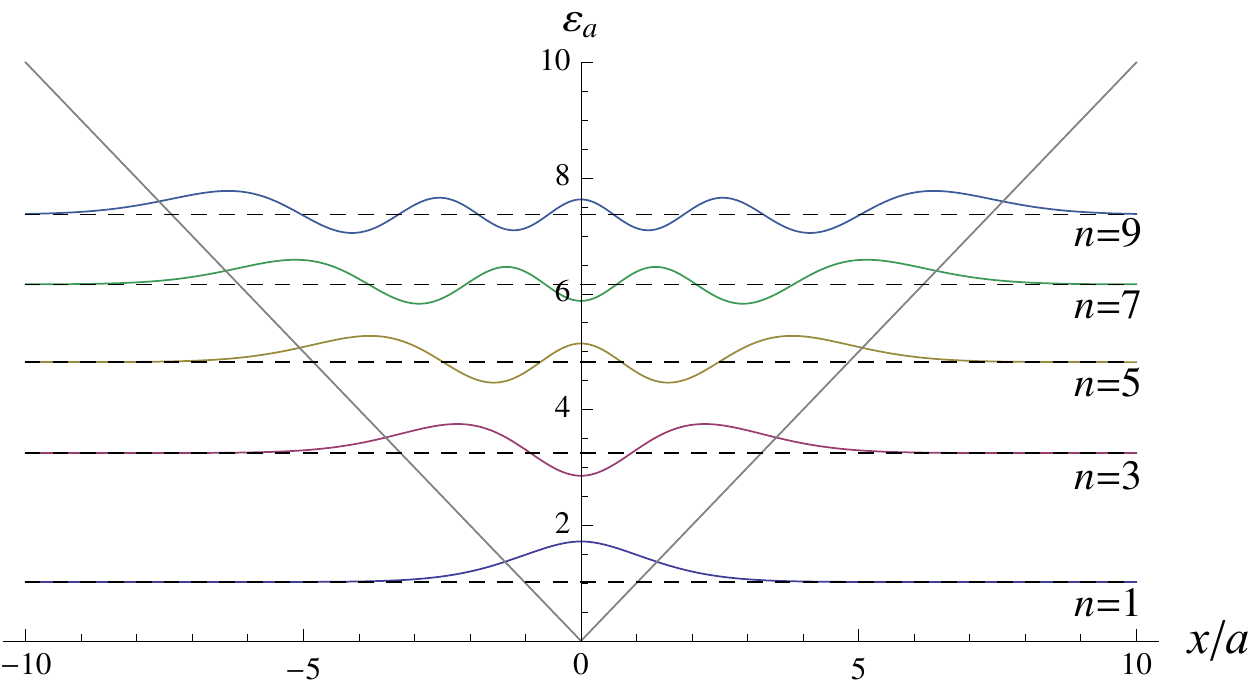}
\caption{The first five parity-even  eigenfunctions of $\hat{H}_1$.}
\label{CanonicalEvenEigenfunctions}
\end{figure}

The boundary condition $\psi'(\chi=0)=0$ translates to
$\psi'(\xi=-\varepsilon_a)=0$, so $-\varepsilon_a$ must be a zero point of $\Ai'(\xi)$.
The graphs of $\Ai(\xi)$ and $\Ai'(\xi)$ are shown in FIG.~\ref{AiryAiAiPrimeGraphs}.
Let $\beta_n$, $n=1,2,3,\cdots$, be the zero-points of $\Ai'(\xi)$ arranged in descending order,
that is:
\begin{equation}
\cdots\;<\;\beta_3\;<\;\beta_2\;<\;\beta_1\;<\;0\;.
\end{equation}
These separate the zeroes of $\Ai(\xi)$:
\begin{equation}
\beta_{n+1} \;<\; \alpha_n \;<\; \beta_{n}\;.
\end{equation}
Thus,
\begin{equation}
\varepsilon_{a,2n-1}\,=\,-\beta_n
\quad\rightarrow\quad
E_{2n-1} 
\,=\, -\dfrac{\hbar^2}{2ma^2}\beta_n
\,=\, -\beta_n Fa
\,.
\end{equation}
The corresponding eigenfunctions are
\begin{equation}
\psi_{2n-1}(\chi)\;\propto\;
\begin{cases}
\Ai(\phantom{-}\chi+\beta_n) & \mbox{for $\chi>0$} \\
\Ai(-\chi+\beta_n) & \mbox{for $\chi<0$}
\end{cases}
\end{equation}
Again, 
using the fact that $\Ai(\xi)$ is the solution to Eq.~(\ref{AiryDiffEq}),
it is straightforward to show that
\begin{equation}
\int_{-\infty}^{\infty} \bigl[\,\Ai(|\chi|+\alpha_n)\,\bigr]^2\,d\chi 
\;=\;
-2\beta_n\bigl[\,\Ai(\beta_n)\,\bigr]^2
\;.
\end{equation}
The normalized eigenfunctions are therefore:
\begin{equation}
\psi_{2n-1}(\chi) \;=\; \dfrac{1}{\sqrt{-2\beta_n}\,|\Ai(\beta_n)|}\;\Ai(|\chi|+\beta_n)\;.
\end{equation}
The first five of these eigenfunctions are plotted in Fig~\ref{CanonicalEvenEigenfunctions}.

\noindent
If we change the variable back to $x = a\chi$, then the wavefunction is
\begin{eqnarray}
\psi_{2n-1}(x) 
& = & \dfrac{1}{\sqrt{-2a\beta_n}\,|\Ai(\beta_n)|}\;\Ai\left(\dfrac{|x|}{a}+\beta_n\right)
\;.
\end{eqnarray}
%

\subsection{Expectation Values and Uncertainties}

From the symmetry of the problem, it is
clear that the expectation values of $\hat{x}$ 
and $\hat{p}$ for all the eigenstates of $\hat{H}_1$ are zero.
To calculate the expectation values of 
$\hat{x}^2$ and $\hat{p}^2$,
we will need the following relations:
%
\begin{eqnarray}
\int_{\alpha_n}^\infty d\xi\;\xi\bigl[\,\Ai(\xi)\,\bigr]^2 
& = &\int_{\alpha_n}^\infty d\xi\;\Ai(\xi)\,\Ai''(\xi) 
\cr
& = & 
\dfrac{\alpha_n}{3}\,\bigl[\,\Ai'(\alpha_n)\,\bigr]^2 
\;,\cr
\int_{\alpha_n}^\infty d\xi\;\xi^2\bigl[\,\Ai(\xi)\,\bigr]^2
& = & \dfrac{\alpha_n^2}{5}\,\bigl[\,\Ai'(\alpha_n)\,\bigr]^2 
\;,\cr
%
%
\int_{\beta_n}^\infty d\xi\;\xi\bigl[\,\Ai(\xi)\,\bigr]^2 
& = & \int_{\beta_n}^\infty d\xi\;\Ai(\xi)\,\Ai''(\xi) 
\cr
& = &
-\dfrac{\beta_n^2}{3}\,\bigl[\,\Ai(\beta_n)\,\bigr]^2 
\;,\cr
\int_{\beta_n}^\infty d\xi\;\xi^2\bigl[\,\Ai(\xi)\,\bigr]^2
& = & \dfrac{1-\beta_n^3}{5}\,\bigl[\,\Ai(\beta_n)\,\bigr]^2 
\;.
\end{eqnarray}
These relations lead to
\begin{eqnarray}
\dfrac{\vev{\hat{x}^2}_{2n}}{a^2}
& = & \dfrac{8}{15}\alpha_n^2\;,
\cr
\dfrac{a^2\vev{\hat{p}^2}_{2n}}{\hbar^2}
& = & -\dfrac{\alpha_n}{3}\;,
\cr 
\dfrac{\vev{\hat{x}^2}_{2n-1}}{a^2}
& = & \dfrac{8\beta_n^3-3}{15\beta_n}\;,
\cr
\dfrac{a^2\vev{\hat{p}^2}_{2n-1}}{\hbar^2}
& = & -\dfrac{\beta_n}{3}\;.
\end{eqnarray}
Therefore:
\begin{eqnarray}
\dfrac{\Delta x_{2n}}{a}
& = & \dfrac{\sqrt{\vev{\hat{x}^2}_{2n}}}{a}
\;=\; \sqrt{\dfrac{8\alpha_n^2}{15}}\;,
\cr
\dfrac{a\,\Delta p_{2n}}{\hbar}
& = & \dfrac{a\sqrt{\vev{\hat{p}^2}_{2n}}}{\hbar}
\;=\; \sqrt{-\dfrac{\alpha_n}{3}}\;,
\cr
\dfrac{\Delta x_{2n-1}}{a}
& = & \dfrac{\sqrt{\vev{\hat{x}^2}_{2n-1}}}{a}
\;=\; \sqrt{\dfrac{8\beta_n^3-3}{15\beta_n}}\;,
\cr
\dfrac{a\,\Delta p_{2n-1}}{\hbar}
& = & \dfrac{a\sqrt{\vev{\hat{p}^2}_{2n-1}}}{\hbar}
\;=\; \sqrt{-\dfrac{\beta_n}{3}}\;,
\end{eqnarray}
and
\begin{eqnarray}
\dfrac{\Delta x_{2n}\,\Delta p_{2n}}{\hbar}
& = & \sqrt{\dfrac{8(-\alpha_n)^3}{45}}
\;,\cr
\dfrac{\Delta x_{2n-1}\,\Delta p_{2n-1}}{\hbar}
& = & \sqrt{\dfrac{8(-\beta_n)^3+3}{45}}
\;.
\end{eqnarray}
For the ground state $n=1$, we have
\begin{equation}
-\beta_1 \;\approx\; 1.01879\;,
\end{equation}
and
\begin{equation}
\dfrac{\Delta x_1\,\Delta p_1}{\hbar}
\;=\; \sqrt{\dfrac{8(-\beta_1)^3+3}{45}}
\;\approx\; 0.5046 \;>\;\dfrac{1}{2}\;.
\end{equation}
Note that:
\begin{equation}
\begin{cases}
a\rightarrow 0      \quad & \mbox{as $mF\rightarrow\infty$}\;, \\
a\rightarrow \infty \quad & \mbox{as $mF\rightarrow 0$}\;.
\end{cases}
\end{equation}
Therefore,
\begin{equation}
\begin{cases}
\Delta x_n\rightarrow 0\;,\;\Delta p_n\rightarrow \infty \quad & \mbox{as $mF\rightarrow\infty$}\;, \\
\Delta x_n\rightarrow \infty\;,\;\Delta p_n\rightarrow 0 \quad & \mbox{as $mF\rightarrow 0$}\;. 
\end{cases}
\end{equation}
%

\bigskip
\section{Sums involving the Bateman Function}\label{BatemanProperties}

The definition of the Bateman function is given in Eq.~(\ref{BatemanFunction}).
Using Eq.~(\ref{exptanFS}), we find
\begin{eqnarray}
e^{i(\mu-\lambda)\tan\theta}
& = & e^{i\mu\tan\theta}e^{-i\lambda\tan\theta}
\vphantom{\Big|}
\cr
& = & 
\left[
\sum_{s=0}^\infty k_{2s}(\mu)\, e^{2is\theta}
\right]
\left[
\sum_{t=0}^\infty k_{2t}(\lambda)\, e^{-2it\theta}
\right]
\cr
& = & \sum_{s=0}^\infty \sum_{t=0}^\infty
k_{2s}(\mu)\,k_{2t}(\lambda)\,e^{2i(s-t)\theta}
\;.
\label{kkprod}
\end{eqnarray}
Setting $\mu=\lambda$ and
integrating both sides of this relation from $\theta=-\pi/2$ to $\theta=\pi/2$, we find
\begin{eqnarray}
\pi & = &  
\sum_{s=0}^\infty \sum_{t=0}^\infty
k_{2s}(\mu)\,k_{2t}(\mu)
\underbrace{\int_{-\pi/2}^{\pi/2} d\theta\;e^{2i(s-t)\theta}}_{\displaystyle \pi\delta_{st}}
\cr
& = & \pi\sum_{s=0}^\infty \bigl[\, k_{2s}(\mu) \,\bigr]^2
\;,
\label{kkint}
\end{eqnarray}
which indicates that
\begin{equation}
\sum_{s=0}^\infty \bigl[\, k_{2s}(\mu) \,\bigr]^2 \;=\; 1\;.
\label{kknorm}
\end{equation}
Next, using Eqs.~(\ref{exptanFS}) and (\ref{exptanprime}),
we find
\begin{eqnarray}
\lefteqn{
\biggl[
\dfrac{\mu}{\cos^2\theta}\,e^{i\mu\tan\theta}
\biggr]
\biggl[
e^{-i\lambda\tan\theta}
\biggr]
}
\cr
& = & 
\biggl[
\sum_{s=0}^{\infty}(2s)\,k_{2s}(\mu)\,e^{2is\theta}
\biggr]
\biggl[
\sum_{t=0}^{\infty}k_{2t}(\lambda)\,e^{-2it\theta}
\biggr]
\cr
& = &
\sum_{s=0}^{\infty}\sum_{t=0}^{\infty}
(2s)\,k_{2s}(\mu)\,k_{2t}(\lambda)\,e^{2i(s-t)\theta}\;.
\end{eqnarray}
Setting $\mu=\lambda$ and integrating from $\theta=-\pi/2$ to
$\theta=\pi/2$, the right-hand-side becomes
\begin{equation}
\pi\sum_{s=0}^\infty (2s)\bigl[\,k_{2s}(\mu)\,\bigr]^2
\;.
\end{equation}
The left-hand-side is however,
\begin{equation}
\mu\int_{-\pi/2}^{\pi/2}\dfrac{d\theta}{\cos^2\theta}
\;=\; \infty\;,
\end{equation}
due to the singularities at $\theta=\pm\pi/2$.
Therefore,
\begin{equation}
\sum_{s=0}^\infty (2s)\bigl[\,k_{2s}(\mu)\,\bigr]^2 \;=\;\infty\;.
\end{equation}
Similarly, we find

\begin{eqnarray}
\lefteqn{
\biggl[
\dfrac{\mu}{\cos^2\theta}\,e^{i\mu\tan\theta}
\biggr]
\biggl[
\dfrac{\lambda}{\cos^2\theta}\,e^{-i\lambda\tan\theta}
\biggr]
}
\cr
& = & 
\biggl[
\sum_{s=0}^{\infty}(2s)\,k_{2s}(\mu)\,e^{2is\theta}
\biggr]
\biggl[
\sum_{t=0}^{\infty}(2t)k_{2t}(\lambda)\,e^{-2it\theta}
\biggr]
\cr
& = &
\sum_{s=0}^{\infty}\sum_{t=0}^{\infty}
(2s)(2t)\,k_{2s}(\mu)\,k_{2t}(\lambda)\,e^{2i(s-t)\theta}\;.
\end{eqnarray}
Setting $\mu=\lambda$ and integrating from $\theta=-\pi/2$ to $\theta=\pi/2$
shows that
\begin{equation}
\sum_{s=0}^\infty (2s)^2\bigl[\,k_{2s}(\mu)\,\bigr]^2 \;=\;\infty\;.
\end{equation}

\bigskip
\section{Benczik's Solution}\label{Sandor}

Here, we review the approach used by Benczik in his Ph.D. thesis \cite{Benczik:2007we}
to solve for the eigenvalues of a particle in the half potential, Eq.~(\ref{HalfPotential}).

Using operators which obey the canonical commutation relation 
$[\,\hat{q},\,\hat{p}\,]=i\hbar$,
the operators which obey Eq.~(\ref{CommutationRelation})
can be expressed as
\begin{eqnarray}
\hat{x} & = & \hat{q}
+ \beta\;\dfrac{\hat{p}^2 \hat{q}+\hat{q}\,\hat{p}^2}{2}
\;,\cr
\hat{p} & = & \hat{p}\;.
\end{eqnarray}
Following Benczik we use the representation
\begin{equation}
\hat{q}\;=\;q\;,\quad \hat{p}\;=\;\dfrac{\hbar}{i}\dfrac{\partial}{\partial q}\;,
\end{equation}
in which case $\hat{x}$ and $\hat{p}$ are represented by
\begin{eqnarray}
\hat{x} & = & q\left(1-\hbar^2\beta\,\dfrac{d^2}{dq^2}\right)
-\hbar^2\beta\,\dfrac{d}{dq}\;,\cr
\hat{p} & = & \dfrac{\hbar}{i}\,\dfrac{\partial}{\partial q}\;.
\end{eqnarray}
The last term in the expression for $\hat{x}$ can be dropped at the expense of
changing the weight function in the definition of the inner product, that is,
we can use
\begin{equation}
\hat{x} \;=\; q\left(1-\hbar^2\beta\,\dfrac{d^2}{dq^2}\right)
\end{equation}
without affecting the energy eigenvalues.
Then the range $x>0$ corresponds to $q>0$
so the Schr\"odinger equation in that range is
\begin{equation}
\hat{H}\psi \;=\; -\dfrac{\hbar^2}{2m}\,\dfrac{d^2\psi}{dq^2}
\,+\, Fq\left(1-\hbar^2\beta\dfrac{d^2}{dq^2}\right)\psi
\;=\; E\,\psi\;,
\end{equation}
or, changing the variable to the dimensionless $\chi = q/a$, we have
\begin{equation}
\left(1+\kappa^2\,\chi\right)\dfrac{d^2\psi}{d\chi^2} - 
\left(\chi-\varepsilon_a\right)\psi
\;=\; 0\;,
\label{zetaEq}
\end{equation}
where $\kappa=\hbar\sqrt{\beta}/a$, 
and $\varepsilon_a=E/Fa$ as in the main text.
When $\chi\gg 1$, this equation is approximately
\begin{equation}
\kappa^2\,\dfrac{d^2\psi}{d\chi^2}-\psi \;\approx\; 0\;,
\end{equation}
to which the solutions are $\psi(\chi)\sim e^{\pm\chi/\kappa}$.
To obtain a normalizable solution, we must demand that the solution behave
asymptotically as $e^{-\chi/\kappa}$.
Change variable again to $s=2(1+\kappa^2\chi)/\kappa^3$. 
Eq.~(\ref{zetaEq}) becomes
\begin{equation}
\dfrac{d^2\psi}{ds^2} +
\left(-\dfrac{1}{4} + \dfrac{\lambda}{s}\right)\psi\;=\;0\;,\qquad
\lambda\;\equiv\;\dfrac{1+\kappa^2\varepsilon}{2\kappa^3}\;.
\label{BenczikEq}
\end{equation}
This is a special form of Whittaker's differential equation which is given by
\begin{equation}
\dfrac{d^2\psi}{ds^2}
+\left(-\dfrac{1}{4}+\dfrac{\lambda}{s}-\dfrac{\mu^2-(1/4)}{s^2}\right)\psi
\;=\;0\;.
\end{equation}
The two linearly independent solutions are known as Whittaker's functions and
denoted $M_{\lambda,\mu}(s)$ and $W_{\lambda,\mu}(s)$. 
They are given by
\begin{eqnarray}
M_{\lambda,\mu}(s) & = & e^{-s/2} s^{\mu+1/2}
\;{}_1F_1\!\left(\mu-\lambda+\frac{1}{2};2\mu+1;s\right)\;,\cr
W_{\lambda,\mu}(s) & = & e^{-s/2} s^{\mu+1/2}
\;U\!\left(\mu-\lambda+\frac{1}{2};2\mu+1;s\right)\;.\cr
& &
\end{eqnarray}
Here, ${}_1F_1(\alpha;\gamma;z)$ is the confluent hypergeometric function of the first kind,
while $U(\alpha;\gamma;z)$ is the confluent hypergeometric function of the
second kind, aka Kummer's function of the second kind:
\begin{eqnarray}
\lefteqn{U(\alpha;\gamma;z)} \cr 
& = &
\dfrac{\pi}{\sin(\pi\gamma)}
\Biggl[
\dfrac{{}_1{F}_1(\alpha;\gamma;z)}{\Gamma(\gamma)\Gamma(\alpha-\gamma+1)}
\cr & & \qquad\qquad
-\dfrac{z^{1-\gamma}\,{}_1{F}_1(\alpha-\gamma+1;2-\gamma;z)}{\Gamma(\alpha)\Gamma(2-\gamma)}
\Biggr]
\;.
\cr & &
\end{eqnarray}
When $\gamma$ is a non-integer
${}_1{F}_1(\alpha;\gamma;z)$ and 
$z^{1-\gamma}\,{}_1{F}_1(\alpha-\gamma+1;2-\gamma;z)$ can be taken to be
the two linearly independent solutions to Kummer's differential equation for
hypergeometric functions. Then $\gamma$ is an integer, however,
they are not independent, and we must use 
${}_1{F}_1(\alpha;\gamma;z)$ and $U(\alpha;\gamma;z)$.\footnote{%
In Mathematica, they are encoded as
\texttt{Hypergeometric1F1[a,b,z]} and
\texttt{HypergeometricU[a,b,z]}.}

The asymptotic forms of $M_{\lambda,\mu}(s)$ and $W_{\lambda,\mu}(s)$
when $s\ll 1$ are
\begin{eqnarray}
M_{\lambda,\mu}(s) & \sim &
\Gamma(2\mu+1)
\Biggl[
\dfrac{e^{i\pi\left(\mu-\lambda+\frac{1}{2}\right)}}
      {\Gamma\!\left(\mu+\lambda+\frac{1}{2}\right)}\;e^{-s/2}\,s^\lambda
\cr & & \qquad\qquad\quad
+\dfrac{1}{\Gamma\left(\mu-\lambda+\frac{1}{2}\right)}\;e^{s/2}\,s^{-\lambda}
\Biggr]
\;,\cr
W_{\lambda,\mu}(s) & \sim &
e^{-s/2}\,s^\lambda\;,
\cr & & 
\end{eqnarray}
so the solution with the correct asymptotic form is $W_{\lambda,\mu}(s)$.
The function $W_{\lambda,\mu}(s)$ has the property
$W_{\lambda,\mu}(s)=W_{\lambda,-\mu}(s)$, so the sign of $\mu$ is not important.
Choosing $\mu=-\frac{1}{2}$ to recover Eq.~(\ref{BenczikEq}), we obtain
\begin{equation}
\psi(s) \;\propto\;
W_{\lambda,-\frac{1}{2}}(s)
\;=\; e^{-s/2} \,U\left(-\lambda;0;s\right)\;.
\label{SolutionU}
\end{equation}
Since $\chi=0$ corresponds to $s=2/\kappa^3$,
the boundary condition we must impose is 
\begin{equation}
U(-\lambda;0;2/\kappa^3)\;=\;0\;.
\label{Uboundary0}
\end{equation}
This will determine the allowed values of $\lambda$, which in turn will determine the
energy eigenvalues.



\end{document}